\pdfoutput=1
\documentclass[aps,prd,twocolumn,superscriptaddress,preprintnumbers,nofootinbib]{revtex4-2}
\usepackage{amsmath}
\usepackage{amssymb}
\usepackage{natbib}
\usepackage{graphicx}
\usepackage{epsf}
\usepackage{subfigure}
\usepackage{color}
\usepackage{threeparttable}
\usepackage{dcolumn}
\usepackage{comment}
\usepackage{epsfig}
\usepackage{xspace}
\usepackage{multirow}
\usepackage[
colorlinks=True,
allcolors=blue,
linktocpage=true,
]{hyperref}
\usepackage{latexsym}
\usepackage{array}
\DeclareGraphicsExtensions{.jpg,.pdf,.png,.eps,.ps}

\newcommand*{\TT}{TT}
\newcommand*{\EE}{E\!E}
\newcommand*{\TE}{T\!E}
\newcommand*{\lcdm}{$\Lambda$CDM}
\newcommand*{\planck}{\textit{Planck}}
\newcommand*{\polarbear}{\textsc{polarbear}}
\newcommand*{\wmap}{\textsc{WMAP}}

\defcitealias{crites15}{C15}
\defcitealias{henning18}{H18}
\defcitealias{hivon02}{H02}

\definecolor{burntorange}{rgb}{0.8, 0.33, 0.0}

\definecolor{amber}{rgb}{1.0, 0.49, 0.0}

\newcommand{\skipt}[1]{}

\begin{document}

\title{Measurements of the $E$-Mode Polarization and Temperature-$E$-Mode Correlation of the CMB from SPT-3G 2018 Data}

% Addresses
\affiliation{Department of Physics, University of Chicago, 5640 South Ellis Avenue, Chicago, IL, 60637, USA}
\affiliation{Kavli Institute for Cosmological Physics, University of Chicago, 5640 South Ellis Avenue, Chicago, IL, 60637, USA}
\affiliation{School of Physics, University of Melbourne, Parkville, VIC 3010, Australia}
\affiliation{School of Physics and Astronomy, Cardiff University, Cardiff CF24 3YB, United Kingdom}
\affiliation{Kavli Institute for Particle Astrophysics and Cosmology, Stanford University, 452 Lomita Mall, Stanford, CA, 94305, USA}
\affiliation{SLAC National Accelerator Laboratory, 2575 Sand Hill Road, Menlo Park, CA, 94025, USA}
\affiliation{Department of Statistics, University of California, One Shields Avenue, Davis, CA 95616, USA}
\affiliation{Fermi National Accelerator Laboratory, MS209, P.O. Box 500, Batavia, IL, 60510, USA}
\affiliation{Department of Astronomy, University of Illinois at Urbana-Champaign, 1002 West Green Street, Urbana, IL, 61801, USA}
\affiliation{Department of Physics, University of California, Berkeley, CA, 94720, USA}
\affiliation{Department of Physics \& Astronomy, University of California, One Shields Avenue, Davis, CA 95616, USA}
\affiliation{High-Energy Physics Division, Argonne National Laboratory, 9700 South Cass Avenue., Argonne, IL, 60439, USA}
\affiliation{California Institute of Technology, 1200 East California Boulevard., Pasadena, CA, 91125, USA}
\affiliation{Institut d'Astrophysique de Paris, UMR 7095, CNRS \& Sorbonne Universit\'{e}, 98 bis boulevard Arago, 75014 Paris, France}
\affiliation{Department of Astronomy and Astrophysics, University of Chicago, 5640 South Ellis Avenue, Chicago, IL, 60637, USA}
\affiliation{Department of Physics, Stanford University, 382 Via Pueblo Mall, Stanford, CA, 94305, USA}
\affiliation{Enrico Fermi Institute, University of Chicago, 5640 South Ellis Avenue, Chicago, IL, 60637, USA}
\affiliation{University of Chicago, 5640 South Ellis Avenue, Chicago, IL, 60637, USA}
\affiliation{Department of Physics and McGill Space Institute, McGill University, 3600 Rue University, Montreal, Quebec H3A 2T8, Canada}
\affiliation{High Energy Accelerator Research Organization (KEK), Tsukuba, Ibaraki 305-0801, Japan}
\affiliation{NIST Quantum Devices Group, 325 Broadway Mailcode 817.03, Boulder, CO, 80305, USA}
\affiliation{Materials Sciences Division, Argonne National Laboratory, 9700 South Cass Avenue, Argonne, IL, 60439, USA}
\affiliation{Canadian Institute for Advanced Research, CIFAR Program in Gravity and the Extreme Universe, Toronto, ON, M5G 1Z8, Canada}
\affiliation{CASA, Department of Astrophysical and Planetary Sciences, University of Colorado, Boulder, CO, 80309, USA }
\affiliation{Department of Physics, University of Illinois Urbana-Champaign, 1110 West Green Street, Urbana, IL, 61801, USA}
\affiliation{Department of Physics and Astronomy, University of California, Los Angeles, CA, 90095, USA}
\affiliation{Department of Physics, Center for Education and Research in Cosmology and Astrophysics, Case Western Reserve University, Cleveland, OH, 44106, USA}
\affiliation{Department of Physics, University of Colorado, Boulder, CO, 80309, USA}
\affiliation{Physics Division, Lawrence Berkeley National Laboratory, Berkeley, CA, 94720, USA}
\affiliation{Department of Physics and Astronomy, Michigan State University, East Lansing, MI 48824, USA}
\affiliation{Three-Speed Logic, Inc., Victoria, B.C., V8S 3Z5, Canada}
\affiliation{Harvard-Smithsonian Center for Astrophysics, 60 Garden Street, Cambridge, MA, 02138, USA}
\affiliation{Dunlap Institute for Astronomy \& Astrophysics, University of Toronto, 50 St. George Street, Toronto, ON, M5S 3H4, Canada}
\affiliation{Department of Astronomy \& Astrophysics, University of Toronto, 50 St. George Street, Toronto, ON, M5S 3H4, Canada}
% Authors
\author{D.~Dutcher}
\affiliation{Department of Physics, University of Chicago, 5640 South Ellis Avenue, Chicago, IL, 60637, USA}
\affiliation{Kavli Institute for Cosmological Physics, University of Chicago, 5640 South Ellis Avenue, Chicago, IL, 60637, USA}
\author{L.~Balkenhol}
\affiliation{School of Physics, University of Melbourne, Parkville, VIC 3010, Australia}
\author{P.~A.~R.~Ade}
\affiliation{School of Physics and Astronomy, Cardiff University, Cardiff CF24 3YB, United Kingdom}
\author{Z.~Ahmed}
\affiliation{Kavli Institute for Particle Astrophysics and Cosmology, Stanford University, 452 Lomita Mall, Stanford, CA, 94305, USA}
\affiliation{SLAC National Accelerator Laboratory, 2575 Sand Hill Road, Menlo Park, CA, 94025, USA}
\author{E.~Anderes}
\affiliation{Department of Statistics, University of California, One Shields Avenue, Davis, CA 95616, USA}
\author{A.~J.~Anderson}
\affiliation{Fermi National Accelerator Laboratory, MS209, P.O. Box 500, Batavia, IL, 60510, USA}
\affiliation{Kavli Institute for Cosmological Physics, University of Chicago, 5640 South Ellis Avenue, Chicago, IL, 60637, USA}
\author{M.~Archipley}
\affiliation{Department of Astronomy, University of Illinois at Urbana-Champaign, 1002 West Green Street, Urbana, IL, 61801, USA}
\author{J.~S.~Avva}
\affiliation{Department of Physics, University of California, Berkeley, CA, 94720, USA}
\author{K.~Aylor}
\affiliation{Department of Physics \& Astronomy, University of California, One Shields Avenue, Davis, CA 95616, USA}
\author{P.~S.~Barry}
\affiliation{High-Energy Physics Division, Argonne National Laboratory, 9700 South Cass Avenue., Argonne, IL, 60439, USA}
\affiliation{Kavli Institute for Cosmological Physics, University of Chicago, 5640 South Ellis Avenue, Chicago, IL, 60637, USA}
% ...
\collaboration{SPT-3G Collaboration}
\noaffiliation

\begin{abstract}
We present measurements of the $E$-mode ($\EE$) polarization power spectrum and temperature-$E$-mode ($\TE$) cross-power spectrum of the cosmic microwave background using data collected by SPT-3G, the latest instrument installed on the South Pole Telescope.
This analysis uses observations of a 1500\,deg$^2$ region at 95, 150, and 220\,GHz taken over a four month period in 2018.
We report binned values of the $\EE$ and $\TE$ power spectra over the angular multipole range $300 \le \ell < 3000$, using the multifrequency data to construct six semi-independent estimates of each power spectrum and their minimum-variance combination.
These measurements improve upon the previous results of SPTpol across the multipole ranges $300\le\ell\le1400$ for $\EE$ and $300\le\ell\le1700$ for $\TE$, resulting in constraints on cosmological parameters comparable to those from other current leading ground-based experiments.
We find that the SPT-3G dataset is well-fit by a \lcdm{} cosmological model with parameter constraints consistent with those from \planck{} and SPTpol data.
From SPT-3G data alone, we find $H_0 = 68.8 \pm 1.5\,\mathrm{km\,s^{-1}\,Mpc^{-1}}$ and $\sigma_8 = 0.789 \pm 0.016$, with a gravitational lensing amplitude consistent with the \lcdm{} prediction ($A_L = 0.98 \pm 0.12$).
We combine the \mbox{SPT-3G} and the \planck\ datasets and obtain joint constraints on the \lcdm{} model.
The volume of the 68\% confidence region in six-dimensional \lcdm{} parameter space is reduced by a factor of $1.5$ compared to \planck{}-only constraints, with only slight shifts in central values.
We note that the results presented here are obtained from data collected during just half of a typical observing season with only part of the focal plane operable, and that the active detector count has since nearly doubled for observations made with SPT-3G after 2018.
\end{abstract}


\maketitle

%%%%%%%%%%%%%%%%%%%%%%%%%%%%%%%%%%%%
% INTRODUCTION
%%%%%%%%%%%%%%%%%%%%%%%%%%%%%%%%%%%%
\section{Introduction}
\label{sec:intro}

The cosmic microwave background (CMB) is a rich source of information about the early universe and its evolution over cosmic time.
Density fluctuations present during the epoch of baryon-photon decoupling at $z \sim 1100$ imprint a faint temperature anisotropy on the CMB,
and measurements of the angular power spectrum of these anisotropies are a pillar of the standard six-parameter \lcdm{} cosmological model.
Satellite measurements of the CMB temperature power spectrum are now cosmic variance-limited from the largest angular scales down to roughly seven arcminutes \citep{planck18-5} (corresponding to angular multipoles $\ell \lesssim 1600$), and ground-based observations extend these measurements to arcminute scales, at which point foregrounds begin to dominate over the primary CMB temperature signal \citep{louis17, reichardt20}.

The CMB anisotropies are linearly polarized at the 10\% level as a result of local quadrupole fluctuations at the surface of last scattering \citep{hu97d}.
The linear polarization map can be decomposed into two components: even-parity, curl-free ``$E$-modes" and odd-parity, divergence-free ``$B$-modes."
To first order, density fluctuations in the early universe only source $E$-modes, while $B$-modes are created by tensor perturbations, such as primordial gravitational waves, or gravitational lensing of the CMB by intervening large-scale structure \citep{seljak97, kamionkowski97b, knox02}.
In this paper we focus on the brighter $E$-mode component of this polarization.
The $E$-mode ($\EE$) polarization power spectrum and the temperature-$E$-mode ($\TE$) cross-power spectrum can provide tighter constraints on cosmological parameters than temperature data alone \citep{galli14}, and they can be measured out to smaller angular scales on account of the low fractional polarization of extragalactic foregrounds \citep{gupta19, datta19, trombetti18}, providing a powerful consistency check of \lcdm{}.

The CMB temperature and polarization power spectra have been measured over a wide range of angular scales by the \planck{} satellite \citep{planck18-5} and ground-based telescopes including the Atacama Cosmology Telescope (ACT) \citep{choi20}, BICEP/\textit{Keck} \citep{bicep2keck18}, \polarbear\ \citep{polarbear19a, polarbear20}, and the South Pole Telescope (SPT)  \citep[][hereafter H18]{henning18} \citep{sayre20}.
Several current and upcoming experiments aim to improve existing power spectrum constraints, including Advanced ACT \citep{henderson16}, BICEP3/BICEP Array \citep{ahmed14, hui18}, \polarbear{-2}/Simons Array\citep{suzuki14},  the Simons Observatory \citep{ade19}, and SPT-3G \citep{benson14}.

While the data are generally well-described by \lcdm{}, there are mild tensions in parameter constraints between small and large angular scales [\citetalias{henning18}, \citealp{aylor17, addison16}] and significant tensions between CMB measurements and late-time cosmological probes, most notably in the value of the Hubble constant $H_0$ \citep{planck18-6, riess19}.
Upcoming measurements of the high-$\ell$ CMB power spectra may shed light on the origin of these tensions.

In this paper, we present the first science results from SPT-3G, the latest survey instrument installed on the South Pole Telescope \citep{carlstrom11}.
We report measurements of the $\EE$ and $\TE$ power spectra over the angular multipole range \mbox{$300 \le \ell < 3000$} from observations of a $\sim$1500\,deg$^2$ region undertaken during a four-month period of 2018, and we present the resulting constraints on cosmological parameters.

The shortened 2018 observing season is the result of telescope downtime at the beginning of the year due to an issue with the telescope drive system, which caused damage to detector readout and rendered approximately half the focal plane inoperable.
We addressed the issue at the close of 2018 and have since seen normal performance during the 2019 and 2020 observing seasons.
Nevertheless, the data collected during 2018 is already sufficient to provide the most sensitive measurements made to date with SPT over the multipole ranges \mbox{$300 \le \ell \le 1400$} for $\EE$ and \mbox{$300 \le \ell \le 1700$} for $\TE$.
The resulting constraints on cosmological parameters from the SPT-3G 2018 power spectra improve upon those set by SPTpol \citepalias{henning18} and are competitive with those from other current leading ground-based experiments \citep{aiola20}.

This paper is organized as follows.
We begin with an overview of the SPT-3G instrument in \S\ref{sec:instrument}.
In \S\ref{sec:data_reduction} we discuss the scanning strategy of the telescope, low-level data processing, and the coadded maps.
In \S\ref{sec:power_spectrum} we detail the absolute calibration of the maps and the procedure used for obtaining unbiased measurements of power spectra.
Tests for systematic error in the data collection or processing steps are discussed in \S\ref{sec:systematics}.
The method for obtaining constraints on cosmological parameters from the power spectra measurements is detailed in \S\ref{sec:mcmc}.
We present final bandpower measurements in \S\ref{sec:bandpowers} and discuss the resulting constraints on cosmological parameters in \S\ref{sec:constraints}.

%%%%%%%%%%%%%%%%%%%%%%%%%%%%%%%%%%%%
% INSTRUMENT
%%%%%%%%%%%%%%%%%%%%%%%%%%%%%%%%%%%%
\section{The SPT-3G Instrument}
\label{sec:instrument}
% Optics and Receiver
Deployed in early 2017, SPT-3G is the third survey camera to be installed on SPT.
SPT-3G is a significant upgrade over the previous instruments, utilizing redesigned wide-field optics to increase the field of view from $\sim$1\,deg$^2$ to 2.8\,deg$^2$ and populating the 3.5$\times$ larger focal plane area with multichroic pixels.
Light rays from the 10\,m primary mirror are redirected by a 2\,m ellipsoidal secondary mirror and 1\,m flat tertiary mirror into the receiver cryostat \citep{sobrin18}, in which three 0.72\,m diameter anti-reflection-coated alumina lenses \citep{nadolski20} re-image the Gregorian focus onto the detectors.
The \mbox{SPT-3G} receiver can be divided functionally into two cryostats that share a common vacuum: an optics cryostat that contains the cold optical elements, and a detector cryostat that contains the detectors and associated readout electronics.
Each cryostat is cooled to 4\,K by its own dedicated pulse tube cooler, and the detectors are further cooled to their operating temperature of 300\,mK by a custom closed-cycle three-stage helium sorption refrigerator manufactured by Chase Research Cryogenics.\footnote{\url{http://www.chasecryogenics.com/}}
With the cooling power required by the SPT-3G instrument, the refrigerator can provide a stable base temperature of 300\,mK for approximately 17 hours before it must be raised to 4\,K for a 4.5 hour recharge cycle.

% Detectors
The 0.43\,m diameter focal plane is populated with $\sim$16,000 transition-edge sensor (TES) bolometers fabricated on ten monolithic 150\,mm silicon wafers.
Each detector wafer contains an array of 269 trichroic dual linearly polarized pixels, with each pixel consisting of a broadband sinuous antenna coupled to six TES bolometers via superconducting microstrip and in-line filters, which define the three observing frequency bands centered at 95, 150, and 220\,GHz.
This pixel architecture was originally developed for \textsc{polarbear}-2 and is also used by the Simons Observatory and LiteBIRD experiments \citep{suzuki12, suzuki18, galitzki18}.
Details of the SPT-3G detector wafer fabrication can be found in \cite{posada15, posada18} and characterization of the 2018 deployed array in \cite{dutcher18}.
The detectors are read out using a 68$\times$ frequency-domain multiplexing system jointly developed by the SPT-3G and \polarbear{-2} collaborations \citep{bender14, bender16}.

%%%%%%%%%%%%%%%%%%%%%%%%%%%%%%%%%%%%
% DATA REDUCTION
%%%%%%%%%%%%%%%%%%%%%%%%%%%%%%%%%%%%
\section{Observations and Data Reduction}
\label{sec:data_reduction}

\subsection{Observations}
\begin{figure}[ht!]
\includegraphics[width=7cm]{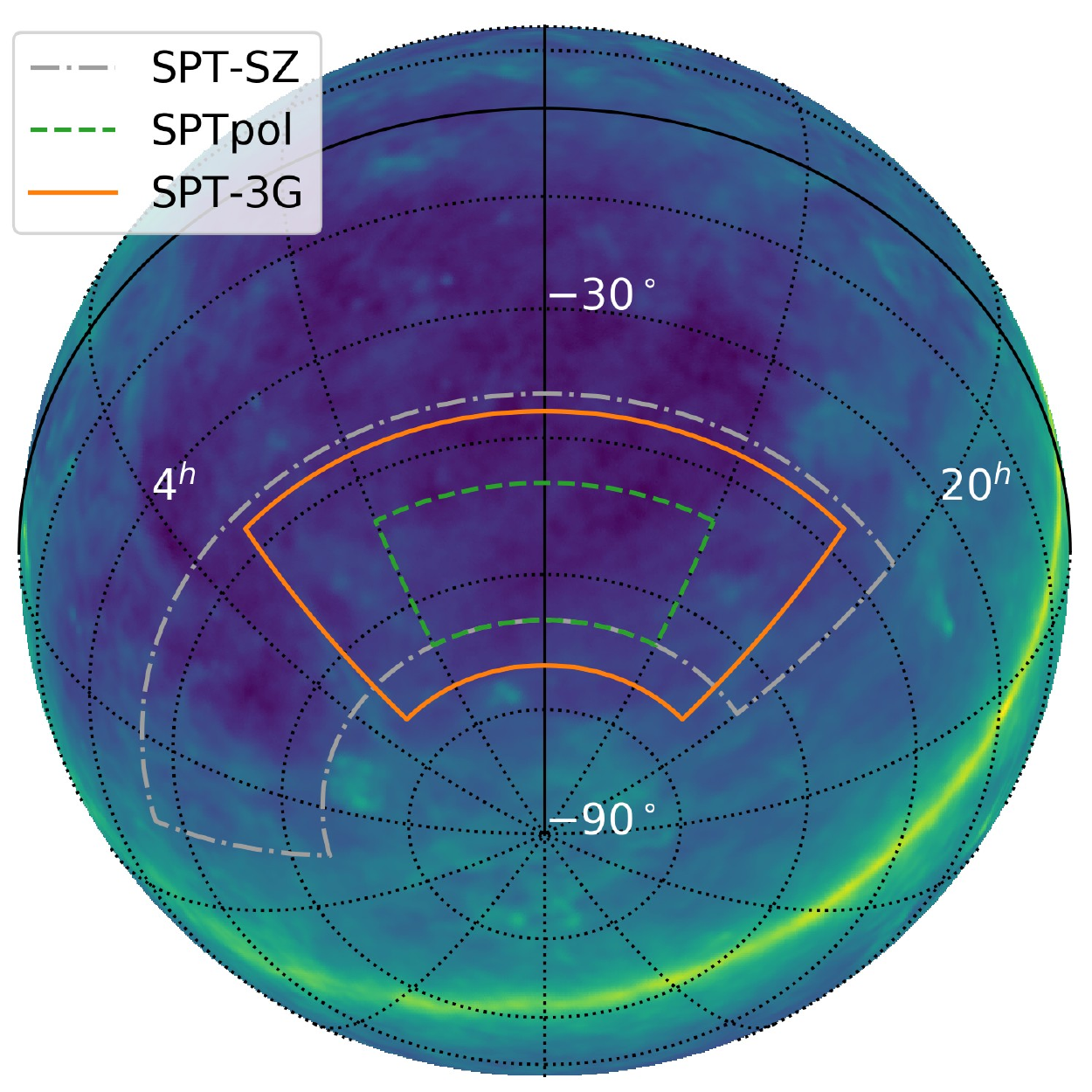}
\centering
\caption{
The SPT-3G 1500~deg$^2$ survey field \textit{(orange, solid)} overlaid on a \planck\ map of thermal dust emission \citep{planck15-10}.
Also shown are the the SPTpol 500\,deg$^2$ field \citepalias{henning18} \textit{(green, dashed)} and the SPT-SZ 2500\,deg$^2$ field \citep{story13} \textit{(gray, dot-dashed)}.
}
\label{fig:footprint}
\end{figure}

The main SPT-3G survey field is a $\sim$1500\,deg$^2$ region extending from $-42^\circ$ to $-70^\circ$ declination and from $20^\textrm{h}40^\textrm{m}0^\textrm{s}$ to $3^\textrm{h}20^\textrm{m}0^\textrm{s}$ right ascension, illustrated in Figure~\ref{fig:footprint}.
This survey footprint also overlaps the regions observed by the BICEP/\textit{Keck} series of experiments \citep{bicep2keck18, hui18}.
We observe the full 1500\,deg$^2$ via four $7.5^\circ$-tall subfields centered at $-44.75^\circ$, $-52.25^\circ$, $-59.75^\circ$, and $-67.25^\circ$ declination, respectively, with each subfield covering the full RA range.
These subfields are chosen so as to maximize telescope scanning efficiency while minimizing fluctuations in detector gain due to changes in atmospheric loading over the course of an observation.

%----------------------------------
% Survey footprints figure
%----------------------------------

As a result of the telescope's unique location at the geographic South Pole, there is nearly a direct correspondence between the local coordinates of azimuth and elevation and the celestial coordinates of right ascension and (negative) declination, respectively.
The telescope observes each subfield in a raster pattern, performing constant-elevation sweeps in azimuth before making a small step in elevation and repeating.
Each sweep of the telescope across the field, referred to as a scan, takes approximately 100 seconds to cover the full azimuth range.
The telescope performs one right-going scan and one left-going scan at each elevation step.
A full subfield observation requires approximately 2.5 hours to complete, and two subfields are each observed three times during one observing day, defined by the combined fridge hold and cycle time.
As the survey field is constantly above the horizon at the South Pole, the start of the observing day is allowed to drift with respect to sidereal time with no penalty to observing efficiency.

\subsection{Relative Calibration}
\label{sec:relcal}

We regularly conduct a series of calibration observations in order to relate the input power on each detector to CMB fluctuation temperature.
This conversion is derived from observations of two Galactic HII regions that serve as relatively compact sources of mm-wave flux, RCW38 and MAT5a (NGC 3576).
RCW38 is located at RA: $8^\textrm{h}59^\textrm{m}5^\textrm{s}$ Dec: $-47^\circ30'36''$ and is used for the two higher-declination fields, while MAT5a is located at RA: $11^\textrm{h}11^\textrm{m}53^\textrm{s}$ Dec: $-61^\circ18'47''$ and is used for the two lower-declination fields.
Dense scans are taken such that each pixel in the focal plane can form a complete map of the source; these per-detector maps are then compared to calibrated maps of RCW38 or MAT5a made by the SPT-SZ experiment.
During 2018, such observations of either RCW38 or MAT5a were nominally performed once per observing day, depending on the pair of subfields to be observed, though in later seasons the cadence has been relaxed to one dense observation per HII region per week.

Temporal calibration shifts on shorter timescales are tracked using detector response to an internal calibration source (``the calibrator") and much shorter ($\sim$10-minute) observations of the HII regions conducted before and after each CMB subfield observation.
The short HII region observations also serve to monitor changes in atmospheric opacity.
This procedure yields a conversion from input power to CMB fluctuation temperature for every detector and every observation, subject to statistical variations in the calibration observations and differences in beam shapes and passbands between \mbox{SPT-3G} and SPT-SZ.
We expect these differences to bias the absolute calibration by less than 10\%, and we correct for this bias by comparing fully coadded maps to \planck{} (see \S\ref{sec:abscal}).

\subsection{TOD Processing}
\label{sec:filtering}
We apply a series of linear processing steps to the detector time-ordered data (TOD) to decrease and flatten the noise in the signal range, which in this analysis corresponds to approximately 0.3--6\,Hz.
To reduce computing requirements, SPT-3G data is stored in a custom streaming file format\footnote{\url{https://github.com/CMB-S4/spt3g\_software}} that enables the data from only one scan of the telescope to be loaded into memory at once, and all TOD processing steps are performed on a scan-by-scan basis.
Only data taken during the constant-velocity portion of each scan is used, and the data taken while the telescope is changing direction is discarded.

To prevent high-frequency noise from aliasing down into the signal band when binning data into map pixels, we apply a Fourier-space filter with functional form $e ^ {(-\ell_x / \ell_0) ^ 6}$ and low-pass cutoff $\ell_0 = 6600$.
The relation between $\ell_x$ and temporal frequency is determined by on-sky scanning speed and is recomputed for each scan of the telescope.
We also high-pass-filter the data to remove the effects of slow signals, such as those caused by atmospheric noise or thermal drifts of the detector cold stage.
To do this, we fit and subtract up to a 19$^\mathrm{th}$-order Legendre polynomial from the TOD and project out Fourier modes corresponding to angular scales below $\ell_x = 300$.
During this step, TOD samples in which a detector was pointed within $5'$ of a point source brighter than 50\,mJy at 150\,GHz are masked in that detector's TOD to prevent filter-induced ringing artifacts in the output map.

We apply one additional filtering step, referred to as the common-mode (CM) filter, in which the signals from detectors in a specified group are averaged together, and the result is then subtracted from each of those detectors' TOD, thereby removing any common signal.
Here we use all detectors in the same frequency band on the same detector wafer to form the common mode, effectively imposing a high-pass filter that removes most of the temperature signal on scales larger than the angular extent of a wafer ($\ell \sim 500$) while largely preserving the polarization signal.
The TOD samples corresponding to point sources brighter than 50\,mJy at 150\,GHz are interpolated over during the CM filter to avoid creating spurious decrements in the map.

\subsection{Data Quality Cuts}
To prevent low-quality data from degrading a map, detectors with abnormal behavior or properties are flagged on a per-scan basis during TOD processing.
If a detector is flagged, its data is dropped from the corresponding scan.
Some of the lower-level reasons to flag a detector include a failure to properly bias or entering a fully superconducting state during an observation, poor calibration data due to noise fluctuations or detector operational issues, and readout errors during data acquisition.
An average of 448 detectors are flagged in each scan for such reasons.
We also flag detectors for irregular TOD features, on average removing an additional 342 detectors per scan due to (1) abrupt, large deviations from a rolling average, or ``glitches", with causes including cosmic ray hits and vibrations within the cryostat, or (2) excess line power in the 8--10\,Hz range, thought to originate from instability in the detector or readout circuit.

In addition to the cuts above, we do not include one of the detector wafers in this analysis, as its TOD are contaminated by a series of noise lines at multiples of 1.0\,Hz and 1.4\,Hz, the latter of which corresponds to the frequency of the pulse-tube cooler used in the cryostat.
This wafer has been replaced for subsequent observing seasons.

After filtering, an inverse-variance weight $w_i$ is computed for each detector based on the noise in its TOD from 1--4\,Hz.
The distribution of weights is examined for outliers, and detectors with weights three sigma above or below the mean are flagged, removing on average another 33 detectors from each scan.
The map for a given observation is constructed as a weighted average of the data from all detectors (after filtering and cuts) using this weight distribution.

Beyond cuts on individual detectors, whole scans are dropped from the observation data if there are errors in the telescope pointing information or if fewer than $\sim50\%$ of active bolometers pass cuts.
Entire observations are cut if there was an error with data acquisition, if all detectors were flagged (e.g., due to a failed calibration observation), or if the helium in the sorption refrigerator ran out during the observation.
After cutting 17 such observations, there are 562 subfield observations remaining, with an approximate average of 6600 active detectors equally distributed among the three frequency bands per observation.

\subsection{Maps}
We use the same mapmaking methodology as implemented for SPTpol analyses \citep{crites15, keisler15, henning18, sayre20} and described in \cite{jones07}, here binning the TOD into $2'$ square pixels using the Lambert azimuthal equal-area projection.

%----------------------------------
% TQU maps figure
%----------------------------------
\begin{figure*}[p!]
\includegraphics[width=16.5cm]{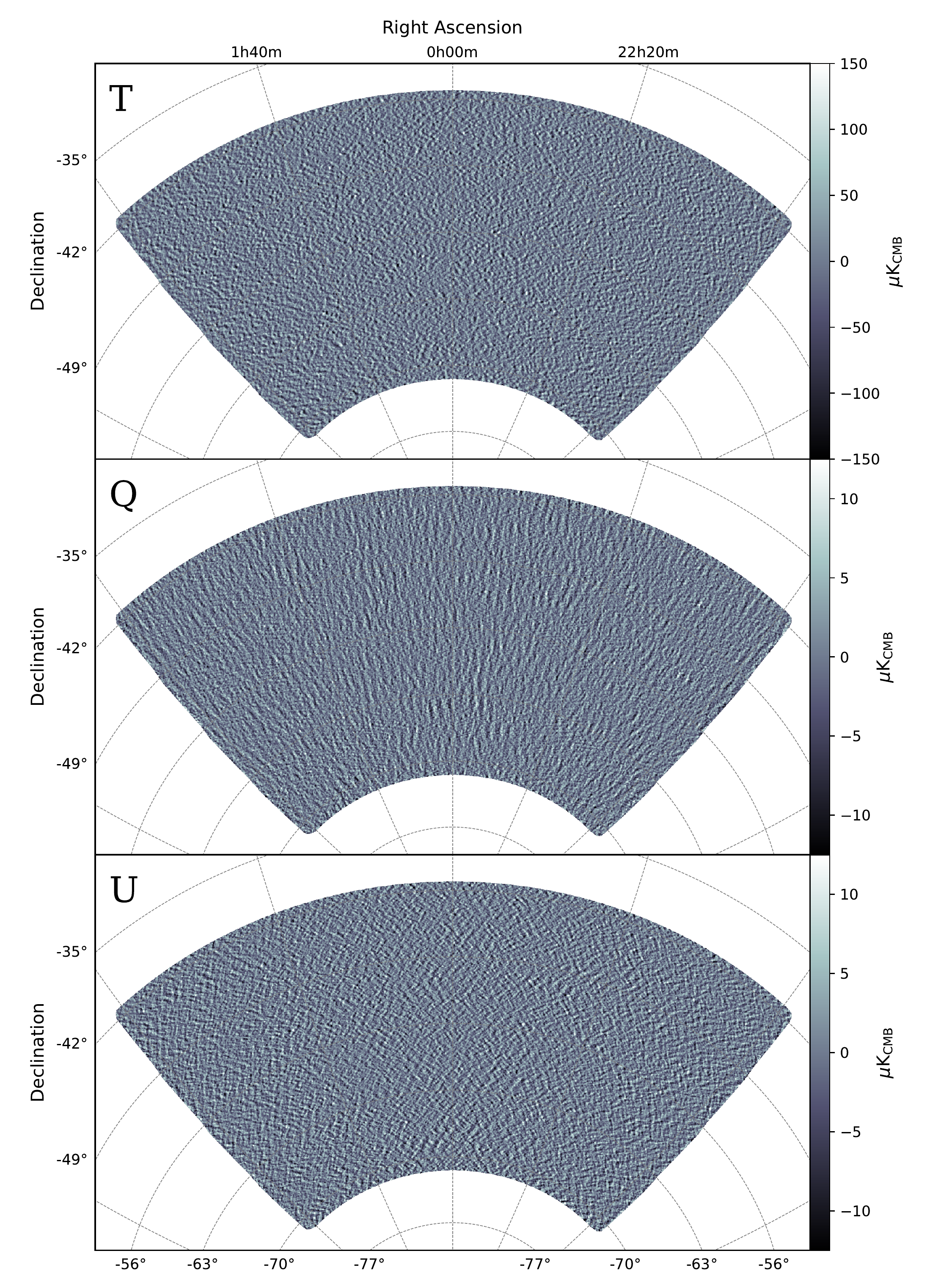}
\centering
\caption{
SPT-3G 2018 150\,GHz temperature \textit{(top)}, Stokes $Q$ \textit{(middle)}, and Stokes $U$ \textit{(bottom)} maps.
Note the factor of ten difference in color scale between temperature and polarization maps.
The data have been filtered to remove features larger than $\sim 0.5^\circ$, and the polarization maps have been smoothed by a $6'$ FWHM Gaussian.
}
\label{fig:coadded_maps}
\end{figure*}

%----------------------------------
% E map figure
%----------------------------------
\begin{figure*}[t!]
\includegraphics[width=16.5cm]{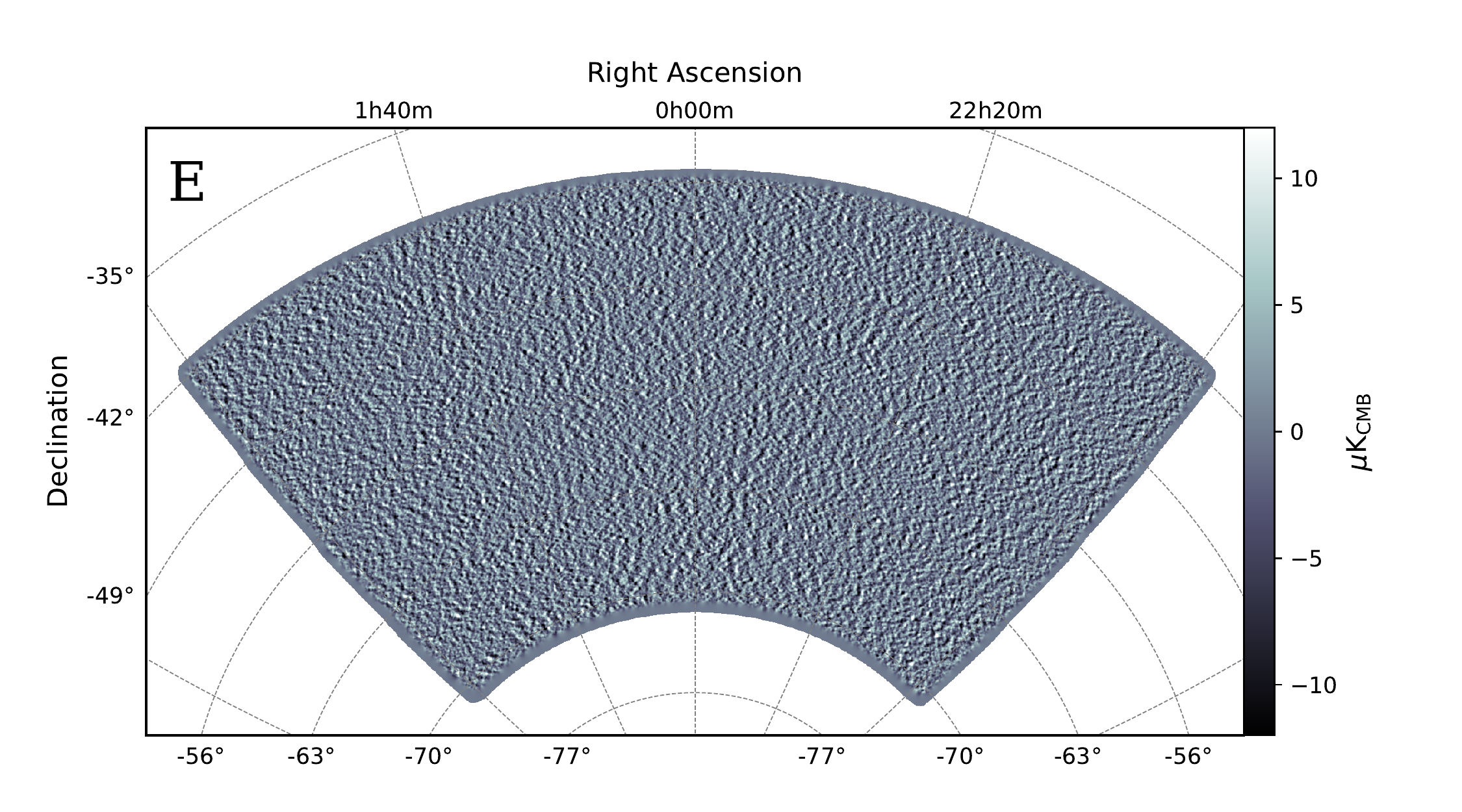}
\centering
\caption{
SPT-3G 2018 150\,GHz $E$-mode polarization map.
The data have been filtered to remove features larger than $\sim 0.5^\circ$, and the map has been smoothed by a $6'$ FWHM Gaussian.
}
\label{fig:emode_map}
\end{figure*}

%----------------------------------
% Map Depths figure
%----------------------------------
\begin{figure*}[t!]
\includegraphics[width=17.2cm]{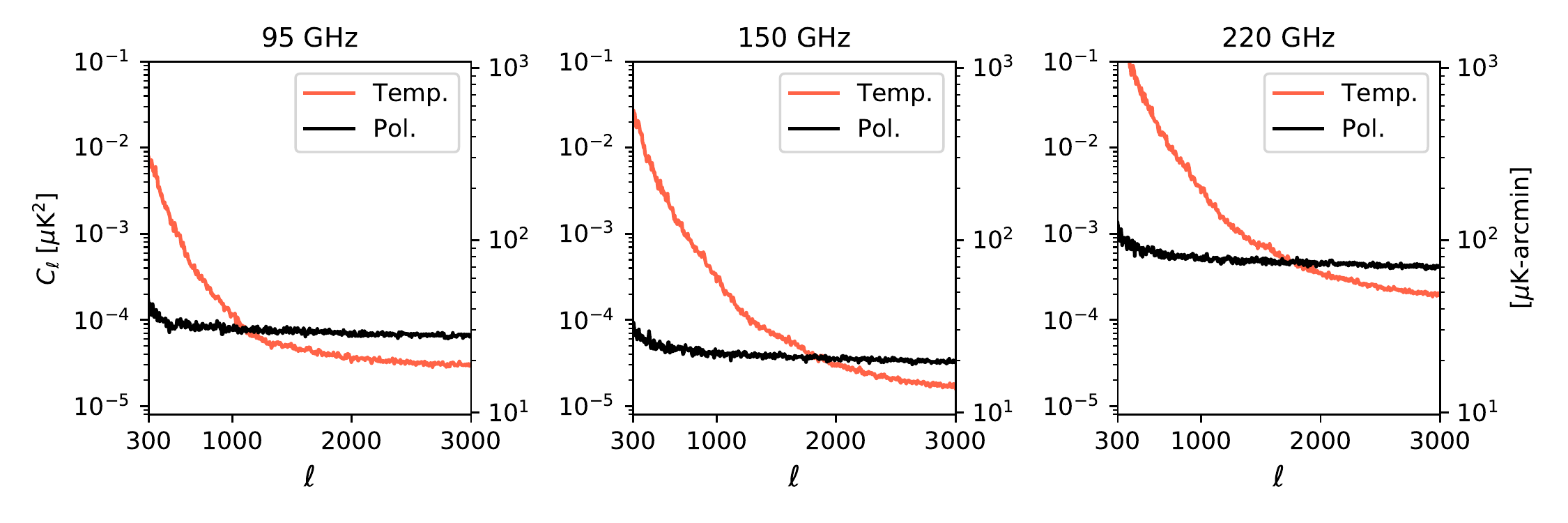}
\centering
\caption{
Temperature and polarization noise power spectra, corrected for the transfer functions of TOD processing.
In each subplot, the left-hand vertical axis displays the noise in units of $\mu$K$^2$, while the right-hand vertical axis displays the equivalent map depth in units of $\mu$K-arcmin.
}
\label{fig:map_depths}
\end{figure*}

The full-season coadded maps of temperature, Stokes $Q$, and Stokes $U$ for 150\,GHz are shown in Figure~\ref{fig:coadded_maps}.
The cross-hatched patterns in the $Q$ and $U$ polarization maps are indicative of measuring $E$-modes at high signal-to-noise.
The $E$-mode polarization map itself is shown in Figure~\ref{fig:emode_map}.
The noise levels in the coadded maps are measured by differencing two half-depth coadded maps and calculating the power spectrum of the result, correcting for the transfer function effects of the TOD filtering described above.
The map depths as a function of $\ell$ for both temperature and polarization data are shown in Figure~\ref{fig:map_depths}; averaged over the range \mbox{$1000 < \ell < 2000$}, the polarized map depths at 95, 150, and 220\,GHz are 29.6, 21.2, and 75\,$\mu$K-arcmin, respectively.

From the 562 subfield observations, we construct subsets of partial-depth full-field maps, or ``bundles", that are then used as the basic inputs to the rest of the analysis.
The bundles are constructed by chronologically coadding observations within each subfield until the combined unpolarized weight approaches $1/(N_\mathrm{bundles})^\mathrm{th}$ of the unpolarized weight in the full-season coadd, typically requiring 3--5 observations.
The coadds from each of the four subfields are then combined to create one full-field bundle.
This approach assures each bundle has approximately equal weight and even coverage of the field, to the extent allowed by the relatively small number of observations.
We chose $N_\mathrm{bundles} = 30$ to balance total number with uniformity across the bundles.

%%%%%%%%%%%%%%%%%%%%%%%%%%%%%%%%%%%%
% POWER SPECTRUM
%%%%%%%%%%%%%%%%%%%%%%%%%%%%%%%%%%%%
\section{Power Spectrum}
\label{sec:power_spectrum}

We calculate power spectra from the maps in the flat-sky approximation, in which we relate the Fourier wave numbers $(k_x, k_y)$ to angular multipole via $|\mathbf{k}| = \ell$.
We rotate curved-sky $Q$ and $U$, defined along the longitudes and latitudes on a sphere, to flat-sky $Q'$ and $U'$, defined along the vertical and horizontal axis of a flat map, by
\begin{equation}
\begin{split}
Q' &= Q\cos(2\psi_\alpha) + U\sin(2\psi_\alpha) \\
U' &= -Q\sin(2\psi_\alpha) + U\cos(2\psi_\alpha) ~,
\end{split}
\end{equation}
where $\psi_\alpha$ is the angle measured from the vertical axis to North for pixel $\alpha$ as defined by the map projection.
The Fourier transforms of the rotated $Q'$ and $U'$ maps are then used to construct the Fourier-space $E$-mode map via \citep{zaldarriaga01}
\begin{equation}
E_{\boldsymbol{\ell}} = Q'_{\boldsymbol{\ell}}\cos2\phi_{\boldsymbol{\ell}} + U'_{\boldsymbol{\ell}}\sin2\phi_{\boldsymbol{\ell}} ~,
\end{equation}
where $\boldsymbol{\ell} = (\ell_x, \ell_y)$ and $\phi_{\boldsymbol{\ell}} = \arctan(-\ell_x / \ell_y)$.

\subsection{Cross-spectra}
Following prior SPT analyses, we use the pseudo-$C_\ell$ method to compute binned power spectrum estimates, or ``bandpowers", and use a cross-spectrum approach \citep{tristram05, polenta05} to eliminate noise bias.
We compute cross-spectra between pairs of bundles by first multiplying each map by an apodization mask $\mathbf{W}$, with the product denoted as $\mathbf{m}^{X, \nu_i}_A$, where $X \in \{T, E\}$, $A$ indexes bundle number, and $i$ indexes frequency band.
We then compute sets of cross-spectra via
\begin{equation}
\tilde{D}_{b, A\!\times\!B} ^ {XY,~\nu_i \times \nu_j}
= \frac{1}{N_b} \sum _{\ell \in b} \frac {\ell(\ell+1)}{2\pi}
\mathrm{Re}\left[m^{(X, \nu_i)}_{\ell, A} m^{(Y, \nu_j)*}_{\ell, B}\right] ~,
\end{equation}
for all bundles $A \ne B$, where $N_b$ is the number of modes in each $\ell$-bin $b$.
The average of all cross-spectra for a given spectrum and frequency combination is then used to obtain the final bandpowers, $\tilde{D}_{b} ^ {XY,~\nu_i \times \nu_j}$.
As is customary, here we report power spectra using the flattened spectrum, defined as 
\begin{equation}
D_\ell \equiv \frac{\ell(\ell+1)}{2\pi}C_\ell.
\end{equation}

\subsection{Unbiased Spectra}
To obtain unbiased estimates of power spectra, we follow the MASTER algorithm \citep[][hereafter H02]{hivon02}, briefly summarized here.
The power spectra of maps constructed as described above yield estimates of the true $C_{\ell}$ that have been biased by TOD- and map-level processing.
These biased or pseudo-$C_\ell$, denoted by $\tilde{C}_\ell$, and the true $C_\ell$ are related via
\begin{equation}
\langle \tilde{C}_{\ell}\rangle = \sum_{\ell'} M_{\ell \ell'} F_{\ell'} B^2_{\ell'} \langle C_{\ell'}\rangle ~,
\label{eqn:pseudoCl}
\end{equation}
in which the brackets denote ensemble averages, $B_{\ell}$ describes the effects of the instrument beam and map pixelization, $F_{\ell}$ is a transfer function encapsulating the effects of TOD filtering, and $M_{\ell\ell'}$ is a matrix describing the mixing of power that results from incomplete sky coverage.

Following \citetalias{hivon02}, we introduce the binning operator $P_{b\ell}$ and its inverse operation $Q_{\ell b}$: if we write the binned equivalent of Eq.~\ref{eqn:pseudoCl} utilizing the shorthand \mbox{$K_{\ell \ell'} \equiv M_{\ell \ell'} F_{\ell'} B^2_{\ell'}$} and $K_{bb'} \equiv P_{b\ell}K_{\ell \ell'}Q_{\ell' b'}$, then an unbiased estimator of the true power spectrum can be calculated from the pseudo spectra via
\begin{equation}
\widehat{C_b} = K^{-1}_{bb'} P_{b' \ell'} \tilde{C}_{\ell}~.
\end{equation}
To compare the unbinned theory $C_{\ell}^{\rm th}$ to our bandpowers, we compute the binned theory spectra as $C_b^{\rm th} = W_{b\ell}C_{\ell}^{\rm th}$, where $W_{b\ell}$ are the bandpower window functions defined as
\begin{equation}
W_{b\ell} = K^{-1}_{bb'}P_{b' \ell'}K_{\ell' \ell}~.
\end{equation}

\subsection{Mask and Mode-Coupling}
Prior to computing their Fourier transforms, we multiply the maps by an apodization mask $\mathbf{W}$ to smoothly roll-off the map edges to zero and remove excess power from bright point sources.
The apodization mask is generated in much the same manner as in \citetalias{henning18}, using the same mask for all map bundles across all frequency bands.
First, a binary mask is created for each bundle by smoothing the coadded bundle weights with a $5'$ Gaussian, then setting to zero any pixels with a weight below 30\% of the median map weight.
The intersection of all the bundle masks is then edge-smoothed with a $30'$ cosine taper.
Point sources detected above 50\,mJy at 150\,GHz are masked with a $5'$ radius disk (the same size mask used during TOD processing), and the cutouts edge-smoothed with a $10'$ cosine taper.
The effective area of the final mask, defined as $\sum \mathbf{W}^2 A_{\alpha}$ where \mbox{$A_{\alpha} = 4$ arcmin$^2$} is the area of each pixel, is equal to 1614\,deg$^2$.
This area is larger than the stated survey size as a result of the inclusion of lower-weight regions along the map boundaries.

Applying a real-space apodization mask, or imposing any survey boundary, convolves the Fourier transform of the effective mask with that of the on-sky signal, coupling power between formerly independent $\ell$-modes.
This effect is encapsulated in the mode-coupling matrix $M_{\ell \ell'}$.
Previous SPT analyses have used an analytic calculation of the mode-coupling matrix in the flat-sky regime, as derived in \citetalias{hivon02} for temperature and the Appendix of \cite[][hereafter C15]{crites15} for polarization (for notational simplicity we omit the $XY$ superscript on $M_{\ell \ell'}$, though separate matrices for $\TE$ and $\EE$ are used in the analysis).
In \citetalias{henning18} this calculation was further verified for the input range $0<\ell<500$ with the use of curved-sky HEALPix\footnote{\url{http://healpix.sf.net/}} \citep{gorski05, zonca19} simulations.

Here we employ an alternate means of simulating $M_{\ell \ell'}$ that additionally captures distortions due to the map projection.
A set of HEALPix skies are generated in a similar manner as in \citetalias{henning18}, with each realization formed from an input spectrum set to zero outside of a selected $\Delta \ell=5$ bin; however, here the curved-sky maps are then reprojected to our flat map projection before applying the apodization mask.
The power spectrum is then computed in the usual manner, revealing to which multipoles the $\Delta \ell=5$ input power has been mixed.
One full realization of the mode-coupling matrix requires 640 individual simulations to cover the range $0<\ell<3200$ in increments of $\Delta \ell=5$, and 150 such realizations are averaged to obtain the final mode-coupling matrix $M_{\ell \ell'}$.

\subsection{Transfer Function}
The filter transfer function $F_\ell$ captures the effects of the filtering steps discussed in \S\ref{sec:filtering}.
$F_\ell$ is obtained through simulations, discussed further in \S\ref{sec:simulations}.
In brief, a known input spectrum $C_{\ell}^{\rm th}$ is used to generate $\mathcal{O}$(100s) of sky realizations and simulated TOD, to which are then applied the same filtering steps as on the real data.
The output spectra are then compared to the input spectra to obtain the effects of TOD filtering.

Solving Eq.~\ref{eqn:pseudoCl} for $F_\ell$ directly would necessitate inverting $M_{\ell \ell'}$, which may be ill-conditioned.
Instead, we iteratively solve for $F_\ell$ using the method prescribed in \citetalias{hivon02}:
\begin{equation}
\begin{split}
F_\ell ^{(0)} &= \frac {\langle \tilde{C}_{\ell}^{\rm sim} \rangle}
                       {w_2 B_\ell ^2 C_{\ell}^{\rm th}}~, \\
F_\ell ^ {(i+1)} &= F_\ell ^{(i)} + \frac{
    \langle \tilde{C}_{\ell}^{\rm sim} \rangle -
        M_{\ell\ell'}F_\ell ^ {(i)} B_\ell ^2 C_{\ell}^{\rm th}}
    {w_2 B_\ell ^2 C_{\ell}^{\rm th}} ~,
\end{split}
\end{equation}
where $w_2 \equiv \frac{1}{\Omega}\int d^2 r \mathbf{W}^2$ and $\Omega$ is the area of the map in steradians.
We find three iterations sufficient to achieve a stable result.

%----------------------------------
% Transfer function figure
%----------------------------------
\begin{figure}[htb!]
\includegraphics[width=8.6cm]{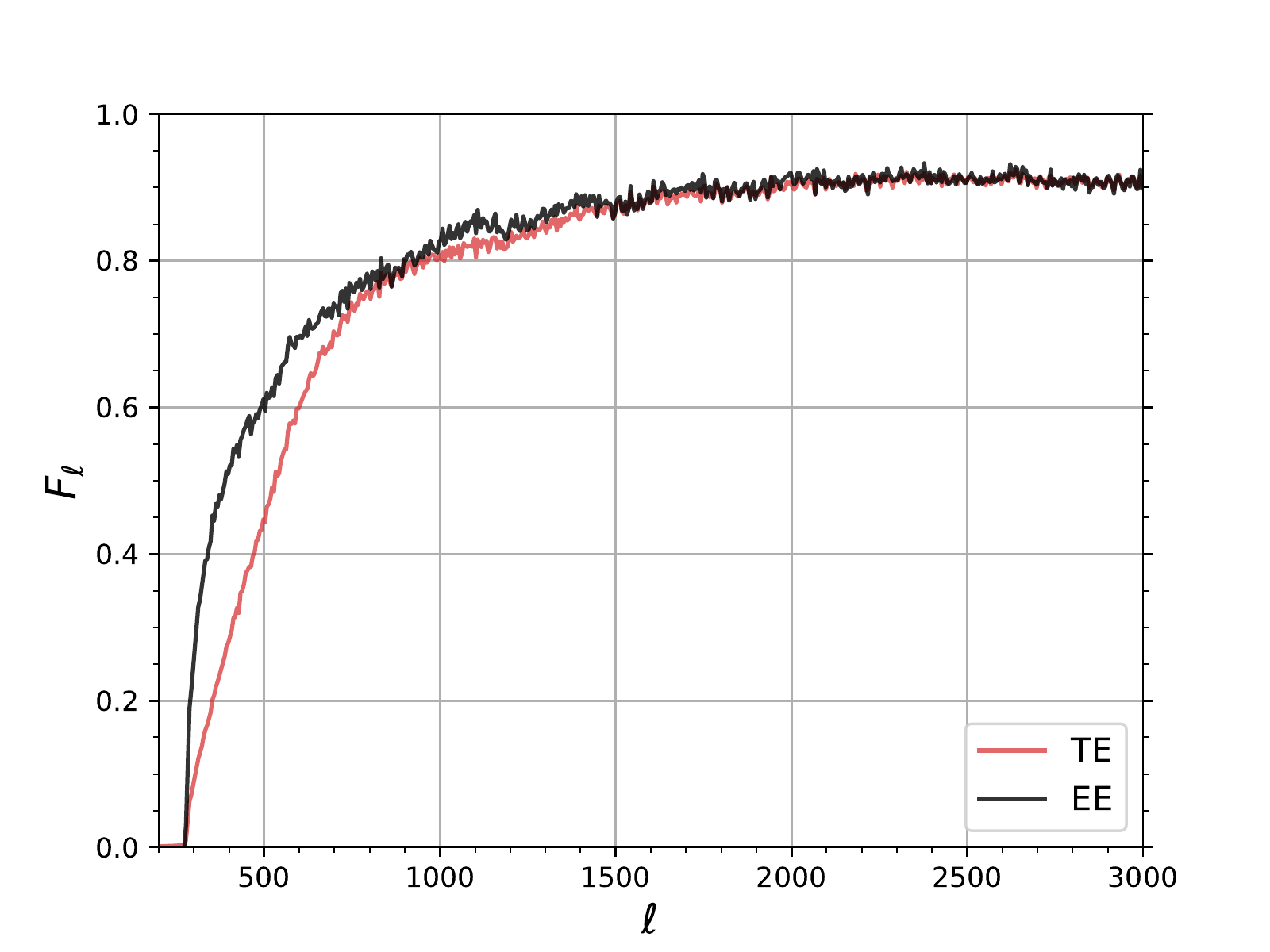}
\centering
\caption{
Filter transfer functions for 150\,GHz $\TE$ and $\EE$ power spectra, computed using 250 TOD simulations of the full SPT-3G 2018 dataset.
The difference between the $\TE$ and $EE$ transfer functions is caused by the common-mode filter.
}
\label{fig:tf}
\end{figure}

The iterative approach is unstable for the $\TE$ power spectrum due to zero crossings, so instead we use the geometric mean of the $\TT$ and $\EE$ transfer functions in the same manner as \citetalias{crites15} and \citetalias{henning18}.
For cross-frequency power spectra, a transfer function is computed directly for each $\nu_i \times \nu_j$ spectrum.
The $\TE$ and $\EE$ transfer functions for 150\,GHz are shown in Figure~\ref{fig:tf}, with similar results found for 95\,GHz and 220\,GHz.
The difference between the $\TE$ and $EE$ transfer functions primarily arises from the CM filter, which removes large-scale power from temperature while preserving it in polarization.
This also causes $~10\%$ differences in $F_\ell$ between the three frequency bands for $\ell<1000$, which diminishes to $<1\%$ at higher multipoles.

\subsubsection{Simulations}
\label{sec:simulations}
To create the simulations used for recovering the effect of TOD- and map-level processing on the data, we first generate 250 Gaussian realizations of the CMB described by the best-fit \lcdm{} model to the \textsc{base\_plikHM\_TTTEEE\_lowl\_lowE\_lensing} \planck{} data set \citep{planck18-6}.
To these we add foreground contributions using two methods.
For foreground components expected to be roughly Gaussian-distributed (such as the thermal and kinetic SZ effects), we create Gaussian realizations of power spectra from \cite{george15}.
These realizations are correlated between frequencies. 
We also add Poisson-distributed foregrounds according to source population models from \cite{dezotti05} for radio galaxies and from \cite{bethermin12} for dusty star-forming galaxies, with polarization fractions from \cite{gupta19} and flux-frequency scaling relations from \cite{everett20}.
We neglect Galactic foregrounds for these simulations, as the expected polarized power from dust within our survey region is 1--2 orders of magnitude smaller than the $E$-mode signal over the multipoles and observing frequencies considered here (Galactic dust is accounted for in the likelihood; see \S\ref{sec:mcmc}).
The $\TE$ power for all simulated foregrounds is set to zero.
These simulated components are then combined in multipole space and multiplied by a Gaussian approximation of the SPT-3G beam (see \S\ref{sec:beam}), with FWHMs of $1.7', 1.4', 1.2'$ at 95, 150, 220\,GHz, respectively, before generating real-space HEALPix sky realizations.
These noiseless mock skies are then used along with recorded telescope pointing information from every 2018 subfield observation to generate simulated detector TOD, which are then processed using the same detector cuts and filtering as applied to the real data.
The resulting ``mock observations" are then bundled and analyzed in exactly the same manner as the real data.

\subsection{Beam}
\label{sec:beam}

The beam describes the instrument response to a point source.
The maps produced are a convolution of the beam with the underlying sky, equivalently described as a multiplication in Fourier space by the beam window function $B_\ell$.
$B_\ell$ is estimated in a similar manner to the composite beam analyses in \cite{story13, crites15, keisler11}, using point sources in the 1500\,deg$^2$ field and five dedicated Mars observations taken during 2018.

The Mars data are convolved with a Gaussian estimate of the telescope pointing jitter (approximately $12"$\,rms) derived from the fitted locations of point sources in individual observations.
The brightness of Mars produces a high signal-to-noise beam template out to tens of arcminutes away from the peak response; however, we observe significant evidence for detector nonlinearity at the peak response in the planet scans.
To avoid this, the Mars maps are first produced individually for left-going and right-going scans, and any data taken in a scan after Mars passes within $\sim 1$ beam FWHM is masked, as the falling edge of the beam response is most prone to contamination from detector nonlinearity.

The hole at the location of the peak planet response is filled in by stitching a coadd of point sources that has been convolved with the Mars disk.
The stitching operation simultaneously fits a relative scale and offset between the two beam observations using an annular region where both measurements have high signal-to-noise.
$B_\ell$ is then taken to be the square-root of the azimuthal average of the 2D power spectrum of the composite map, after correcting for the planet disk and pixel window functions.
The normalization of the beam response is defined by the map calibration procedure described in \S\ref{sec:subfield_cal}.

\begin{figure}[ht!]
\includegraphics[width=8.6cm]{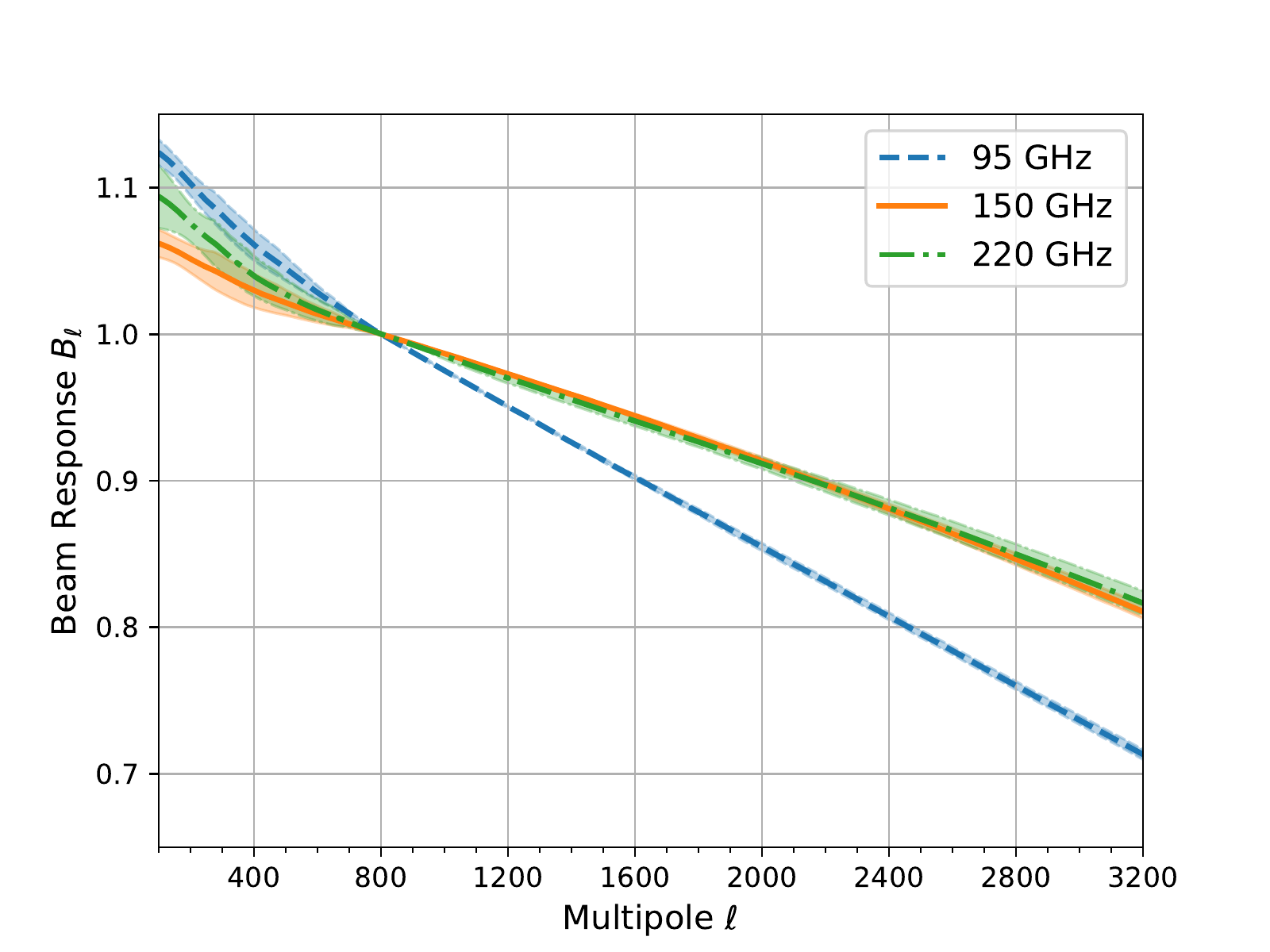}
\centering
\caption{
One-dimensional multipole-space representation of the measured instrument beam, $B_\ell$,
with uncertainties indicated by the shaded regions.
The data are normalized to unity at $\ell = 800$.}
\label{fig:b_ell}
\end{figure}

$B_\ell$ and uncertainties for the three frequencies are shown in Figure~\ref{fig:b_ell}.
Over the range of multipoles relevant for this analysis, the fractional beam uncertainty is less than 1.5\%.
The beam covariance is derived from a set of alternate $B_\ell$ curves produced by varying the subfield from which the field sources are drawn, varying which of the five planet observations is used, and sampling from the nominal covariance of the stitching scaling and offset parameters.
The beam covariance is then added to the bandpower covariance matrix, discussed in \S\ref{sec:cov}.

%----------------------------------
% Beam profile figure
%----------------------------------

\subsection{Absolute Calibration}
\label{sec:abscal}

\subsubsection{Subfield calibration}
\label{sec:subfield_cal}
As this work references separate HII regions for calibrating different halves of the survey field, we calculate and apply a temperature calibration factor for each subfield individually before coadding observations from the four subfields into a single map.
To set the individual temperature calibrations, we compute cross-spectra between our subfield temperature maps and the \planck{} PR3 maps\footnote{\url{https://pla.esac.esa.int/}} of the nearest frequency channel, using 100\,GHz, 143\,GHz, and 217\,GHz for our  95\,GHz, 150\,GHz, and 220\,GHz bands, respectively.

The \planck{} maps are mock-observed with TOD filtering identical to the real data, though with larger masked regions around point sources to account for the larger \planck\ beam.
An apodization mask with larger point source cut-outs is applied to both the mock-\planck{} and SPT maps, and the corresponding mode-coupling matrix $M^{\rm ps}_{\ell, \ell'}$ is used.
We compute the \planck{}-only and SPT-only power spectra using cross-spectra between half-depth maps from the respective experiments, and we compute the cross-spectra between the two experiments using full-depth maps.
We divide out the binned mode-mixing matrix to account for the cut sky and source masking, and compute the binned ratio of the power spectra
\begin{equation}
\epsilon_b =
\frac{P_{b,\ell}\, B^{\rm Planck}_\ell (P_{b,\ell} M^{\rm ps}_{\ell, \ell'}Q_{\ell', b'})^{-1}\, \tilde{D}_{b'} ^ {\rm SPT_1 \times SPT_2}}
{P_{b,\ell}\, B^{\rm SPT}_\ell (P_{b,\ell} M^{\rm ps}_{\ell, \ell'}Q_{\ell', b'})^{-1}\, \tilde{D}_{b'} ^ {\mathrm {SPT} \times \mathrm{Planck}}}.
\end{equation}
The average of this ratio over $400 \le \ell \le 1500$ is used to set the relative temperature calibration between subfields.
All subfield calibration factors are within $\lesssim 7\%$ of unity, consistent with the expected accuracy of the calibration procedure described in \S{\ref{sec:relcal}.

We establish uncertainties on the above ratio by combining a single \lcdm\ sky realization with FFP10 noise simulations for \planck\ and sign-flip noise realizations for SPT, generated by coadding real SPT-3G data maps with random signs.
We compute several similar ratios using other combinations of \planck\ and SPT data to form the cross-spectra as a data systematics and pipeline consistency check.
We find agreement to $\lesssim 1\%$ in the ratios across different data spectra inputs over the multipole range considered.
The beam measured in this manner also serves as cross-check of our low-$\ell$ beams; while the results are consistent with the position-space measurement, they are less sensitive as a result of the \planck\ beam size and map noise, and are therefore not used to constrain the shape of the beam response.

\subsubsection{Full-field calibration}

We determine the final calibration of the \mbox{SPT-3G} temperature and $E$-mode maps by comparing the measured SPT-3G $\TT$ and $\EE$ power spectra to the full-sky, foreground-corrected \planck{} power spectra.
Note that while the map calibration described above is expected to be accurate at the percent level, that procedure does not address the absolute amplitude of the $Q$ and $U$ polarization maps.
This motivates the $\EE$ power spectrum comparison.
While not strictly necessary, we also adjust the temperature calibration to be based on the power spectrum comparison for symmetry.

We calculate calibration factors for each frequency band for the temperature (e.g., $T_{\rm cal}^{\rm \,95\,GHz}$) and $E$-mode (e.g., $E_{\rm cal}^{\rm \,95\,GHz}$) maps.
The cross-spectra calibration factors are then $\TE \propto (T_\mathrm{cal} E_\mathrm{cal})$ and $\EE \propto (E_\mathrm{cal} E_\mathrm{cal})$.
The calibration factors are constructed based on comparing the \planck{} combined CMB-only power spectra to the SPT-3G $95\!\times\!95$, $150\!\times\!150$, and $220\!\times\!220$ bandpowers over the angular multipole range $300 \le \ell \le 1500$ using the \planck{} bin-width of $\Delta\ell=30$.
We apply the \mbox{SPT-3G} bandpower window functions to the unbinned \planck{} spectra for this comparison.
For temperature, we also account for foreground contamination by subtracting from the SPT-3G bandpowers the best-fit foreground model from \cite{reichardt20} with additional radio galaxy power from the different point source mask threshold calculated according to the model in \cite{dezotti05}.
The foreground corrections are negligible for the $\EE$ spectra.
We account for the uncertainties on the bandpower measurements in this comparison using the covariance described in \S\ref{sec:cov} as well as the uncertainties on the \planck{} spectra.
We also include the correlated uncertainties in the calibration factors due to the overall \planck{} absolute calibration uncertainty (taken to be 0.25\% at the map level) and the common sample variance and \planck{} noise uncertainty across the three frequencies for the $\EE$ and $\TT$ comparisons.

The adjustments to the $T_{\rm cal}$ factors recomputed in this manner are all within $\sim$1\% of unity, while the $E_{\rm cal}$ factors, which may be thought of as the inverse of the effective polarization efficiencies, are 1.028, 1.057, and 1.136 for 95, 150, and 220\,GHz, respectively.
That $E_{\rm cal}$ is a larger correction than $T_{\rm cal}$ is to be expected, as we do not have per-detector measurements of polarization properties, and instead rely on the as-designed values.
We note that despite this, the polarization calibration factors found here are of roughly the same size as those required for SPTpol in \citetalias{crites15} and \citetalias{henning18}, which did make use of such per-detector polarization information.

The calibration factors are applied to the maps before calculation of the final bandpowers, and we include all six calibration parameters as nuisance parameters in the likelihood when fitting for cosmology, using priors centered on unity and with widths based on the calculated covariance matrix.
The uncertainties on the six calibration parameters are given alongside those of other nuisance parameters in \S\ref{sec:mcmc}.

\subsection{T-to-P Leakage}
\subsubsection{Monopole deprojection}
Polarization data can be contaminated by leaked temperature signal caused by a variety of factors, including mismatched gain between detectors in a polarization pair and differential beam shapes.
As in \citetalias{crites15} and \citetalias{henning18}, we perform a monopole deprojection, in which a scaled copy of the $T$ map is removed from the $Q$ and $U$ maps.
We neglect higher-order leakage terms, as they typically become relevant near the beam scale ($\ell \sim 11000$), while this analysis extends only to $\ell=3000$.

In both \citetalias{crites15} and \citetalias{henning18}, the monopole leakage coefficients $\epsilon^{P}$, where $P \in \{Q, U\}$, were calculated by directly comparing the respective $C^{T\!P}_\ell$ to $C^{\TT}_\ell$ over some range of $\ell$, and the deprojected maps obtained via $P' = P-\epsilon^{P}T$.
The same method used in this analysis would be biased by the high-pass TOD filter, due to the following effect.
In the 2D Fourier plane, $QQ$ power is oriented along the $\ell_x$ and $\ell_y$ axes while $UU$ power is oriented at 45$^\circ$.
As the temperature signal is uncorrelated with $Q$ and $U$ across the sky, the azimuthal average of the $TQ$ and $TU$ correlations should be zero (i.e., at each $\ell$, the orthogonal lobes of power in the 2D Fourier plane are of equal magnitude but opposite sign).
However, as the telescope scanning direction is along $\ell_x$, the high-pass filter removes power from low-$\ell_x$ modes, leaving a residual signal in the $TQ$ azimuthal average that is highly correlated with $\TE$.
As $TU$ modes are oriented primarily at 45$^\circ$ in the 2D Fourier plane, the loss of $\ell_x < 300$ power does not change their net-zero azimuthal average.

To account for the correlation with $\TE$, we fit each of $TQ$ and $TU$ to a linear combination of $\TE$ and $TT$ according to:
\begin{equation}
C_\ell ^ {T\!P} = \epsilon^{P, TT}C_\ell ^ {TT} + \epsilon^{P, \TE}C_\ell^{\TE}~.
\label{eq:tp_leakge}
\end{equation}
The $\epsilon^{P, TT}$ coefficients are then used for monopole deprojection in the usual fashion, while the $\epsilon^{P, \TE}$ values are discarded.

Two tests of this deprojection method are performed before application to data.
First we check that the $\epsilon^{P, TT}$ coefficients are consistent with zero in noiseless mock observations.
Then, a known amount of $T$-to-$P$ leakage is injected in the simulations to verify it can be recovered.
After passing both of these checks, we calculate the leakage coefficients from real data, obtaining the values in Table~\ref{tab:tp_leakage}.
We perform the deprojection on the data, though the resulting shift in bandpowers is entirely negligible given the reported bandpower uncertainties.
We accordingly neglect the error on the monopole leakage terms.

\begin{table}[ht]
\def\arraystretch{1.5}
\setlength{\tabcolsep}{7pt}
\centering
\begin{tabular}{c  c  c  c }
\hline\hline
 & 95\,GHz & 150\,GHz & 220\,GHz\\
\hline
$\epsilon^{Q, TT}$ & 0.006 $\pm$ 0.002 & 0.005 $\pm$ 0.002 & 0.008 $\pm$ 0.010 \\
$\epsilon^{U, TT}$ & 0.008 $\pm$ 0.002 & 0.013 $\pm$ 0.002 & 0.015 $\pm$ 0.010 \\
\hline
\end{tabular}
\caption{
$T$-to-$P$ monopole leakage coefficients.
}
\label{tab:tp_leakage}
\end{table}

\subsubsection{Leakage from the common-mode filter}
Another form of $T$-to-$P$ leakage results from the CM filter.
As the polarized power is measured using the difference in signal between orthogonally polarized detectors, subtracting the same common mode from all detectors should not affect the measured polarization.
However, here we have not enforced explicit pair-differencing when making polarized maps, allowing the polarized signal in a given map pixel to be formed from detectors in physically distant focal plane pixels.
The CM filter generally removes a different amount of power from two such detectors, thereby affecting the polarization signal.
While the CM filter is empirically seen to reduce polarization noise, it also directly injects some fraction of the $\ell \sim 500$ (corresponding to the angular extent of a detector wafer) temperature power into the polarization maps.
To quantify this leakage, we mock-observe a set of $T$-only simulations and measure the power leaked into $\EE$ and $\TE$.
We find the leakage to depend on the particular configuration of detectors used to form the CM, differing in both sign and magnitude across the three frequency bands, with maximum amplitudes near $\ell=500$ of 0.20\,$\mu$K$^2$ for $\EE$ and 10\,$\mu$K$^2$ for $\TE$.

This CM filter-induced $T$-to-$P$ leakage is also present in the simulations used to obtain the filter transfer function.
Although $F_\ell$ is a multiplicative correction, and this $T$-to-$P$ leakage is an additive bias, to first order $F_\ell$ already removes this leakage;
when reconstructing the input $D_{\ell, \mathrm{th}}^{\EE}$ from simulated $\tilde{D}_{\ell}^{\EE}$ using Eq.~\ref{eqn:pseudoCl}, no residual bias is seen.
As will be discussed in \S\ref{sec:systematics}, realistic changes to the input spectra used for the simulations do not significantly affect $F_\ell$, so this bias will already be reduced to a negligible level for $\EE$ data.

The leakage in $\TE$ is not handled so easily, however, as $F_\ell ^{\TE}$ is not constructed specifically from $\TE$ spectra, but rather as the geometric mean of $F_\ell ^{TT}$ and $F_\ell ^{\EE}$.
When reconstructing the input $D_{\ell, \mathrm{th}}^{\TE}$ from simulated $\tilde{D}_{\ell}^{\TE}$ using Eq.~\ref{eqn:pseudoCl}, a residual bias remains.
The same set of simulations for obtaining $F_\ell$ is used to calculate the following residual $\TE$ bias, which is then subtracted from the data:
\begin{equation}
\TE_\mathrm{bias} = \tilde{D}_{\ell, sim} ^ {\TE} - \sum_{\ell'} M_{\ell \ell'} F_{\ell'}^{\TE}B_{\ell'}^2 D_{\ell, \mathrm{th}} ^ {\TE} ~.
\end{equation}
In addition to the check against varying input simulation spectra discussed below, $T$-only
\planck{} maps corresponding to the SPT-3G coverage region are mock-observed to verify
the leakage bias in $\TE$ to be expected from the real sky, with excellent agreement
found between those results and those from the standard set of simulations.

\subsection{Bandpower Covariance Matrix}
\label{sec:cov}
The bandpower covariance matrix captures the uncertainty in individual bandpowers and their correlations as well as the correlations between different spectra and different frequency bands.
This covariance matrix includes contributions from noise and sample variance.
We estimate the noise variance from the set of measured cross-spectra and the sample variance from the set of 250 signal-only simulations.
In a final step, the uncertainty from the beam measurement is added.

The calculation of the covariance matrix follows the general procedure outlined in the Appendix of \cite{lueker10}.
The three frequency bands are used to form three auto-frequency spectra and three cross-frequency spectra for both $\EE$ and $\TE$, giving the covariance matrix a 12$\times$12 block structure.
The estimate of the covariance is noisy given the finite number of simulations and observations; we therefore ``condition" the covariance matrix to reduce noise in both the diagonal and off-diagonal elements.

For the diagonal elements, we expect a fractional uncertainty of $\sqrt{2/n_\mathrm{obs}}$; for the 30 data bundles in this analysis, this is 26\%.
To mitigate this, we extract the effective number of modes in each $\ell$-bin from the signal-only simulations detailed in \S\ref{sec:simulations}, which allows us to compare the poor noise variance estimates to their expectation values.
This comparison yields an estimate of the noise spectra, which we smooth with a Gaussian kernel and use to assemble an improved estimate of the noise variance.
We add the sample variance contribution to the noise variance to obtain conditioned diagonals for all covariance blocks.

To ameliorate the noise of off-diagonal elements, we condition the underlying correlation matrices.
We average the estimated correlation matrices of all 12 on-diagonal blocks and inspect band-diagonal slices (i.e., elements the same distance away from the diagonal).
To account for the widening of the mode-coupling matrix over the multipole range, we fit second-order polynomials to these elements.
We replace off-diagonal elements with these fits and set elements further than \mbox{$\Delta \ell >100$} from the main diagonal to zero as correlations become negligible.
The correlation matrix conditioned in this way is then combined with the previously calculated diagonal elements of each block to construct the conditioned covariance matrix.

The uncertainty from the beam measurement is added to the bandpower covariance matrix described above using the same procedure as in [\citenum{keisler11, story13}, \citetalias{crites15}].
First, we construct a ``beam correlation matrix"
\begin{equation}
\mathbf{\rho}_{bb'}^{\rm beam} = \left(\frac{\delta D_b}{D_b}\right) \left(\frac{\delta D_{b'}}{D_{b'}}\right),
\end{equation}
where
\begin{equation}
\frac{\delta D_b}{D_b} = 1-\left(1+ \frac{\delta B_b}{B_b} \right)^{-2}
\end{equation}
represents the effect of the beam uncertainty $\delta B_b$ on the power spectrum.
Model bandpowers $D_b$ are then used to generate a covariance from the beam correlation matrix:
\begin{equation}
\mathbf{C}_{bb'}^{\rm beam} = \mathbf{\rho}_{bb'}^{\rm beam}  D_b D_{b'}.
\end{equation}
Our final results are robust with respect to the beam covariance assumed, with no effect on cosmological constraints after increasing the covariance by a factor of four.

%%%%%%%%%%%%%%%%%%%%%%%%%%%%%%%%%%%%
% SYSTEMATIC ERRORS
%%%%%%%%%%%%%%%%%%%%%%%%%%%%%%%%%%%%
\section{Tests for Systematic Errors}

\begin{table*}[htb!]
\centering
\setlength{\tabcolsep}{10pt}
\begin{tabular}{ l || c c | c c| c c || c}
\hline\hline
 & \multicolumn{2}{c|}{95\,GHz}
 &  \multicolumn{2}{c|}{150\,GHz}
 &  \multicolumn{2}{c||}{220\,GHz}
 & Row Fisher \\
 & $\TE$ & $\EE$ & $\TE$ & $\EE$ & $\TE$ & $\EE$ & PTE \\
\hline
Azimuth & 0.5974 & 0.4939 & 0.1969 & 0.0054 & 0.9023 & 0.8598 & 0.1636\\
First-Second & 0.3131 & 0.6800 & 0.2594 & 0.9825 & 0.6745 & 0.4779 & 0.7779\\
Left-Right & 0.3207 & 0.2285 & 0.6895 & 0.6761 & 0.3906 & 0.5617 & 0.6346\\
Moon Up-Down & 0.8127 & 0.9954 & 0.7333 & 0.4974 & 0.9175 & 0.7619 & 0.9943\\
Saturation & 0.0962 & 0.8606 & 0.1186 & 0.4727 & 0.6097 & 0.4083 & 0.3320\\
Wafer & 0.1091 & 0.0038 & 0.4806 & 0.0432 & 0.6597 & 0.5993 & 0.0140\\
\hline
\end{tabular}
\caption{
Individual null test PTE values and the combined PTE value for each test across all frequencies and spectra.
}
\label{tab:null_PTE_table}
\end{table*}

\label{sec:systematics}
We perform two primary tests on the data and analysis pipeline; the first using null tests to probe for systematic effects in the data, and the second verifying the robustness of the pseudo-spectrum debiasing pipeline against changes to the input power spectrum.

\subsection{Null Tests}

To check that the data are free of systematics above the noise level, we perform a series of null tests, in which the data are divided based on a possible source of systematic error, and the groups of data are then differenced to form a collection of null maps.
The cross-spectra of the null maps are then compared to the expected null spectrum if that systematic were absent.
The expectation spectra are calculated using the same noiseless mock observations detailed in \S\ref{sec:simulations} used for obtaining $F_\ell$.
The expected null spectra are typically consistent with zero, although differences in e.g.\ live detector counts can cause non-zero expectation spectra.

We perform the following null tests, most of which have also been explored in prior SPT analyses:
\begin{description}
\item [Azimuth] We test for sensitivity to ground signals by ordering the data based on the average azimuth of the observation.
We divide azimuth according to the direction of the Dark Sector Laboratory, the building connected to the telescope, which we expect to be the dominant source of any ground-based pickup.
\item [First-Second] This tests for time-dependent effects by ordering the data chronologically into the beginning and end of the season.
For 2018, this is degenerate with splitting the data based on if the Sun was below or above the horizon, and therefore tests for both Sun contamination and long time-scale drifts.
\item [Left-Right] This divides each observation into left-going scans and right-going scans, and is intended to test for asymmetric scanning or effects due to the elevation steps.
\item [Moon up - Moon down] We test for additional beam sidelobe pickup by dividing the data based on whether the Moon was above or below the horizon.
\item [Saturation] We test for effects of decreased array responsivity by ordering the data based on the average number of detectors flagged as saturated during an observation.
\item [Wafer] We test for effects due to differing detector properties by dividing the wafers into two groups based on optical response to the calibrator and bolometer saturation power.
Separate maps for each observation are made from the two sets of wafers.
\end{description}

With the exception of the Azimuth test, the null tests use the same chronological bundles as used in the cross-spectrum calculation.
For the Left-Right test, each bundle is separated into left-going and right-going scans, and these are differenced to create the null maps.
An analogous procedure is used for the Wafer null test.
For the First-Second, Moon Up-Moon Down, and Saturation tests, each observation is assigned a value based on the susceptibility of that observation to the potential source of systematic error, and the bundles are then rank-ordered by the average of this value across their constituent observations.
The halves of the rank-ordered list are then subtracted (i.e., bundle 1 from bundle 16, bundle 2 from bundle 17, ..., bundle 15 from bundle 30) to form the null maps.
For the Azimuth test, the normal chronological bundles would average down any potential systematic, as the observing cadence of the telescope effectively randomizes the azimuthal range over which the field is observed.
The observations are therefore re-bundled according to the separation between their mean azimuth and the azimuth corresponding to the Dark Sector Laboratory.

For each null test, we use the average and distribution of all null cross-spectra to compute the chi-square compared to the null expectation spectrum, and we then compute the probability to exceed (PTE) this chi-square value given the degrees of freedom.
An exceedingly low PTE or a preponderance of low PTEs indicate the data are in larger disagreement with expectation than random chance would allow.
We perform three checks on the collection of PTEs: (1) the entire table of PTE values is consistent with a uniform distribution between 0 and 1 with a Kolmogorov-Smirnov (KS) test p-value $> 0.05$, (2) individual PTE values are larger than $0.05/N_\mathrm{tests}$, and (3) the combination of PTEs in each row using Fisher's method has a PTE above $0.05/N_\mathrm{rows}$.
We neglect correlations between PTE values when performing these tests, which has the effect of strengthening the KS and Fisher tests while weakening the multiple-comparisons-corrected individual PTE test.
These tests and significance thresholds were agreed upon before looking at the collection of final PTEs to avoid confirmation bias.

The null test PTEs are collected in Table~\ref{tab:null_PTE_table}.
The distribution of PTEs is consistent with a uniform distribution with a KS test p-value of 0.76.
With 36 tests and six rows, the individual PTE threshold is 0.0014, and the row threshold is 0.0083;
although the Azimuth test for 150\,GHz $\EE$ and Wafer test for 95\,GHz $\EE$ are marginal, all of the tests pass the agreed-upon criteria, and we conclude the listed systematics do not affect the data in a statistically significant way.

\subsection{Sensitivity to Cosmological Model}
Any corrections to the data based on simulations, such as $F_\ell$ or additive bias corrections, should be robust against the chosen input cosmology to the simulations.
The simulations in \S\ref{sec:simulations} were constructed to match the true sky as closely as possible, so we can be confident that the resulting simulations will yield valid results; however, we still want to test that the pipeline is stable against small variations to the input power spectra.

We create an additional set of simulations with a contrived cosmology chosen to be $\sim5\sigma$ discrepant with the results found in \citetalias{henning18}, with parameter values $\Omega_b h^2 = 0.02$, $\Omega_c h^2 = 0.14$, $H_0 = 61\,\mathrm{km\,s^{-1} Mpc^{-1}}$, $\ln(10^{10}A_s) = 3.12$, $n_s = 0.9$, and $\tau = 0.06$.
Additionally, the foreground power is doubled in comparison to the standard set of simulations.
Fifty noiseless realizations of this cosmology are supplied to the mock-observing pipeline, and the resulting $\tilde{C}_\ell$ are debiased using the transfer function and $\TE$ bias corrections derived from the standard set of simulations.
The input spectra are recovered to well within the uncertainties on the reported data bandpowers, and we therefore find no measurable bias due to $F_\ell$ or the $\TE_\mathrm{bias}$ correction.

%%%%%%%%%%%%%%%%%%%%%%%%%%%%%%%%%%%%
% PARAMETER FITTING METHODOLOGY
%%%%%%%%%%%%%%%%%%%%%%%%%%%%%%%%%%%%
\section{Parameter Fitting and Modeling}
\label{sec:mcmc}

\begin{table}[ht!]
\def\arraystretch{1.5}
\small
\setlength{\tabcolsep}{10pt}
\centering
\begin{tabular}{l D{+}{\,\pm\,}{-1}}
\hline\hline
 Parameter & \multicolumn{1}{c}{\;\;\;\;\;Prior}\\
 \hline
 $\tau$ & 0.0543+0.0073\\
 $100\kappa$ & 0+0.045\\
 $A^{\EE}_{80}$ & 0.095+0.012\\
 $\alpha_{\EE}$ & -2.42+0.02\\
 $A^{\TE}_{80}$ & 0.184+0.072\\
 $\alpha_{\TE}$ & -2.42+0.02\\
 $D^\mathrm{ps,~95\times95}_{3000}$ &  0.041+0.012\\
 $D^\mathrm{ps,~150\times150}_{3000}$ & 0.0115+0.0034\\
 $D^\mathrm{ps,~220\times220}_{3000}$ & 0.048+0.014\\
 $D^\mathrm{ps,~95\times150}_{3000}$ & 0.0180+0.0054\\
 $D^\mathrm{ps,~95\times220}_{3000}$ & 0.0157+0.0047\\
 $D^\mathrm{ps,~150\times220}_{3000}$ & 0.0190+0.0057\\
 $T_{\rm cal}^{\rm \,95\,GHz}$ & 1.0+0.0049\\
 $T_{\rm cal}^{\rm \,150\,GHz}$ & 1.0+0.0050\\
 $T_{\rm cal}^{\rm \,220\,GHz}$ & 1.0+0.0067\\
 $E_{\rm cal}^{\rm \,95\,GHz}$ & 1.0+0.0087\\
 $E_{\rm cal}^{\rm \,150\,GHz}$ & 1.0+0.0081\\
 $E_{\rm cal}^{\rm \,220\,GHz}$ & 1.0+0.016\\
\hline
\end{tabular}
\caption[
Priors used for the MCMC fit.
]{
Gaussian priors used for the MCMC fit, including the optical depth to reionization $\tau$, mean-field lensing convergence $\kappa$, the amplitude $A^{XY}_{80}$ (in $\mu$K$^2$) at 150\,GHz and spectral index $\alpha^{XY}_{80}$ of polarized Galactic dust, the $\EE$ power of Poisson-distributed point sources $D_{3000}^{\mathrm{ps,~\nu_i\times\nu_j}}$ (in  $\mu$K$^2$), absolute temperature calibration factor $T_{\rm cal}^{\, \nu_i}$, and absolute polarization calibration factor $E_{\rm cal}^{\, \nu_i}$.
}
\label{tab:priors_table}
\end{table}

We obtain cosmological parameter constraints using the Markov Chain Monte Carlo (MCMC)
package \textsc{CosmoMC} \citep{lewis02b}.\footnote{\url{https://cosmologist.info/cosmomc/}}
The theoretical CMB spectra are calculated using \textsc{camb} \citep{lewis00}\footnote{\url{https://camb.info/}}, and are modified to account for the effects of instrumental calibration, aberration due to relative motion with respect to the CMB rest frame \citep{jeong14}, and super-sample lensing \citep{manzotti14}.
We also add terms representing Galactic dust emission and polarized dusty and radio galaxies.

We parameterize the \lcdm\ model as follows: the density of cold dark matter $\Omega_c h^2$; the baryon density $\Omega_b h^2$; the amplitude of primordial density perturbations, $A_s$, and the tilt of their power spectrum, $n_s$, defined at a pivot scale of $0.05\mathrm{Mpc^{-1}}$; the optical depth to reionization $\tau$; and \textsc{CosmoMC}'s internal proxy for the angular scale of the sound horizon at decoupling, $\theta_{MC}$.
For the range of angular multipoles considered here, $\tau$ is degenerate with $A_\mathrm{s}$.
We therefore use large-scale polarization information from \planck\ to inform a Gaussian prior of $\tau = 0.0543 \pm 0.0073$ \citep{planck18-6}, and we report constraints on the combined amplitude parameter $10^9 A_{\rm{s}} e^{-2\tau}$ in this work.
Widening the prior to $\tau = 0.065 \pm 0.015$ based on a recent analysis of \planck\ and \wmap\ data by \citep{natale20} has no significant effect on cosmological parameter constraints.

We account for aberration in a manner similar to \citetalias{henning18} and \cite{louis17} by modifying the theory spectrum as
\begin{equation}
C_{\ell} \rightarrow C_\ell - C_\ell \frac{d \ln C_\ell}{d \ln \ell} \beta \langle \cos \theta \rangle ,
\end{equation}
where $\beta = 1.23\times10^{-3}$ is the velocity of the Local Group with respect to the rest frame of the CMB, and \mbox{$\langle\cos \theta \rangle = -0.39$} is the mean angular separation between the CMB dipole and the SPT-3G survey field.
For super-sample lensing, we follow the procedure laid out by \citetalias{crites15} and \citetalias{henning18}, modifying the CMB spectrum resulting from a set of parameters \textbf{p} as
\begin{equation}
\hat{C}_\ell^{XY}(\mathbf{p}; \kappa)
= C_\ell^{XY}(\mathbf{p})
  - \frac{\partial\ell^2 C_\ell^{XY}(\mathbf{p})}{\partial \ln \ell}
  \frac{\kappa}{\ell^2} ,
\end{equation}
where the nuisance parameter $\kappa$ quantifies the mean lensing convergence across the survey field.
We apply a Gaussian prior on $\kappa$ centered on zero with standard deviation $\sigma_{\kappa} = 4.5\times10^{-4}$, with the uncertainty estimated from the survey size \citep{manzotti14}.

The power from Galactic dust is assumed to follow a modified blackbody spectrum with $T_\mathrm{dust} = 19.6$~K and $\beta_\mathrm{dust} = 1.59$ and is modeled according to the relation from \cite{planck12-30,planck18-11}:
\begin{equation}
D^{XY}_{\ell, \mathrm{dust}} = A^{XY}_{80} \left(\frac{\ell}{80}\right)^ {\alpha_{XY} + 2},
\label{eq:fgnd_param}
\end{equation}
where $A^{XY}_{80}$ is the amplitude of the spectrum at $\ell = 80$ at 150\,GHz, and $\alpha_{XY}$ is the angular power dust spectral index.
Based on \cite{planck12-30}, we apply a Gaussian prior on $\alpha_{XY}$ with a central value of -2.42 and uncertainty 0.02.
We estimate the properties of polarized Galactic dust on the SPT-3G 1500\,deg$^2$ field using \planck{} observations in the frequency bands 100\,GHz, 143\,GHz, 217\,GHz, and 353\,GHz.
We assume the aforementioned spectral energy distribution and fit to the amplitude using the ten cross-frequency spectra obtained from an optimal combination of all possible half-mission map cross-spectra.
Taking into account \planck{} color corrections \citep{planck18-11}, pessimistic calibration errors and assuming the \planck{} best fit cosmology, we constrain the amplitude of polarized Galactic dust to be $A^{EE}_{80} = 0.095 \pm 0.012$ and $A^{TE}_{80} = 0.184 \pm 0.072$, which we adopt as Gaussian priors in our MCMC analysis.
We further check that the constraints remain stable when also fitting for $\beta_\mathrm{dust}$ and $\alpha_{EE}$, the fit values of which are in good agreement with our chosen values.

The $\EE$ power spectrum of the emission from a Poisson distribution of partially polarized synchrotron and dusty galaxies can be described as
\begin{equation}
D_\ell = D^{\rm ps}_{\rm 3000}  \left(\frac{\ell}{3000}\right)^2.
\end{equation}
The $\TE$ signal from these galaxies is expected to be zero, as the polarization angles are uncorrelated between galaxies.
In the baseline case, we apply Gaussian priors to the six $D^{\mathrm{ps,\,} \nu_i \times \nu_j}_{\rm 3000}$ parameters based on the temperature values from \cite{reichardt20}, which we adjust for our flux cut following the model of \citep{dezotti05} and scale by the polarization fractions reported by \cite{gupta19}.
The prior width is dominated by uncertainty in the mean squared polarization fraction, which we conservatively double to yield 30\%.

We find that our cosmological parameter constraints are insensitive to the details of the foreground priors, with no significant shifts in the results when the Poisson terms or the polarized Galactic dust amplitudes are doubled or set to zero.
We conclude that over our multipole range the bandpowers are largely insensitive to both of these foreground sources.
The priors discussed in this section are summarized in Table~\ref{tab:priors_table}.

We verify that our likelihood is unbiased by analyzing a set of 100 simulated spectra.
Mock bandpowers are created by adding random noise realizations based on our data covariance matrix to the latest \planck{} best-fit model.
We use the likelihood to obtain the best-fit model for each realization, and we find that for all cosmological parameters, the mean of the ensemble of simulations lies within one standard error of the input value.

%%%%%%%%%%%%%%%%%%%%%%%%%%%%%%%%%%%%
% BANDPOWERS
%%%%%%%%%%%%%%%%%%%%%%%%%%%%%%%%%%%%

\section{The SPT-3G 2018 Power Spectra}
\label{sec:bandpowers}

\subsection{Bandpowers}

\begin{table}[ht!]
\small
\setlength{\tabcolsep}{5pt}
\centering
\begin{tabular}{c | c  D{.}{.}{4.1}  D{.}{.}{2.2} | c  D{.}{.}{3.2}  D{.}{.}{2.2}}
\hline\hline
\rule{0pt}{3ex} $\ell$ Range & $\ell_\mathrm{eff}^{\TE}$ & \multicolumn{1}{r}{$D_b^{\TE}$} & \multicolumn{1}{r}{$\sigma^{\TE}$} & $\ell_\mathrm{eff}^{\EE}$ & \multicolumn{1}{r}{$D_b^{\EE}$} & \multicolumn{1}{r}{$\sigma^{\EE}$} \\[2pt]
%\rule{0pt}{3ex} $\ell$ Range & $\ell_\mathrm{eff}^{\TE}$ & D_\ell^{\TE} & \sigma^{\TE} & $\ell_\mathrm{eff}^{\EE}$ & D_\ell^{\EE} & \sigma^{\EE} \\[2pt]
\hline
300 -- 349 & 326 & 103.7 & 11.3 & 325 & 14.1 & 1.0 \\
350 -- 399 & 376 & 39.8 & 8.4 & 375 & 20.4 & 1.2 \\
400 -- 449 & 426 & -47.8 & 7.0 & 425 & 19.0 & 1.1 \\
450 -- 499 & 475 & -72.1 & 6.0 & 475 & 12.0 & 0.6 \\
500 -- 549 & 523 & -35.1 & 4.7 & 524 & 7.2 & 0.4 \\
550 -- 599 & 574 & 10.2 & 5.6 & 575 & 11.6 & 0.6 \\
600 -- 649 & 625 & 23.6 & 6.6 & 624 & 29.7 & 1.1 \\
650 -- 699 & 675 & -63.7 & 7.3 & 674 & 39.0 & 1.3 \\
700 -- 749 & 725 & -120.8 & 6.8 & 725 & 34.5 & 1.2 \\
750 -- 799 & 774 & -121.2 & 6.6 & 774 & 20.7 & 0.9 \\
800 -- 849 & 824 & -49.2 & 4.7 & 824 & 13.5 & 0.6 \\
850 -- 899 & 874 & 38.0 & 5.0 & 874 & 17.1 & 0.7 \\
900 -- 949 & 924 & 56.6 & 4.9 & 924 & 31.6 & 1.0 \\
950 -- 999 & 974 & 13.3 & 4.8 & 974 & 40.6 & 1.3 \\
1000 -- 1049 & 1024 & -52.3 & 5.2 & 1024 & 38.5 & 1.3 \\
1050 -- 1099 & 1075 & -74.0 & 4.7 & 1075 & 26.2 & 1.0 \\
1100 -- 1149 & 1124 & -54.2 & 3.8 & 1124 & 15.0 & 0.6 \\
1150 -- 1199 & 1174 & -10.0 & 3.3 & 1174 & 12.4 & 0.6 \\
1200 -- 1249 & 1224 & 4.4 & 3.3 & 1224 & 21.9 & 0.9 \\
1250 -- 1299 & 1274 & -15.9 & 3.3 & 1275 & 29.2 & 1.1 \\
1300 -- 1349 & 1324 & -47.8 & 3.4 & 1325 & 31.1 & 1.1 \\
1350 -- 1399 & 1374 & -61.7 & 3.4 & 1374 & 22.7 & 0.9 \\
1400 -- 1449 & 1424 & -42.0 & 3.0 & 1424 & 12.8 & 0.7 \\
1450 -- 1499 & 1474 & -11.9 & 2.7 & 1474 & 10.6 & 0.6 \\
1500 -- 1549 & 1524 & 9.1 & 2.5 & 1524 & 14.4 & 0.7 \\
1550 -- 1599 & 1574 & -0.4 & 2.5 & 1574 & 21.4 & 0.9 \\
1600 -- 1649 & 1624 & -14.7 & 2.4 & 1624 & 20.2 & 0.9 \\
1650 -- 1699 & 1674 & -32.4 & 2.2 & 1674 & 18.2 & 0.8 \\
1700 -- 1749 & 1724 & -24.9 & 2.2 & 1724 & 10.3 & 0.7 \\
1750 -- 1799 & 1775 & -15.2 & 2.0 & 1775 & 8.8 & 0.7 \\
1800 -- 1849 & 1824 & -9.4 & 1.9 & 1825 & 8.9 & 0.7 \\
1850 -- 1899 & 1874 & -3.5 & 1.9 & 1874 & 10.0 & 0.8 \\
1900 -- 1949 & 1924 & -11.3 & 1.8 & 1924 & 12.3 & 0.8 \\
1950 -- 1999 & 1975 & -16.3 & 1.8 & 1975 & 11.1 & 0.8 \\
2000 -- 2099 & 2050 & -14.2 & 0.9 & 2049 & 6.4 & 0.4 \\
2100 -- 2199 & 2151 & -4.8 & 0.9 & 2148 & 5.3 & 0.5 \\
2200 -- 2299 & 2250 & -5.6 & 0.8 & 2248 & 6.8 & 0.5 \\
2300 -- 2399 & 2349 & -9.2 & 0.8 & 2348 & 3.5 & 0.5 \\
2400 -- 2499 & 2450 & -3.6 & 0.8 & 2448 & 3.7 & 0.6 \\
2500 -- 2599 & 2549 & -3.7 & 0.8 & 2548 & 2.6 & 0.6 \\
2600 -- 2699 & 2649 & -3.5 & 0.8 & 2648 & 1.9 & 0.7 \\
2700 -- 2799 & 2749 & -2.1 & 0.8 & 2748 & 1.7 & 0.8 \\
2800 -- 2899 & 2849 & -0.5 & 0.8 & 2848 & 1.2 & 0.9 \\
2900 -- 2999 & 2949 & -2.3 & 0.8 & 2948 & -0.1 & 1.0 \\
\hline
\end{tabular}
\caption{
Minimum-variance bandpowers $D_b$ and their associated uncertainties $\sigma$ for the $\TE$ and $\EE$ power spectra.
We also report the bandpower window function-weighted multipole $\ell_\mathrm{eff}$ for each $\ell$-range.
The bandpowers and errors are quoted in units of $\mu$K$^2$.
The reported uncertainties are the square root of the diagonal elements of the covariance matrix and do not include beam or calibration uncertainties.
}
\label{tab:bandpowers_table}
\end{table}

We present bandpowers and uncertainties for the six $\EE$ and $\TE$ cross-frequency power spectra, plotted in Figure~\ref{fig:autocross_bp} and listed in full in the \mbox{\hyperref[app:appendix]{Appendix}}.
The bandpowers span the multipole range $300 \le \ell < 3000$, with bin widths of $\Delta \ell = 50$ for $\ell < 2000$ and $\Delta \ell = 100$ for $\ell > 2000$.
The 44 bandpowers for each spectrum are measured with each of the six frequency combinations of 95, 150, and 220\,GHz data, resulting in 528 bandpower values in total.

With $150\!\times\!150$\,GHz alone, we measure the first seven acoustic peaks of the $\EE$ spectrum with \mbox{3--4} bandpowers per peak and signal-to-noise $\ge 6.4$ on each bandpower.
The bandpowers are sample variance-dominated at \mbox{$\ell < 1275$} for $\EE$ and $\ell < 1425$ for $\TE$.

%----------------------------------
% EETE bandpowers figure
%----------------------------------
\begin{figure*}[ht!]
\includegraphics[width=17.2cm]{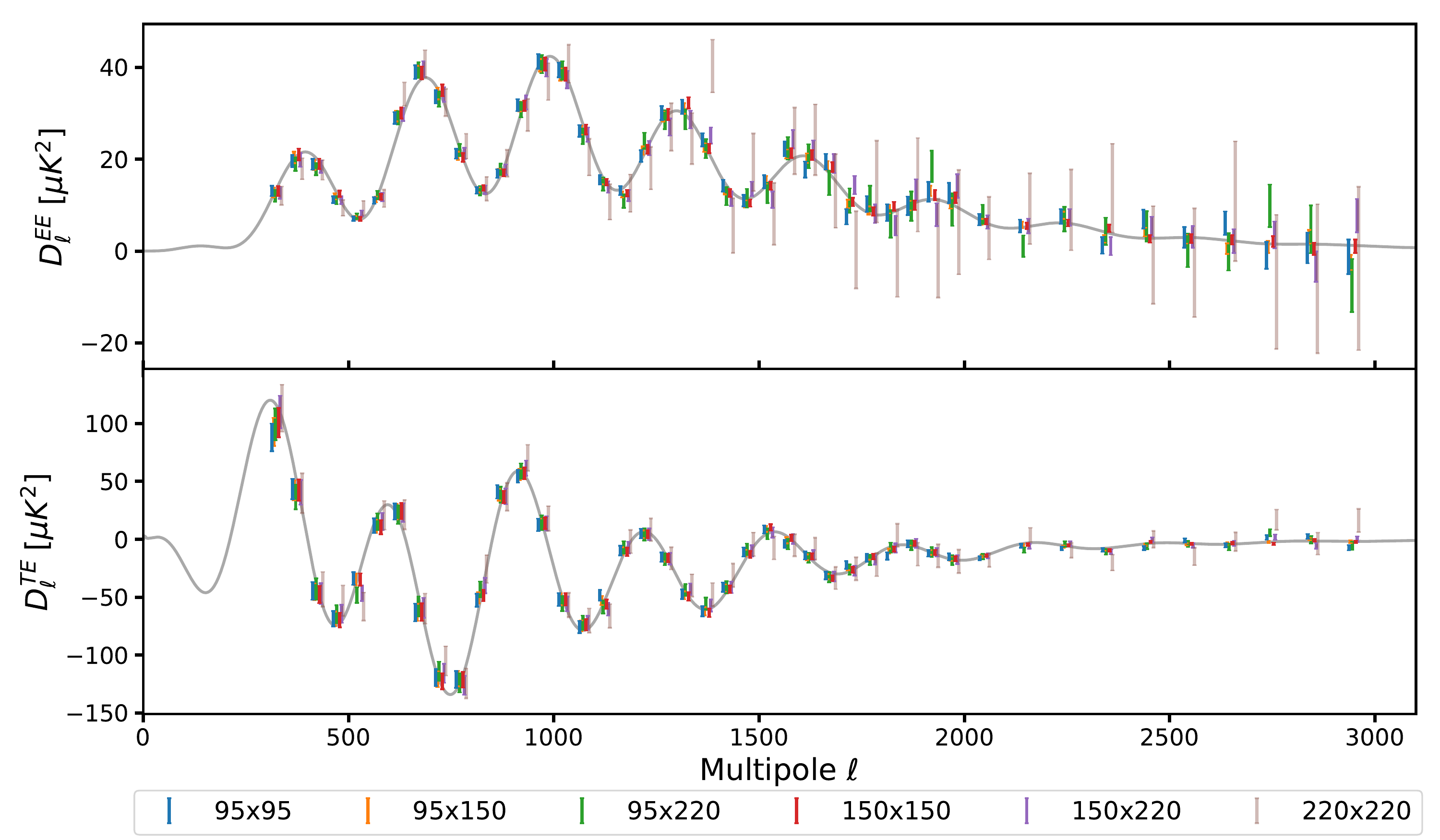}
\centering
\caption{
SPT-3G $\EE$ and $\TE$ bandpower measurements from the six auto- and cross-frequency power spectra overlaid on the \planck{} best-fit \lcdm\ model.
The plotted uncertainties are the square root of the diagonal elements of the covariance matrix and do not include beam or calibration uncertainties.
A small $\ell$ offset has been applied to each point for plotting purposes.
}
\label{fig:autocross_bp}
\end{figure*}

We also construct a set of minimum-variance bandpowers.
Following \citep{mocanu19}, the minimum-variance bandpowers $D^{\rm MV}$ can be expressed as:
\begin{equation}
D^{\rm MV}= \mathbf{(X^\intercal C^{-1} X)^{-1}X^\intercal C^{-1}}D~.
\end{equation}
Here, $D$ and $\mathbf{C}$ are the multifrequency bandpowers and covariance matrix, and $\mathbf{X}$ is a 528$\times$88 design matrix, in which each column is equal to 1 in the six elements corresponding to a power spectrum measurement in that $\ell$-space bin and zero elsewhere.
In this construction, we have made the simplifying assumption that the polarized foreground power is negligible within the bandpower uncertainties.
Relative to the most-sensitive single-frequency band, the $150\!\times\!150$\,GHz bandpowers, the minimum-variance bandpowers have uncertainties 5\%--10\% smaller at $\ell<1000$ and 20\%--30\% smaller at $\ell>2000$.

The minimum-variance $\EE$ and $\TE$ bandpowers and associated errors are summarized in Table~\ref{tab:bandpowers_table} and plotted in Figure~\ref{fig:all_bp} along with measurements from several recent experiments.
These minimum-variance bandpowers, measured using only four months of SPT-3G data with slightly over half the number of detectors relative to subsequent observing seasons, are already the most constraining measurements made to date by an instrument on SPT over the multipole ranges \mbox{$300 \le \ell \le 1400$} for $\EE$ and \mbox{$300 \le \ell \le 1700$} for $\TE$, and are competitive with other current leading measurements.

\subsection{Internal Consistency}
\label{sec:int_consistency}

The minimum-variance construction above assumes the multifrequency bandpowers are measuring the same underlying signal and that polarized foregrounds are negligible.
We test this assumption by examining the chi-square of the multifrequency bandpowers to the minimum-variance bandpowers,
\begin{equation}
\chi^2 = (D - M)^\intercal \mathbf{C}^{-1} (D-M)~,
\end{equation}
where $M = \mathbf{X} D^{\rm MV}$.
We find a $\chi^2$ of 438.1 for 440 degrees of freedom (528 multifrequency bandpowers $-$ 88 minimum-variance bandpowers).
The PTE for this $\chi^2$ is 0.52.
If the $\EE$ and $\TE$ bandpowers are evaluated separately, the PTEs are 0.18 and 0.71, respectively.
This indicates that the measurements from different frequency bands and their cross-correlations are consistent with a common signal, with no evidence for significant contamination due to foregrounds or unmodeled systematics.

%----------------------------------
% Compilation bandpowers figure
%----------------------------------
\begin{figure*}[ht!]
\includegraphics[width=17.2cm]{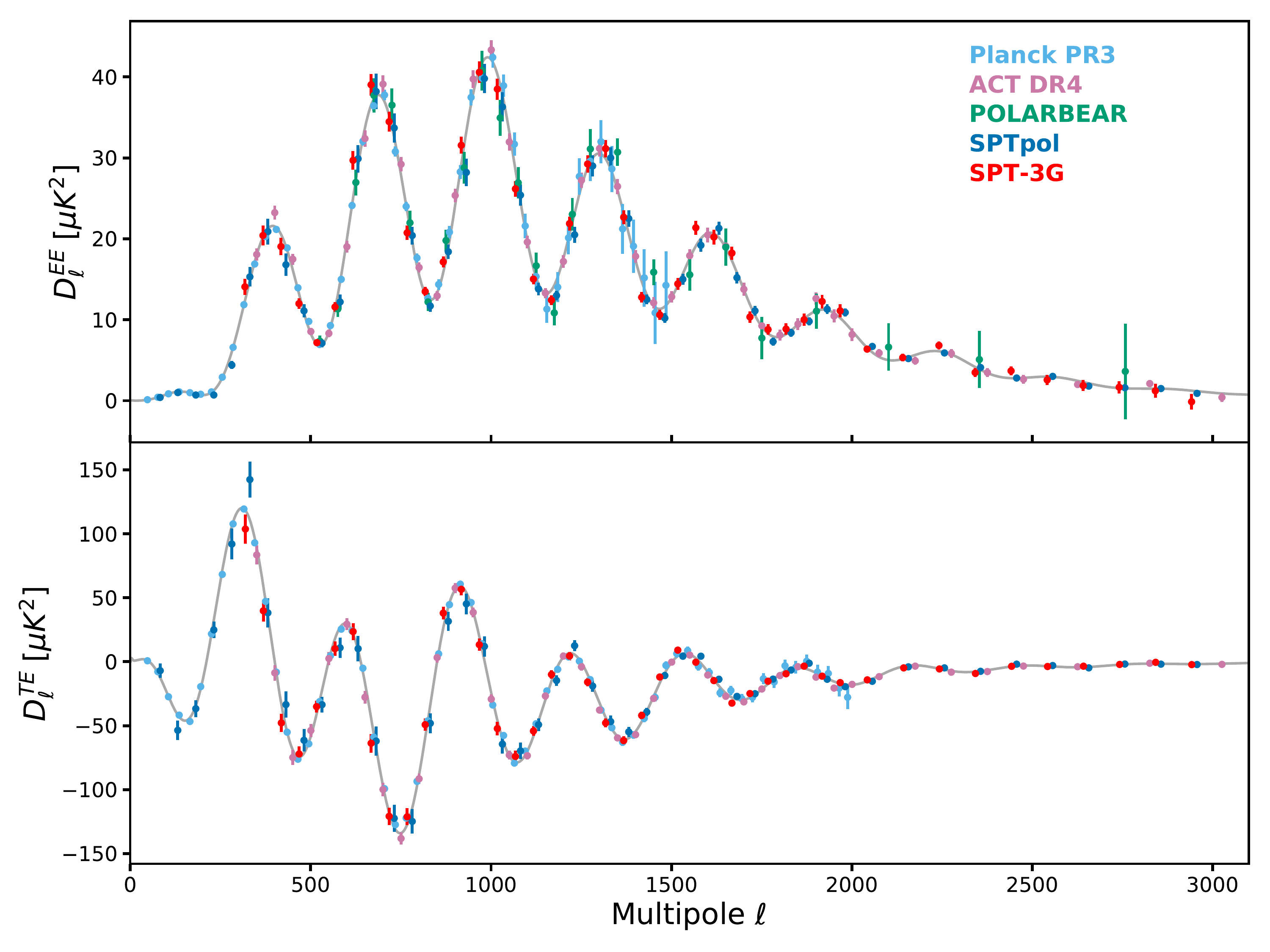}
\centering
\caption{
The minimum-variance SPT-3G $\EE$ and $\TE$ bandpowers \textit{(red)} overlaid on the \planck{} best-fit \lcdm\ model, 
along with the recent measurements from \planck{} \citep{planck18-5}, ACT \citep{choi20},  \polarbear{} \citep{polarbear20}, and SPTpol \citepalias{henning18}.
The \planck{} $\EE$ bandpowers are restricted to $\ell < 1500$.
The uncertainties shown for the SPT-3G bandpowers are the square root of the diagonal elements of the covariance matrix and do not include beam or calibration uncertainties.
}
\label{fig:all_bp}
\end{figure*}

We further investigate the internal consistency of the SPT-3G 2018 $\EE/\TE$ dataset by subdividing it and examining the parameter constraints from each of the seven data splits: the 95, 150, and 220\,GHz auto-frequency spectra, the $\ell<1000$ and $\ell>1000$ data, and the $\EE$ and $\TE$ spectra individually.
We quantify the consistency of each subset with respect to the full model by calculating the parameter-level $\chi^2$ and associated PTEs in Table \ref{tab:datasplit_chisq}, following the methodology of \cite{aylor17}:
\begin{equation}
\chi^2 = \mathbf{\Delta p}^\intercal \mathbf{C}_\mathrm{p}^{-1} \mathbf{\Delta p},
\end{equation}
where $\mathbf{\Delta p}$ is the vector of parameter differences between the full dataset and a given subset.
Following \cite{gratton19}, $\mathbf{C}_\mathrm{p}$ is the difference of the associated parameter covariance matrices, whereby we account for the correlation between the full dataset and the subset.
The comparison is carried out over the parameters $(\Omega_b h^2, \Omega_c h^2, \theta_{MC}, 10^9 A_s e^{-2\tau}, n_s)$.

All seven data splits are firmly within the central 95\% confidence interval $[2.5\%,97.5\%]$ and we conclude that there is no evidence for significant internal tension in the dataset.
We will return to these data splits in \S\ref{sec:params_spt}, when we look at the effect of each subset on the cosmological constraints of the ensemble.

\begin{table}[ht!]
\def\arraystretch{1.5}
\small
\setlength{\tabcolsep}{10pt}
\centering
\begin{tabular}{l c  c}
\hline\hline
 Subset & $\chi^2$ & PTE\\
 \hline
$\EE$ & $4.69$ & $45.45\%$\\
$\TE$ & $8.96$ & $11.06\%$\\
$\ell\le1000$ & $7.82$ & $16.64\%$\\
$\ell>1000$ & $7.70$ & $17.34\%$\\
95\,GHz & $6.68$ & $24.57\%$\\
150\,GHz & $3.75$ & $58.54\%$\\
220\,GHz & $2.35$ & $79.92\%$\\
\hline
\end{tabular}
\caption[
Parameter-level $\chi^2$ difference and PTE between subsets of the data and the full dataset.
]{
Parameter-level $\chi^2$ difference and PTE between subsets of the data and the full dataset.
We do the comparison in the five-dimensional parameter space, $(\Omega_b h^2, \Omega_c h^2, \theta_{MC}, 10^9 A_s e^{-2\tau}, n_s)$, due to the common $\tau$ prior.
}
\label{tab:datasplit_chisq}
\end{table}

%%%%%%%%%%%%%%%%%%%%%%%%%%%%%%%%%%%%
% PARAMETERS
%%%%%%%%%%%%%%%%%%%%%%%%%%%%%%%%%%%%
\section{Cosmological Constraints}
\label{sec:constraints}

\subsection{SPT-3G}
\label{sec:params_spt}
The cosmological parameter constraints from the 2018 SPT-3G $\EE$ and $\TE$ multifrequency bandpowers are summarized in Table~\ref{tab:lcdm_param_table}.
We present the 1D and 2D marginalized posterior probabilities for \lcdm{} parameters and $H_0$ in Figure \ref{fig:triangle_spt}.
Constraints on nuisance parameters are driven by the priors discussed in \S \ref{sec:mcmc}, with all central values well within $1\sigma$ of their respective prior.

\begin{table*}[ht!]
\def\arraystretch{1.2}
\footnotesize
\setlength{\tabcolsep}{12pt}
\centering
\begin{tabular}{c @{\hskip 20pt} D{+}{\,\pm\,}{10} @{\hskip 40pt} D{+}{\,\pm\,}{14} D{+}{\,\pm\,}{14} D{+}{\,\pm\,}{6}}
\hline\hline
& \multicolumn{1}{l}{\text{\;\;\;\;\;\;\;SPT-3G}}
& \multicolumn{1}{l}{\text{\;\;SPT-3G + BAO}}
& \multicolumn{1}{c}{\text{SPT-3G + \planck\ }}
& \multicolumn{1}{c}{\text{\planck\ }} \\
\hline
\multicolumn{5}{l}{Free}\\
$\Omega_b h^2$ &
  0.02242+0.00033\;(0.02243)&
  0.02240+0.00032\;(0.02241)&
  0.02241+0.00013\;(0.0224)&
  0.02236+0.00015\\
$\Omega_c h^2$ &
  0.1150+0.0037\;(0.115)&
  0.1162+0.0015\;(0.1162)&
  0.1196+0.0013\;(0.1195)&
  0.1202+0.0014\\
$100\theta_{\rm MC}$ &
  1.03961+0.00071\;(1.03964)&
  1.03951+0.00066\;(1.03952)&
  1.04074+0.00028\;(1.04073)&
  1.04090+0.00031\\
$10^9 A_s e^{-2\tau}$ &
  1.819+0.038\;(1.821)&
  1.826+0.036\;(1.826)&
  1.879+0.011\;(1.877)&
  1.884+0.012\\
$n_s$ &
  0.999+0.019\;(0.999)&
  0.996+0.018\;(0.996)&
  0.9666+0.0042\;(0.9672)&
  0.9649+0.0044\\
\hline
\multicolumn{3}{l}{Derived}\\
$\Omega_\Lambda$ &
  0.708+0.020\;(0.708)&
  0.7011+0.0083\;(0.7014)&
  0.6867+0.0077\;(0.6871)&
  0.6834+0.0084\\
$H_0$ &
  68.8+1.5\;(68.8)&
  68.27+0.63\;(68.29)&
  67.48+0.55\;(67.49)&
  67.27+0.60\\
$\sigma_8$ &
  0.789+0.016\;(0.789)&
  0.7935+0.0099\;(0.7933)&
  0.8084+0.0069\;(0.8095)&
  0.8120+0.0073\\
$S_8$ &
  0.779+0.041\;(0.779)&
  0.792+0.018\;(0.791)&
  0.826+0.015\;(0.827)&
  0.834+0.016\\
${\rm{Age}}/{\rm{Gyr}}$ &
  13.808+0.051\;(13.807)&
  13.819+0.038\;(13.818)&
  13.797+0.022\;(13.798)&
  13.800+0.024\\
\hline
\end{tabular}
\caption[
$\Lambda$CDM parameter constraints.
]{
Marginalized constraints and 68\% errors of $\Lambda$CDM free and derived parameters from SPT-3G with and without the addition of BAO measurements, from SPT-3G + \planck{}, and from \planck\ alone \citep{planck18-6}.
Best-fit values are given in parentheses.
Note that SPT-3G alone does not constrain the optical depth to reionization $\tau$, but uses a \planck-based Gaussian prior of $0.0543 \pm 0.0073$.
}
\label{tab:lcdm_param_table}
\end{table*}

%----------------------------------
% 3G-pol-Planck triangle plot figure
%----------------------------------
\begin{figure*}[ht!]
\includegraphics[width=17.2cm]{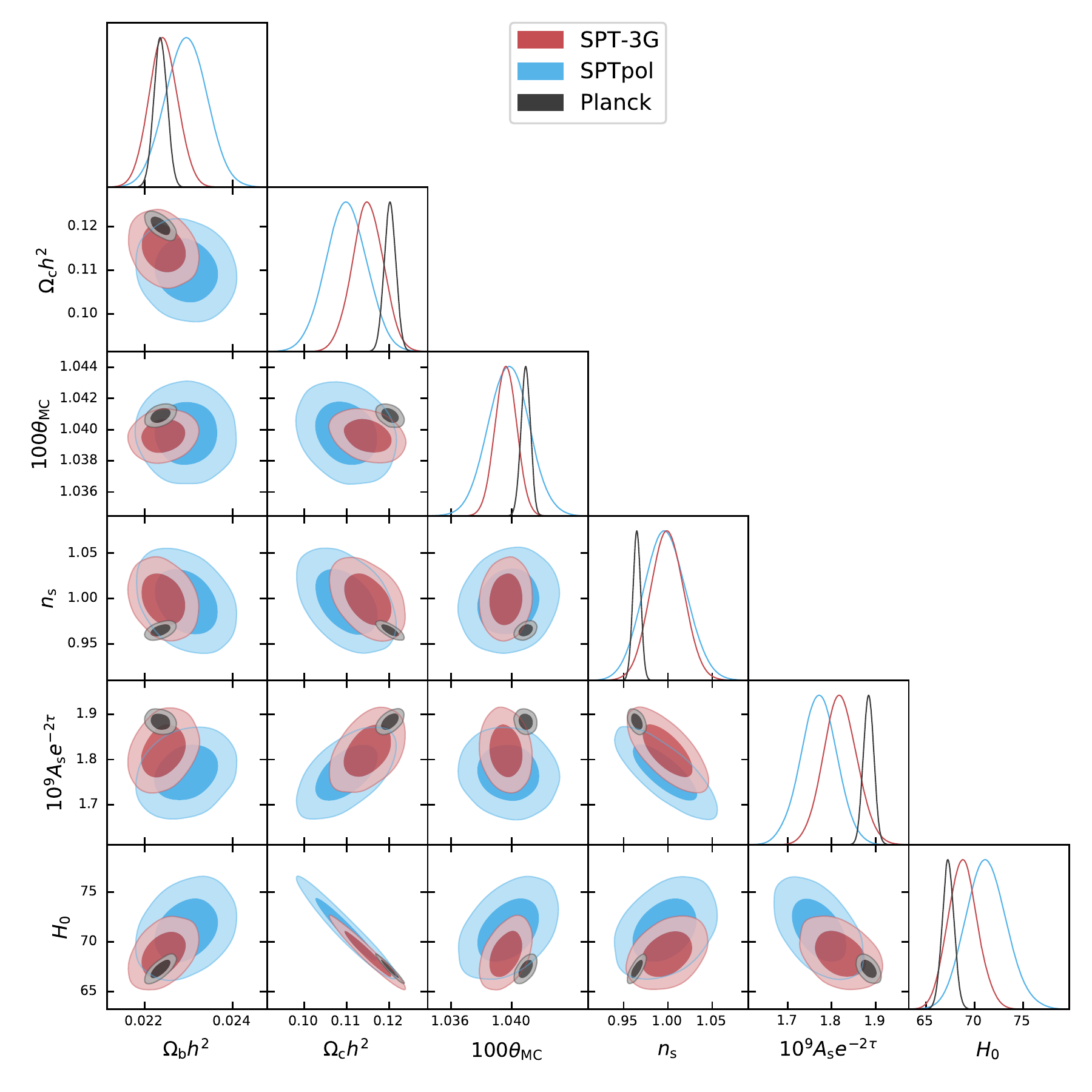}
\centering
\caption{
Marginalized constraints on \lcdm{} parameters and the Hubble constant for the SPT-3G 2018 $\EE/\TE$, SPTpol \citepalias{henning18}, and \planck{} \citep{planck18-5} datasets.
SPT-3G produces consistently tighter constraints than SPTpol.
The results from SPT-3G are statistically consistent with the findings of \planck{}.
}
\label{fig:triangle_spt}
\end{figure*}

We find the value of the Hubble parameter at present day to be
\begin{equation}
H_0 = 68.8 \pm 1.5\,\mathrm{km/s/Mpc},
\end{equation}
in good agreement with other CMB and \lcdm{}-based measurements \citep{planck18-6, choi20} as well as with local distance ladder measurements calibrated using the tip of the red giant branch (TRGB) \citep{freedman19}.
Conversely, this value disagrees at $2.5\sigma$ with the value of \mbox{$H_0 = 74.03 \pm 1.42$\,km/s/Mpc} found by \cite{riess19} using Cepheid-calibrated distance ladder measurements. It is also $1.8\sigma$ and $0.9\sigma$ lower than the value of the Hubble constant measured via the time delays of gravitationally lensed quasars by \cite{wong19} and \cite{birrer20}, respectively.
Our result represents yet another CMB-based measurement, largely independent of \planck{} and also relying on CMB polarization information, that prefers a low value of $H_0$ relative to local measurements.

We find the root mean square fluctuation in the linear matter density field on 8 Mpc/$h$ scales at present day, $\sigma_8$, to be
\begin{equation}
\sigma_8 = 0.789 \pm 0.016.
\end{equation}
This is $1.3\sigma$ lower than the most recent \planck{} result and $0.3\sigma$ higher than the joint constraint from the latest SPTpol lensing power spectrum and BAO data \citep{bianchini20}, though we expect a mild correlation with the latter result due to the partially shared sky area of the surveys.
The SPT-3G 2018 value is in good agreement with local structure measurements: it is $1.0\sigma$ higher than the latest constraints from the Kilo-Degree Survey (KiDS) \citep{heymans20}, $0.5\sigma$ lower than the Dark Energy Survey (DES) Year 1 results \citep{des17} and $0.2\sigma$ higher than the SZ-selected galaxy cluster measurement from the SPT-SZ survey \citep{bocquet19}.
This agreement also holds true for the combined growth structure parameter.
SPT-3G 2018 infers $S_8 = \sigma_8 \sqrt{\Omega_m/0.3} = 0.779 \pm 0.041$, which is within $0.3\sigma$, $0.1\sigma$ and $1.3\sigma$ of the KiDS, DES, and \planck{} results, respectively.
Adjusting the definition of $S_8$ to match the findings of \cite{bocquet19} based on SZ-clusters, we find the values to agree within $0.5\sigma$.

Adding information from baryon acoustic oscillation (BAO) measurements \citep{alam17, blomqvist19} does not shift the best-fit values of \lcdm\ parameters appreciably.
However, it tightens the constraint on the density of cold dark matter by a factor of $2.4$.
This translates into a refined measurement of the Hubble constant of $H_0 = 68.27 \pm 0.63$\,km/s/Mpc, which is comparable to the precision of \planck{} and disfavors an expansion rate at present day greater than $70$\,km/s/Mpc at $2.8\sigma$.
The constraints on matter clustering are similarly improved through the inclusion of BAO data by a factor of $1.6$ to $0.794 \pm 0.010$ for $\sigma_8$ and by a factor of $2.2$ for $S_8$ to $0.792 \pm 0.018$.
The joint SPT-3G and BAO constraint on $\sigma_8$ is within $1.2\sigma$ of the latest result of KiDS, $0.4\sigma$ of DES, $0.3\sigma$ of SZ-clusters, and $1.5\sigma$ of \planck{}.
Furthermore, this result is consistent with the joint SPTpol lensing and BAO constraint on $\sigma_8$ at $0.6\sigma$.
The joint SPT-3G and BAO constraint on $S_8$ is within $1.0\sigma$ of the latest result of KiDS, $0.6\sigma$ of DES, $1.0\sigma$ of SZ-clusters, and $1.7\sigma$ of \planck{}.

From SPT-3G data alone, we constrain $n_s = 0.999 \pm 0.019$.
While this is slightly higher than the \planck{} result, a $1.8\sigma$ offset is not statistically anomalous, especially when analyzed in the context of the full five-dimensional parameter space.
Nevertheless, we point out that other ground-based CMB experiments have observed similar trends: the constraints from SPTpol 500\,deg$^2$ and ACT DR4 lie $1.3\sigma$ and $1.1\sigma$ above the \planck{} value, respectively \citep{henning18, aiola20}.
We explore this facet of the data further in \S \ref{sec:datasplit_params}.

More generally, our results match those of other contemporary CMB experiments.
Given the small shared sky area between SPT-3G 2018 and \planck{}, we neglect correlations and quantify the difference across the five independent \lcdm{} model parameters.
We obtain $\chi^2 = 8.8$, which corresponds to a PTE of $0.12$ and indicates that the two datasets are consistent.

We confirm that the SPT-3G 2018 dataset is consistent with the \lcdm{} model by comparing the full set of multifrequency $\EE$ and $\TE$ bandpowers to the best-fit \lcdm{} model.
We quantify the goodness of fit by calculating the associated $\chi^2$ over the 528 bandpowers, finding $\chi^2=513.0$.
Since nuisance parameters are dominated by their priors, we account for the five free \lcdm\ parameters in translating this to the PTE of $0.61$.
Comparing the best-fit model to the $\EE$ ($\TE$) bandpowers individually we find $\chi^2=273.2\,(224.2)$.
We conclude that the \lcdm{} model provides a good fit to the SPT-3G 2018 dataset.
The $\EE$ and $\TE$ minimum-variance bandpowers and residuals to the best-fit model are shown in Figure~\ref{fig:ee_mv} and Figure~\ref{fig:te_mv}, respectively.

%----------------------------------
% EE MV bandpowers figure
%----------------------------------
\begin{figure*}[ht!]
\includegraphics[width=17.2cm]{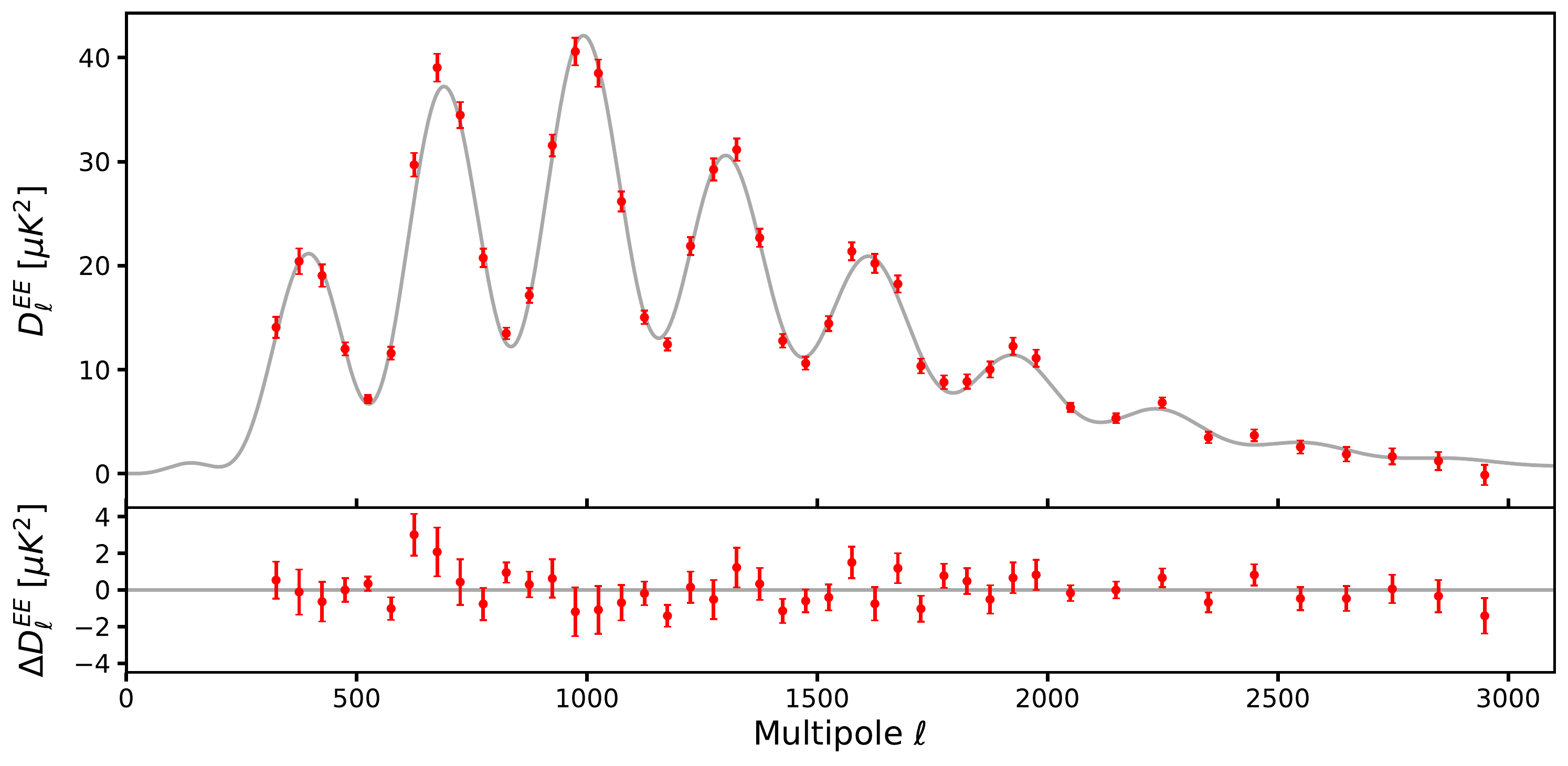}
\centering
\caption{
Minimum-variance $\EE$ bandpowers formed from the six auto- and cross-frequency power spectra and the residuals against the SPT-3G best-fit \lcdm\ model.
Uncertainties are the square root of the diagonal elements of the covariance matrix and do not include beam or calibration uncertainties.
}
\label{fig:ee_mv}
\end{figure*}

%----------------------------------
% TE MV bandpowers figure
%----------------------------------
\begin{figure*}[ht!]
\includegraphics[width=17.2cm]{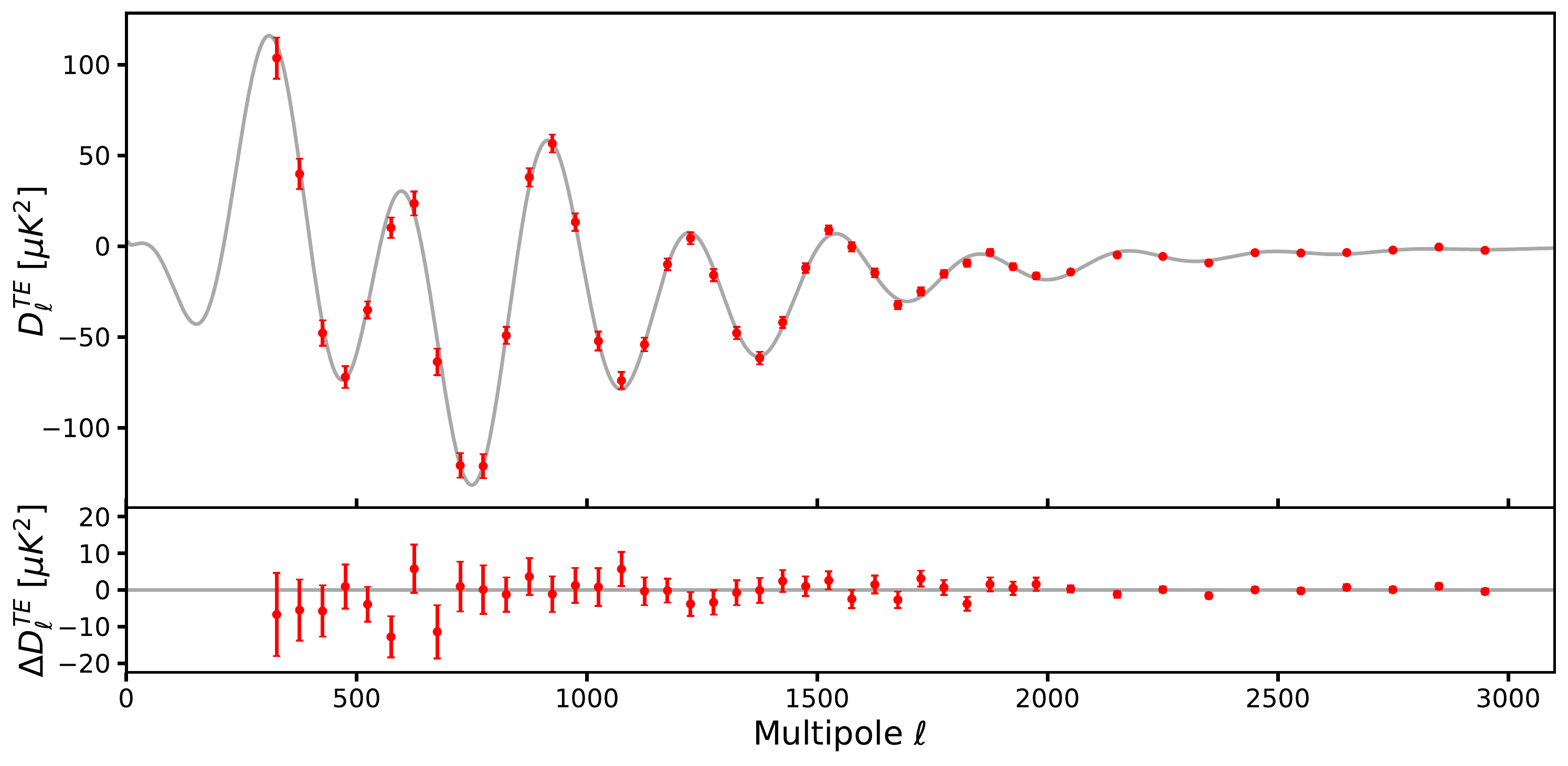}
\centering
\caption{
Minimum-variance $\TE$ bandpowers formed from the six auto- and cross-frequency power spectra and the residuals against the SPT-3G best-fit \lcdm\ model.
Uncertainties are the square root of the diagonal elements of the covariance matrix and do not include beam or calibration uncertainties.
}
\label{fig:te_mv}
\end{figure*}

\subsection{Gravitational Lensing and $\boldsymbol{A_L}$}
\label{sec:Alens}

Our view of the $z=1100$ universe is distorted by the gravitational lensing of CMB photons due to intervening matter between us and the surface of last scattering.
This adds information about the low-redshift universe and results in a smoothing of the acoustic peaks of the CMB power spectra.
The magnitude of this effect is determined by the power spectrum of the lensing potential, which is derived from the six \lcdm{} parameters in the standard cosmological model.
When allowing for a free scaling of the lensing power spectrum, represented by the parameter $A_L$ \citep{calabrese08}, CMB power spectra from \planck\ have shown a preference for lensing $2.8\sigma$ beyond the \lcdm\ prediction of unity with $A_L = 1.180 \pm 0.065$ \citep{planck18-6}.
\citetalias{henning18} report an $A_L$ value below unity at $1.4\sigma$ with $A_L = 0.81 \pm 0.14$.

Introducing the lensing amplitude as a free parameter in our analysis, the SPT-3G 2018 $\EE/\TE$ dataset produces the constraints summarized in Table \ref{tab:lcdm_alens_param_table}.
The core \lcdm\ model parameters do not shift appreciably, and we report a lensing amplitude of
\begin{equation}
A_L = 0.98 \pm 0.12.
\end{equation}
We conclude that the SPT-3G 2018 $\EE/\TE$ dataset is consistent with the level of gravitational lensing expected by the standard model.
The reported lensing amplitude falls within $1.5\sigma$ and $0.9\sigma$ of the aforementioned \planck\ and \citetalias{henning18} results, respectively.

\begin{table}[ht!]
\def\arraystretch{1.2}
\footnotesize
\setlength{\tabcolsep}{3pt}
\centering
\begin{tabular}{c D{+}{\,\pm\,}{15}}
\hline\hline
\multicolumn{2}{c}{SPT-3G}\\
\hline
\multicolumn{2}{l}{Free}\\
$\Omega_b h^2$ &
  0.02242+0.00033 (0.02242)\\
$\Omega_c h^2$ &
  0.1161+0.0056 (0.1165)\\
$100\theta_{\rm MC}$ &
  1.03956+0.00081 (1.03949)\\
$10^9 A_s e^{-2\tau}$ &
  1.827+0.045 (1.83)\\
$n_s$ &
  0.995+0.024 (0.993)\\
$A_{L}$ &
  0.98+0.12 (0.96)\\
\hline
\multicolumn{2}{l}{Derived}\\
$\Omega_\Lambda$ &
  0.701+0.032 (0.699)\\
$H_0$ &
  68.4+2.3 (68.2)\\
$\sigma_8$ &
  0.793+0.022 (0.795)\\
$S_8$ &
  0.792+0.062 (0.795)\\
${\rm{Age}}/{\rm{Gyr}}$ &
  13.814+0.062 (13.82)\\
\hline
\end{tabular}
\caption[
$\Lambda$CDM+$A_L$ parameter constraints.
]{
Marginalized $\Lambda$CDM+A$_L$ parameter constraints and 68\% errors from SPT-3G. Best-fit values are given in parentheses.
}
\label{tab:lcdm_alens_param_table}
\end{table}

\subsection{Interpretation of Data Split Preferences}
\label{sec:datasplit_params}

One motivation for studying the CMB polarization anisotropies is that comparing results from the temperature and polarization power spectra yields a stringent test of the \lcdm{} cosmological model.
Thus while we did not find the parameter differences between subsets of the SPT-3G data to be statistically significant in \S\ref{sec:int_consistency}, it is still interesting to examine these parameter shifts for possible hints of physics beyond the standard cosmological model.
We show the parameter constraints from each data split in Figure \ref{fig:datasplit_param_fluct}.
We continue to quantify the significance of parameter shifts as introduced in \S\ref{sec:int_consistency}, by using the difference of the parameter covariances of the full dataset and the given data split.

Examining the best-fit \lcdm{} parameters of the different subsets of the SPT-3G 2018 $\EE/\TE$ dataset reveals two interesting features.
First, the high-$\ell$ dataset prefers a scalar spectral index above unity, $n_s = 1.048 \pm 0.031$, which corresponds to a $2.0\sigma$ shift from the full dataset.
With $n_s = 1.053 \pm 0.052$, the $\EE$ spectra prefer a higher scalar spectral index than the high-$\ell$ dataset.
However, due to their comparatively poor constraining power for this parameter, the $\EE$ constraint is only offset by $1.1\sigma$ from the full dataset.
The higher value of $n_s$ lowers the combined amplitude parameter, as the two are mildly degenerate over the limited $\ell$-range: the \mbox{high-$\ell$} data prefers $10^9 A_s e^{-2\tau} = 1.750 \pm 0.055$.
These values lie $2.0\sigma$ and $1.8\sigma$ away from the baseline constraints, respectively.
Focusing on the scalar spectral index and the combined amplitude parameter individually, the probability of a shift of the observed size or larger from the full dataset constraint is $2.4\%$ and $3.7\%$, respectively.
We repeat that fluctuations of this size are statistically not uncommon, especially when viewed in the context of the full five-dimensional parameter space.

A raised scalar spectral index corresponds to a power increase in the damping tail compared to intermediate angular scales.
The damping tail is sensitive to an array of interesting physics beyond the standard model, such as extra energy injection in the early universe.
This can be explored by allowing the number of relativistic species at recombination, $N_{\rm eff}$, to vary from the standard model prediction, breaking big-bang nucleosynthesis consistency by changing the primordial helium abundance, $Y_{\rm P}$, or both.
We explore the constraints the SPT-3G 2018 $\EE/\TE$ dataset places on these \lcdm\ model extensions in a forthcoming paper.

The second interesting feature of the data splits is a preference in the $\EE$ spectra for a lower cold dark matter density, $\Omega_c h^2 = 0.0987 \pm 0.0084$, than the $\TE$ spectra, $\Omega_c h^2 = 0.1259 \pm 0.0063$.
These values are $2.2\sigma$ and $2.1\sigma$ away from the full dataset constraints, respectively.
Consequently, different constraints of the Hubble constant are obtained: $H_0=76.4 \pm 4.1\,\mathrm{km/s/Mpc}$ from the $\EE$ spectra and $H_0=65.0 \pm 2.1\,\mathrm{km/s/Mpc}$ from the $\TE$ spectra.
Adding BAO information regularizes the matter density fluctuations and consequently the Hubble constant values: $\EE$ spectra then prefer $H_0=68.7 \pm 1.0\,\mathrm{km/s/Mpc}$ and $\TE$ spectra $H_0=67.82 \pm 0.66\,\mathrm{km/s/Mpc}$.
While this signals that solutions to the Hubble tension are difficult to achieve within the \lcdm{} model, model extensions may reconcile the discrepancy between high- and low-redshift probes \citep{knox19}.

A different way of reconciling the matter content inferred by $\EE$ and $\TE$ spectra, and through this their constraints on the Hubble constant, is by allowing for a free amplitude of the lensing power spectrum.
The matter content implies the strength of lensing-induced acoustic-peak smoothing, which results in a mild degeneracy between the matter density and $A_L$.
This effect was seen in \citetalias{henning18}, where differences in constraints on cosmological parameters to \planck{} were alleviated through this model extension.
Indeed, we find for \mbox{SPT-3G} 2018 that the $\EE$ spectra prefer $A_L = 0.71^{+0.31}_{-0.30}$ and the $\TE$ spectra $A_L = 0.99 \pm 0.29$, while constraints on $\Omega_c h^2$ are brought closer together.
This is mirrored by the Hubble constant, which is constrained to $H_0=68.1 \pm 9.3\,\mathrm{km/s/Mpc}$ and $H_0=64.6 \pm 3.9\,\mathrm{km/s/Mpc}$ by the $\EE$ and $\TE$ spectra, respectively.

Similar trends for low- and high-multipole data as well as $\EE$ and $\TE$ spectra were reported by \citetalias{henning18} and \cite{aiola20}.
We compile the different Hubble constant measurements in Figure \ref{fig:H0_EETE}.
While the statistical evidence is currently too low, if future polarization measurements amplify this potential tension with cosmological parameters inferred from the temperature anisotropies, these trends may be signs for physics beyond the standard model of cosmology.

%----------------------------------
% Datasplit constraints figure
%----------------------------------
\begin{figure*}[ht!]
\includegraphics[width=17.2cm]{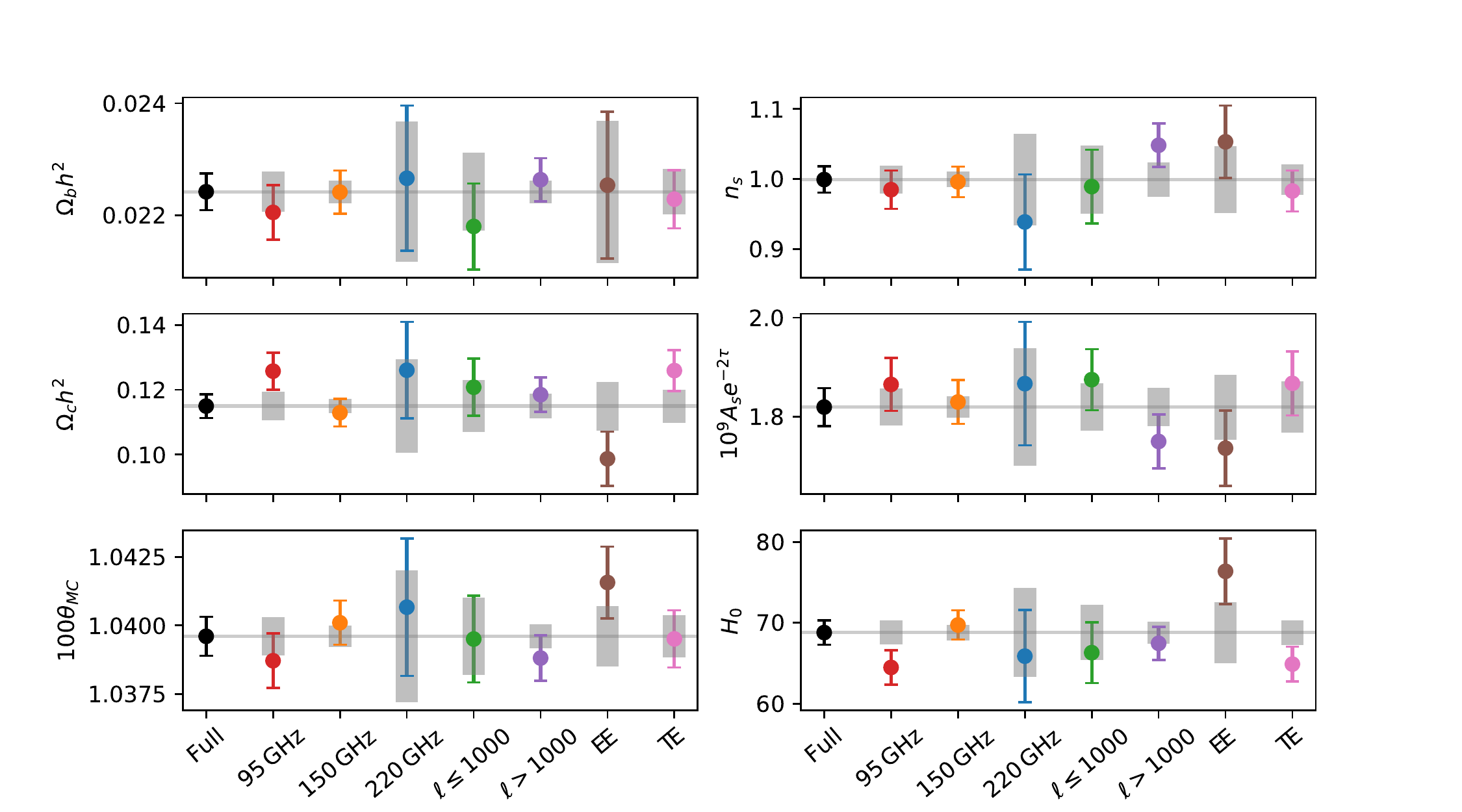}
\centering
\caption{
Parameter constraints from various subsets of the SPT-3G 2018 $\EE/\TE$ dataset.
The gray boxes correspond to the expected level of statistical fluctuation \citep{gratton19}.
}
\label{fig:datasplit_param_fluct}
\end{figure*}

%----------------------------------
% H0 and omega_c figure
%----------------------------------
\begin{figure}[ht!]
\includegraphics[width=8.6cm]{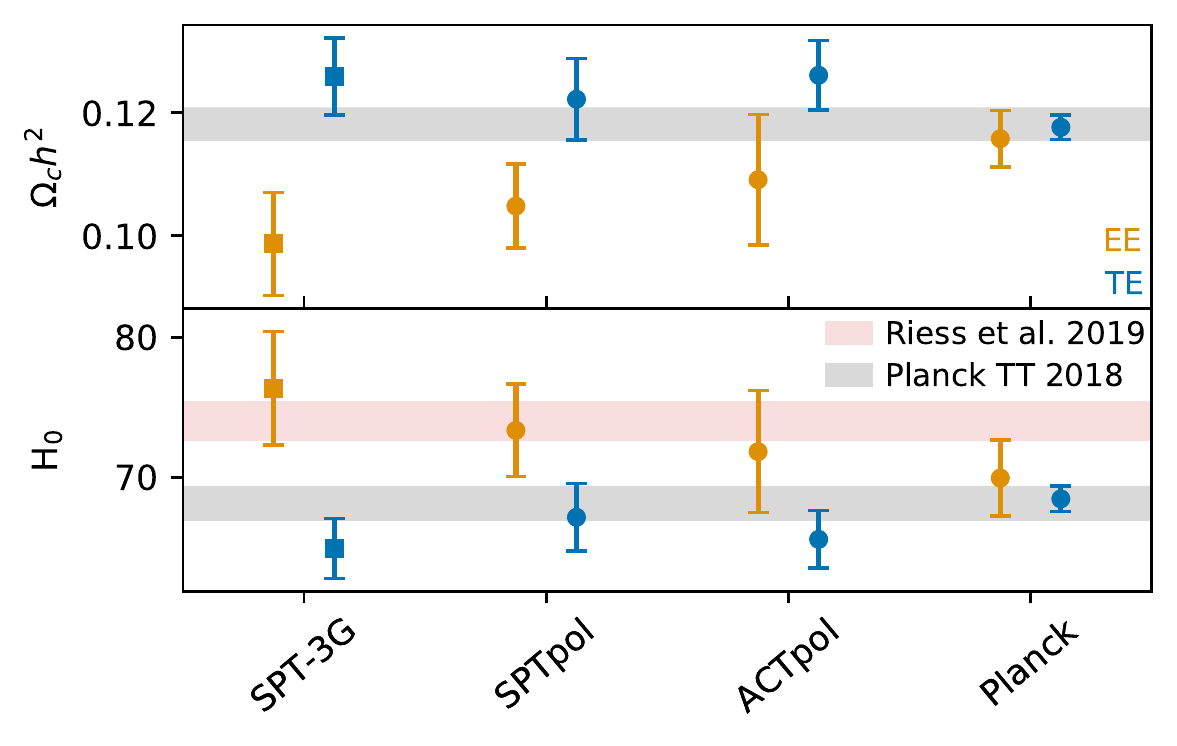}
\centering
\caption{
Constraints on the Hubble constant and cold dark matter density from contemporary CMB experiments.
For each experiment, the constraints from $\EE$ and $\TE$ power spectra are shown in orange and in blue, respectively.
The results highlighted here are from this work, \citetalias{henning18}, \citep{choi20} and \cite{planck18-6}.
We also show the $1\sigma$ constraints on $H_0$ from the most recent Cepheid-calibrated distance ladder measurement \textit{(red band)} \citep{riess19} and the latest \planck\ $\TT$-based constraints \textit{(gray band)} \citep{planck18-6} for reference.
}
\label{fig:H0_EETE}
\end{figure}

\subsection{SPT-3G + Planck}
\label{sec:params_spt_planck}

The \planck{} dataset provides the most precise measurement of the temperature and polarization anisotropies of the CMB on large angular scales, while the \mbox{SPT-3G} 2018 $\EE/\TE$ dataset provides sensitive information on intermediate and small angular scales.
The two datasets thus naturally complement each other, and we may obtain joint constraints by combining them at the likelihood level.
Given the small area shared by the two surveys, we expect correlations to be negligible.

We report joint constraints on \lcdm{} parameters from the \textsc{base\_plikHM\_TTTEEE\_lowl\_lowE} \planck{} and SPT-3G 2018 $\EE/\TE$ datasets in Table \ref{tab:lcdm_param_table}.
We present associated the 1D and 2D marginalized posteriors in Figure \ref{fig:triangle_spt_planck}.
The inclusion of SPT-3G data does not alter the \planck{} best-fit values significantly.

We use the determinants of the \lcdm{} parameter covariance matrices as a measure of the marginalized parameter-space volume.
The ratio of the matrix determinants for SPT-3G 2018 $\EE/\TE$ combined with \planck{} to \planck{}-alone is $0.46$.
This corresponds to a reduction of the 68\% confidence region in six-dimensional \lcdm{} parameter space by a factor of $1.5$.

%----------------------------------
% 3G + Planck triangle plot figure
%----------------------------------
\begin{figure*}[ht!]
\includegraphics[width=17.2cm]{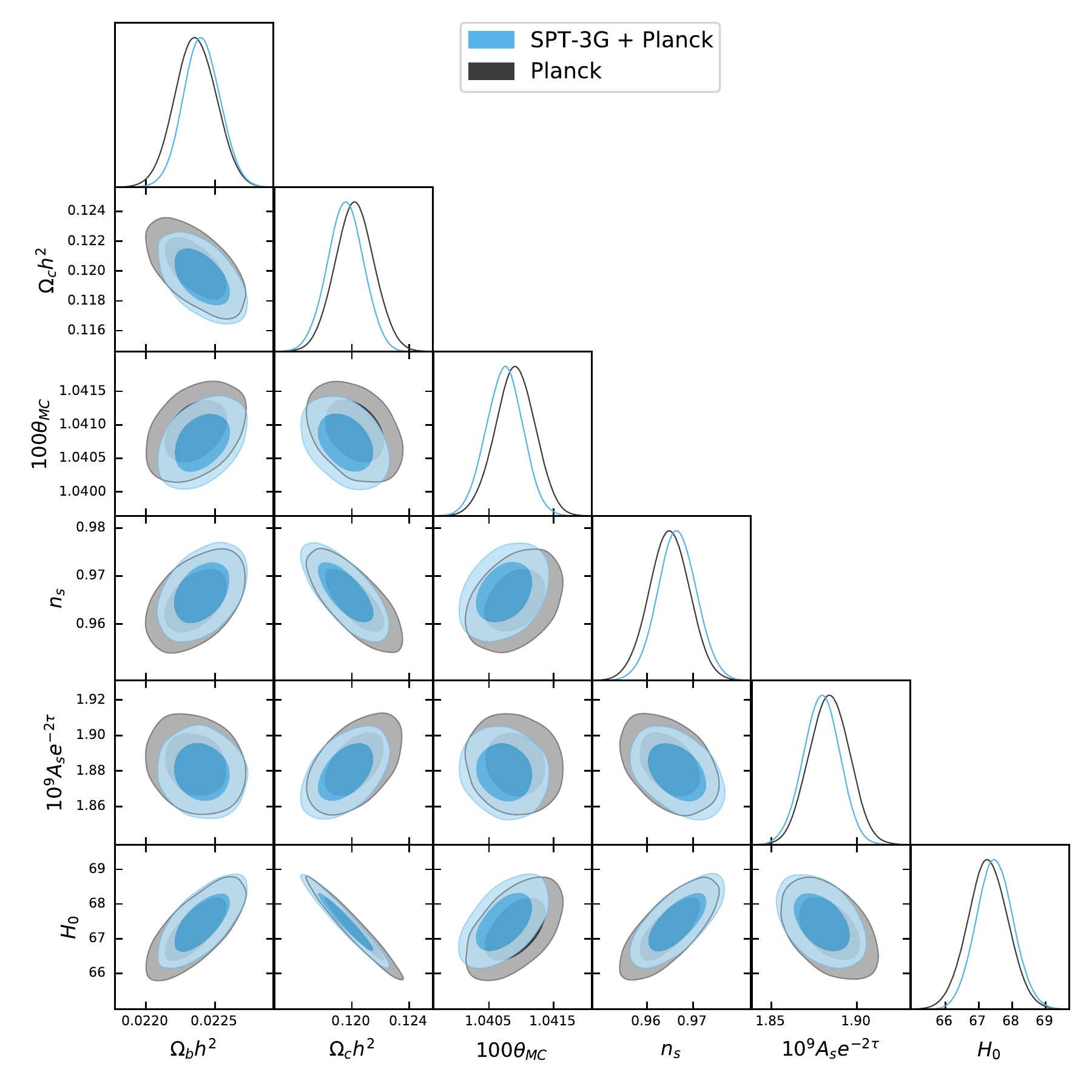}
\centering
\caption{
Joint marginalized constraints on \lcdm{} parameters and the Hubble constant from the SPT-3G 2018 $\EE/\TE$ + \planck{} \citep{planck18-5} datasets.
\planck{}-only constraints are shown for comparison.
}
\label{fig:triangle_spt_planck}
\end{figure*}

%%%%%%%%%%%%%%%%%%%%%%%%%%%%%%%%%%%%
% CONCLUSION
%%%%%%%%%%%%%%%%%%%%%%%%%%%%%%%%%%%%
\section{Conclusion}
\label{sec:conclusion}
\medskip
In this work we have presented the first results from SPT-3G data.
Analyzing 2018 data alone, we have produced high-precision measurements of the CMB $E$-mode angular auto-power and temperature-$E$-mode cross-power spectra over the multipole range \mbox{$300 \le \ell < 3000$}.
The reported bandpowers are the first multifrequency $\EE$ and $\TE$ measurements produced by an instrument on SPT, and they improve upon previous SPT measurements across the multipole ranges \mbox{$300 \le \ell \le 1400$} for $\EE$ and \mbox{$300 \le \ell \le 1700$} for $\TE$, resulting in tighter constraints on cosmological parameters.

\smallskip
The SPT-3G 2018 $\EE/\TE$ dataset is consistent with the \lcdm{} model.
Analyzing constraints from the 95, 150, and 220\,GHz auto-frequency spectra, the $\ell<1000$ versus $\ell>1000$ data, and the $\EE$ and $\TE$ spectra individually, we find no signs of significant internal tension.

\smallskip
The constraints on \lcdm{} model parameters generally agree with other contemporary CMB experiments.
We report a value of the Hubble constant of \mbox{$H_0 = 68.8 \pm 1.5\,\mathrm{km/s/Mpc}$}, in line with the CMB-based measurements of \planck{} and ACT, as well as TRGB-calibrated local distance ladder data.
This is in contrast with the higher values found by Cepheid-calibrated distance ladder data and time-delay measurements from gravitationally lensed quasars.
However, we note an interesting trend in CMB-based constraints from several experiments, including our own, which have consistently found high values of the Hubble constant when analyzing $\EE$ polarization spectra.
The current level of tension between polarization- and temperature-based constraints is not statistically significant, but presents an interesting direction for further investigation.
The SPT-3G 2018 dataset constrains matter-clustering to \mbox{$\sigma_8 = 0.789 \pm 0.016$}, \mbox{$S_8 = 0.779 \pm 0.041$}, which is consistent with other CMB-based measurements and low-redshift probes.

\smallskip
Expanding the \lcdm{} model to allow for a modified amplitude of the lensing power spectrum does not shift parameter constraints appreciably.
With \mbox{$A_L = 0.98 \pm 0.12$}, the SPT-3G 2018 dataset is consistent with the standard model prediction.

\smallskip
By combining the SPT-3G 2018 $\EE/\TE$ and \planck{} datasets at the likelihood level, we mildly improve the marginalized 1D constraints over \planck{} data alone.
The volume of the 68\% confidence region is reduced by a factor of $1.5$ in six-dimensional \lcdm{} parameter space.

\smallskip
Lastly, we note that the high-precision measurements presented in this work use only one half of one observing season of data, which was taken with nearly half the number of currently operating detectors not contributing.
With SPT-3G operating at its full capacity since the start of 2019, we now have data from two full observing seasons on disk, with combined map depths 3--4$\times$ deeper than what was used in this analysis.
Future SPT-3G results will measure the CMB polarization power spectra with exquisite sensitivity on intermediate and small angular scales, constraining physics beyond the standard model with unprecedented precision.

\acknowledgments
The South Pole Telescope program is supported by the National Science Foundation (NSF) through grants PLR-1248097 and OPP-1852617.
Partial support is also provided by the NSF Physics Frontier Center grant PHY-1125897 to the Kavli Institute of Cosmological Physics at the University of Chicago, the Kavli Foundation, and the Gordon and Betty Moore Foundation through grant GBMF\#947 to the University of Chicago.
Argonne National Laboratory's work was supported by the U.S. Department of Energy, Office of High Energy Physics, under contract DE-AC02-06CH11357.
Work at Fermi National Accelerator Laboratory, a DOE-OS, HEP User Facility managed by the Fermi Research Alliance, LLC, was supported under Contract No. DE-AC02-07CH11359.
The Cardiff authors acknowledge support from the UK Science and Technologies Facilities Council (STFC).
The CU Boulder group acknowledges support from NSF AST-0956135.  
The IAP authors acknowledge support from the Centre National d'\'{E}tudes Spatiales (CNES).
JV acknowledges support from the Sloan Foundation.
The Melbourne authors acknowledge support from an Australian Research Council Future Fellowship (FT150100074).
The McGill authors acknowledge funding from the Natural Sciences and Engineering Research Council of Canada, Canadian Institute for Advanced Research, and the Fonds de recherche du Qu\'ebec Nature et technologies.
NWH acknowledges support from NSF CAREER grant AST-0956135.
The UCLA and MSU authors acknowledge support from NSF AST-1716965 and CSSI-1835865.
This research was done using resources provided by the Open Science Grid \citep{pordes07, sfiligoi09}, which is supported by the National Science Foundation award 1148698, and the U.S. Department of Energy's Office of Science.
This research used resources of the National Energy Research Scientific Computing Center (NERSC), a U.S. Department of Energy Office of Science User Facility operated under Contract No. DE-AC02-05CH11231.
Some of the results in this paper have been derived using the healpy and HEALPix packages.
The data analysis pipeline also uses the scientific python stack \citep{hunter07, jones01, vanDerWalt11}.
\vspace*{-3mm}
\appendix*
\section{\textit{EE} and \textit{TE} Bandpower Tables}
\label{app:appendix}
\vspace*{-2mm}
The $\EE$ and $\TE$ bandpowers from the six sets of cross-frequency power spectra are presented in Table~\ref{tab:ee_bandpowers_table} and Table~\ref{tab:te_bandpowers_table}, respectively.
\begin{table*}[h!]
\small
\setlength{\tabcolsep}{4pt}
\def\arraystretch{1.2}
\centering
\begin{tabular}{c c | D{.}{.}{1} D{.}{.}{-1} | D{.}{.}{1} D{.}{.}{-1} | D{.}{.}{1} D{.}{.}{-1} | D{.}{.}{1} D{.}{.}{-1} | D{.}{.}{1} D{.}{.}{-1} | D{.}{.}{1} D{.}{.}{-1} }
\hline\hline
\rule{0pt}{3ex} \multirow{2}{*}{$\ell$ Range} & \multirow{2}{*}{$\ell_\mathrm{eff}$} & \multicolumn{2}{c|}{$\mathrm{95\times95\,GHz}$} & \multicolumn{2}{c|}{$\mathrm{95\times150\,GHz}$} & \multicolumn{2}{c|}{$\mathrm{95\times220\,GHz}$} & \multicolumn{2}{c|}{$\mathrm{150\times150\,GHz}$} & \multicolumn{2}{c|}{$\mathrm{150\times220\,GHz}$} & \multicolumn{2}{c}{$\mathrm{220\times220\,GHz}$} \\\cline{3-14}
 & & D_b & \sigma & D_b & \sigma & D_b & \sigma & D_b & \sigma & D_b & \sigma & D_b & \sigma \\[2pt]
\hline
300 -- 349 & 325 & 13.1 & 1.1 & 12.8 & 1.1 & 12.1 & 1.3 & 13.2 & 1.1 & 12.7 & 1.3 & 12.1 & 2.0 \\
350 -- 399 & 375 & 19.7 & 1.3 & 20.5 & 1.3 & 19.0 & 1.5 & 21.1 & 1.3 & 19.9 & 1.5 & 18.0 & 2.3 \\
400 -- 449 & 425 & 19.0 & 1.2 & 18.8 & 1.1 & 17.9 & 1.3 & 19.1 & 1.1 & 18.4 & 1.3 & 17.7 & 2.1 \\
450 -- 499 & 475 & 11.2 & 0.7 & 12.0 & 0.7 & 11.1 & 0.9 & 12.5 & 0.7 & 11.1 & 0.9 & 9.4 & 1.7 \\
500 -- 549 & 524 & 7.1 & 0.5 & 7.3 & 0.4 & 7.6 & 0.7 & 7.0 & 0.4 & 8.3 & 0.6 & 9.4 & 1.6 \\
550 -- 599 & 575 & 11.1 & 0.7 & 11.3 & 0.6 & 12.2 & 0.9 & 11.8 & 0.7 & 11.8 & 0.9 & 11.5 & 1.9 \\
600 -- 649 & 624 & 29.0 & 1.3 & 29.4 & 1.2 & 29.1 & 1.5 & 30.1 & 1.2 & 29.8 & 1.4 & 34.3 & 2.6 \\
650 -- 699 & 674 & 39.0 & 1.5 & 39.1 & 1.3 & 39.5 & 1.7 & 38.9 & 1.4 & 39.7 & 1.7 & 40.9 & 2.9 \\
700 -- 749 & 725 & 33.6 & 1.4 & 34.4 & 1.3 & 33.1 & 1.7 & 35.0 & 1.3 & 34.1 & 1.6 & 32.4 & 3.0 \\
750 -- 799 & 774 & 21.2 & 1.1 & 20.8 & 0.9 & 22.0 & 1.3 & 20.4 & 0.9 & 21.3 & 1.2 & 22.8 & 2.7 \\
800 -- 849 & 824 & 13.2 & 0.8 & 13.3 & 0.6 & 13.2 & 1.0 & 13.7 & 0.6 & 13.4 & 0.9 & 13.6 & 2.6 \\
850 -- 899 & 874 & 16.9 & 0.9 & 17.2 & 0.7 & 17.8 & 1.2 & 17.0 & 0.8 & 17.7 & 1.1 & 19.1 & 2.9 \\
900 -- 949 & 924 & 31.8 & 1.3 & 31.4 & 1.1 & 30.8 & 1.7 & 31.6 & 1.1 & 32.3 & 1.6 & 29.6 & 3.5 \\
950 -- 999 & 974 & 41.2 & 1.6 & 40.4 & 1.4 & 40.7 & 2.0 & 40.7 & 1.4 & 39.9 & 1.9 & 36.9 & 4.0 \\
1000 -- 1049 & 1024 & 39.4 & 1.6 & 38.4 & 1.3 & 39.3 & 2.0 & 38.5 & 1.4 & 37.3 & 1.9 & 40.7 & 4.2 \\
1050 -- 1099 & 1075 & 26.1 & 1.3 & 26.3 & 1.0 & 24.9 & 1.7 & 26.4 & 1.1 & 25.3 & 1.6 & 20.4 & 4.0 \\
1100 -- 1149 & 1124 & 15.5 & 1.0 & 15.2 & 0.7 & 14.6 & 1.4 & 15.0 & 0.7 & 13.9 & 1.2 & 10.7 & 3.9 \\
1150 -- 1199 & 1174 & 13.1 & 1.0 & 12.3 & 0.7 & 10.8 & 1.5 & 12.6 & 0.7 & 12.1 & 1.2 & 12.6 & 4.1 \\
1200 -- 1249 & 1224 & 20.6 & 1.3 & 21.8 & 0.9 & 23.9 & 1.8 & 22.1 & 1.0 & 22.3 & 1.6 & 18.0 & 4.6 \\
1250 -- 1299 & 1275 & 29.9 & 1.5 & 29.2 & 1.1 & 28.5 & 2.1 & 29.6 & 1.2 & 26.9 & 1.9 & 26.9 & 5.2 \\
1300 -- 1349 & 1325 & 31.2 & 1.6 & 30.9 & 1.1 & 28.5 & 2.2 & 32.1 & 1.2 & 28.5 & 1.9 & 24.4 & 5.5 \\
1350 -- 1399 & 1374 & 24.1 & 1.4 & 22.4 & 1.0 & 22.2 & 2.1 & 22.2 & 1.0 & 25.0 & 1.8 & 40.0 & 5.7 \\
1400 -- 1449 & 1424 & 14.1 & 1.3 & 13.0 & 0.8 & 11.9 & 1.9 & 12.6 & 0.8 & 11.3 & 1.6 & 5.5 & 5.9 \\
1450 -- 1499 & 1474 & 10.9 & 1.3 & 10.2 & 0.7 & 11.4 & 2.0 & 10.4 & 0.8 & 13.4 & 1.6 & 19.2 & 6.2 \\
1500 -- 1549 & 1524 & 15.0 & 1.4 & 15.4 & 0.8 & 12.6 & 2.2 & 14.1 & 0.9 & 11.1 & 1.8 & 8.0 & 6.7 \\
1550 -- 1599 & 1574 & 22.1 & 1.6 & 20.9 & 1.0 & 22.1 & 2.4 & 21.1 & 1.0 & 24.1 & 2.0 & 23.8 & 7.2 \\
1600 -- 1649 & 1624 & 17.6 & 1.7 & 20.0 & 1.1 & 20.4 & 2.6 & 20.7 & 1.1 & 21.7 & 2.1 & 24.0 & 7.6 \\
1650 -- 1699 & 1674 & 19.2 & 1.7 & 18.4 & 1.0 & 14.7 & 2.6 & 18.1 & 1.0 & 18.9 & 2.0 & 12.9 & 8.0 \\
1700 -- 1749 & 1724 & 7.4 & 1.7 & 10.2 & 0.9 & 10.8 & 2.6 & 10.6 & 0.9 & 14.2 & 2.0 & 0.3 & 8.3 \\
1750 -- 1799 & 1775 & 10.1 & 1.7 & 8.7 & 0.9 & 11.3 & 2.7 & 8.5 & 0.9 & 8.0 & 2.0 & 14.9 & 8.8 \\
1800 -- 1849 & 1825 & 8.3 & 1.8 & 9.0 & 0.9 & 5.8 & 2.9 & 9.6 & 0.9 & 5.4 & 2.1 & -0.4 & 9.4 \\
1850 -- 1899 & 1874 & 9.7 & 2.0 & 9.8 & 1.0 & 9.6 & 3.2 & 9.8 & 1.0 & 13.1 & 2.3 & 14.2 & 10.0 \\
1900 -- 1949 & 1924 & 12.7 & 2.1 & 12.9 & 1.1 & 18.2 & 3.3 & 12.0 & 1.1 & 7.8 & 2.4 & 0.6 & 10.6 \\
1950 -- 1999 & 1975 & 12.4 & 2.2 & 10.2 & 1.1 & 8.9 & 3.5 & 11.4 & 1.1 & 13.9 & 2.5 & 6.2 & 11.2 \\
2000 -- 2099 & 2049 & 6.7 & 1.2 & 6.3 & 0.6 & 7.9 & 2.0 & 6.3 & 0.6 & 6.2 & 1.4 & 4.9 & 6.7 \\
2100 -- 2199 & 2148 & 5.3 & 1.4 & 5.6 & 0.7 & 1.1 & 2.3 & 5.4 & 0.7 & 5.4 & 1.6 & 9.0 & 7.6 \\
2200 -- 2299 & 2248 & 7.3 & 1.6 & 7.6 & 0.8 & 6.8 & 2.6 & 6.0 & 0.7 & 7.2 & 1.8 & 8.7 & 8.6 \\
2300 -- 2399 & 2348 & 1.2 & 1.8 & 2.6 & 0.8 & 4.2 & 2.9 & 4.9 & 0.8 & 1.0 & 1.9 & 13.3 & 9.4 \\
2400 -- 2499 & 2448 & 6.8 & 2.0 & 4.0 & 0.9 & 5.2 & 3.2 & 2.6 & 0.8 & 5.2 & 2.1 & -0.8 & 10.4 \\
2500 -- 2599 & 2548 & 2.9 & 2.2 & 2.5 & 1.0 & 0.2 & 3.5 & 2.6 & 0.9 & 3.0 & 2.3 & -2.5 & 11.5 \\
2600 -- 2699 & 2648 & 5.9 & 2.5 & 0.5 & 1.1 & -0.1 & 4.0 & 2.3 & 1.0 & 2.1 & 2.5 & 10.5 & 12.6 \\
2700 -- 2799 & 2748 & -0.9 & 2.8 & 0.8 & 1.3 & 9.5 & 4.5 & 2.0 & 1.1 & 3.4 & 2.8 & -6.4 & 14.1 \\
2800 -- 2899 & 2848 & 0.6 & 3.2 & 3.0 & 1.4 & 4.5 & 5.0 & 0.5 & 1.3 & -3.2 & 3.2 & -5.8 & 15.7 \\
2900 -- 2999 & 2948 & -1.2 & 3.6 & -2.4 & 1.6 & -7.2 & 5.6 & 1.0 & 1.4 & 7.4 & 3.5 & -3.6 & 17.1 \\
\hline
\end{tabular}
\caption{
$\EE$ bandpowers $D_b$ for the six cross-frequency power spectra, along with angular multipole range, bandpower window function-weighted multipole $\ell_\mathrm{eff}$, and associated uncertainty, $\sigma$.
The bandpowers and errors are quoted in units of $\mu$K$^2$.
The reported uncertainties are the square root of the diagonal elements of the covariance matrix and do not include beam or calibration uncertainties.
}
\label{tab:ee_bandpowers_table}
\end{table*}
\begin{table*}[h!]
\small
\setlength{\tabcolsep}{4pt}
\def\arraystretch{1.2}
\centering
\begin{tabular}{c c | D{.}{.}{1} D{.}{.}{-1} | D{.}{.}{1} D{.}{.}{-1} | D{.}{.}{1} D{.}{.}{-1} | D{.}{.}{1} D{.}{.}{-1} | D{.}{.}{1} D{.}{.}{-1} | D{.}{.}{1} D{.}{.}{-1} }
\hline\hline
\rule{0pt}{3ex} \multirow{2}{*}{$\ell$ Range} & \multirow{2}{*}{$\ell_\mathrm{eff}$} & \multicolumn{2}{c|}{$\mathrm{95\times95\,GHz}$} & \multicolumn{2}{c|}{$\mathrm{95\times150\,GHz}$} & \multicolumn{2}{c|}{$\mathrm{95\times220\,GHz}$} & \multicolumn{2}{c|}{$\mathrm{150\times150\,GHz}$} & \multicolumn{2}{c|}{$\mathrm{150\times220\,GHz}$} & \multicolumn{2}{c}{$\mathrm{220\times220\,GHz}$} \\\cline{3-14}
 & & D_b & \sigma & D_b & \sigma & D_b & \sigma & D_b & \sigma & D_b & \sigma & D_b & \sigma \\[2pt]
\hline
300 -- 349 & 326 & 88.4 & 12.0 & 93.2 & 12.1 & 99.8 & 13.7 & 101.1 & 12.7 & 110.5 & 14.0 & 113.7 & 20.3 \\
350 -- 399 & 376 & 43.6 & 8.8 & 42.4 & 8.7 & 36.6 & 10.5 & 42.7 & 9.2 & 40.8 & 10.7 & 40.1 & 17.2 \\
400 -- 449 & 426 & -44.7 & 7.6 & -45.6 & 7.3 & -43.0 & 9.0 & -47.8 & 7.5 & -47.1 & 9.0 & -43.4 & 15.0 \\
450 -- 499 & 475 & -68.8 & 6.7 & -68.9 & 6.2 & -65.0 & 7.8 & -70.0 & 6.4 & -64.5 & 7.7 & -53.2 & 13.2 \\
500 -- 549 & 523 & -34.0 & 5.5 & -34.6 & 5.0 & -48.2 & 6.7 & -34.8 & 5.2 & -46.7 & 6.7 & -58.2 & 12.2 \\
550 -- 599 & 574 & 11.8 & 6.2 & 11.2 & 5.8 & 15.2 & 7.4 & 10.5 & 6.1 & 15.6 & 7.3 & 20.8 & 12.4 \\
600 -- 649 & 625 & 24.1 & 7.0 & 23.8 & 6.7 & 21.5 & 8.1 & 24.5 & 7.0 & 23.1 & 8.1 & 21.4 & 12.8 \\
650 -- 699 & 675 & -63.3 & 7.7 & -63.3 & 7.4 & -58.0 & 8.7 & -63.1 & 7.5 & -59.2 & 8.6 & -60.0 & 13.0 \\
700 -- 749 & 725 & -119.5 & 7.3 & -120.9 & 6.9 & -114.0 & 8.2 & -122.8 & 7.0 & -116.0 & 8.1 & -105.2 & 12.7 \\
750 -- 799 & 774 & -121.2 & 7.2 & -120.4 & 6.7 & -124.1 & 8.3 & -121.3 & 6.8 & -126.2 & 8.1 & -124.6 & 12.9 \\
800 -- 849 & 824 & -52.6 & 5.6 & -50.5 & 4.8 & -43.2 & 6.8 & -48.6 & 5.0 & -40.0 & 6.7 & -25.6 & 12.1 \\
850 -- 899 & 874 & 41.0 & 5.8 & 38.5 & 5.1 & 38.5 & 6.9 & 36.6 & 5.3 & 37.2 & 6.8 & 36.7 & 11.9 \\
900 -- 949 & 924 & 54.5 & 5.5 & 56.0 & 4.9 & 58.9 & 6.6 & 56.9 & 5.1 & 61.5 & 6.5 & 70.4 & 11.3 \\
950 -- 999 & 974 & 12.4 & 5.3 & 13.1 & 4.8 & 14.4 & 6.3 & 13.9 & 5.0 & 13.8 & 6.2 & 18.0 & 10.6 \\
1000 -- 1049 & 1024 & -52.0 & 5.6 & -51.8 & 5.2 & -55.5 & 6.5 & -51.7 & 5.4 & -55.8 & 6.4 & -56.7 & 10.6 \\
1050 -- 1099 & 1075 & -75.6 & 5.3 & -74.6 & 4.7 & -71.9 & 6.2 & -73.7 & 4.9 & -72.1 & 6.1 & -70.1 & 10.4 \\
1100 -- 1149 & 1124 & -48.3 & 4.6 & -52.7 & 3.9 & -58.4 & 5.6 & -55.9 & 4.1 & -60.3 & 5.5 & -66.0 & 10.2 \\
1150 -- 1199 & 1174 & -9.7 & 4.2 & -10.1 & 3.4 & -6.9 & 5.3 & -10.8 & 3.6 & -7.1 & 5.1 & -1.9 & 10.0 \\
1200 -- 1249 & 1224 & 4.9 & 4.1 & 4.3 & 3.4 & 4.2 & 5.1 & 4.3 & 3.6 & 4.3 & 5.0 & 8.3 & 9.8 \\
1250 -- 1299 & 1274 & -15.4 & 4.1 & -15.7 & 3.4 & -17.2 & 5.0 & -16.0 & 3.6 & -16.7 & 4.9 & -16.4 & 9.6 \\
1300 -- 1349 & 1324 & -47.1 & 4.2 & -48.1 & 3.5 & -43.6 & 5.1 & -49.1 & 3.7 & -42.9 & 4.9 & -39.7 & 9.6 \\
1350 -- 1399 & 1374 & -61.8 & 4.3 & -61.8 & 3.5 & -55.3 & 5.3 & -63.0 & 3.7 & -56.8 & 5.1 & -47.5 & 10.0 \\
1400 -- 1449 & 1424 & -41.0 & 4.1 & -41.8 & 3.1 & -41.2 & 5.2 & -42.8 & 3.3 & -41.1 & 5.0 & -30.8 & 10.2 \\
1450 -- 1499 & 1474 & -10.9 & 3.8 & -11.8 & 2.8 & -8.6 & 5.0 & -13.0 & 3.0 & -9.9 & 4.8 & -4.2 & 10.1 \\
1500 -- 1549 & 1524 & 8.4 & 3.6 & 9.0 & 2.6 & 4.8 & 4.7 & 10.2 & 2.8 & 5.9 & 4.5 & -7.4 & 9.8 \\
1550 -- 1599 & 1574 & -3.8 & 3.5 & -0.8 & 2.6 & -4.2 & 4.5 & 1.1 & 2.8 & 0.3 & 4.3 & -5.1 & 9.5 \\
1600 -- 1649 & 1624 & -13.9 & 3.4 & -15.4 & 2.5 & -15.8 & 4.3 & -14.5 & 2.7 & -13.3 & 4.1 & -8.0 & 9.4 \\
1650 -- 1699 & 1674 & -31.0 & 3.3 & -32.0 & 2.4 & -32.4 & 4.3 & -33.1 & 2.5 & -31.7 & 4.0 & -33.1 & 9.5 \\
1700 -- 1749 & 1724 & -21.9 & 3.3 & -24.0 & 2.3 & -25.9 & 4.4 & -25.9 & 2.5 & -26.7 & 4.1 & -25.1 & 9.8 \\
1750 -- 1799 & 1775 & -15.7 & 3.3 & -15.1 & 2.2 & -17.6 & 4.4 & -14.7 & 2.4 & -17.4 & 4.1 & -21.5 & 10.0 \\
1800 -- 1849 & 1824 & -14.1 & 3.2 & -10.0 & 2.1 & -7.1 & 4.3 & -8.4 & 2.2 & -7.3 & 3.9 & 3.4 & 9.9 \\
1850 -- 1899 & 1874 & -3.8 & 3.0 & -3.3 & 2.0 & -5.1 & 4.1 & -3.4 & 2.2 & -3.3 & 3.8 & -12.6 & 9.8 \\
1900 -- 1949 & 1924 & -11.8 & 3.0 & -11.2 & 2.0 & -10.8 & 4.1 & -11.3 & 2.2 & -11.0 & 3.7 & -14.0 & 9.8 \\
1950 -- 1999 & 1975 & -15.0 & 3.0 & -16.4 & 2.0 & -17.8 & 4.1 & -16.3 & 2.1 & -17.3 & 3.7 & -18.7 & 10.1 \\
2000 -- 2099 & 2050 & -16.0 & 1.7 & -14.2 & 1.0 & -14.6 & 2.3 & -13.8 & 1.1 & -14.0 & 2.1 & -17.6 & 5.8 \\
2100 -- 2199 & 2151 & -5.4 & 1.6 & -4.7 & 1.0 & -9.1 & 2.3 & -4.3 & 1.1 & -5.8 & 2.1 & 3.7 & 6.1 \\
2200 -- 2299 & 2250 & -7.6 & 1.6 & -6.3 & 1.0 & -3.9 & 2.3 & -5.0 & 1.0 & -3.6 & 2.0 & -9.2 & 6.4 \\
2300 -- 2399 & 2349 & -8.9 & 1.6 & -8.8 & 1.0 & -10.6 & 2.4 & -9.3 & 1.0 & -10.5 & 2.0 & -19.6 & 6.7 \\
2400 -- 2499 & 2450 & -7.4 & 1.7 & -4.7 & 0.9 & -5.8 & 2.4 & -2.3 & 1.0 & -0.4 & 2.0 & 0.1 & 7.0 \\
2500 -- 2599 & 2549 & -0.9 & 1.7 & -4.2 & 0.9 & -4.0 & 2.5 & -3.6 & 1.0 & -5.1 & 2.0 & -14.3 & 7.4 \\
2600 -- 2699 & 2649 & -5.0 & 1.8 & -3.3 & 1.0 & -6.5 & 2.7 & -3.2 & 1.0 & -3.5 & 2.1 & -2.0 & 7.9 \\
2700 -- 2799 & 2749 & 1.5 & 1.9 & -2.1 & 1.0 & 5.5 & 2.9 & -3.8 & 1.0 & 1.9 & 2.2 & 16.3 & 8.5 \\
2800 -- 2899 & 2849 & 2.4 & 2.1 & 0.2 & 1.1 & -0.3 & 3.1 & -0.7 & 1.0 & -5.5 & 2.3 & -3.6 & 9.2 \\
2900 -- 2999 & 2949 & -6.9 & 2.3 & -1.8 & 1.1 & -5.3 & 3.3 & -2.1 & 1.1 & 0.2 & 2.4 & 15.6 & 9.7 \\
\hline
\end{tabular}
\caption{
$\TE$ bandpowers $D_b$ for the six cross-frequency power spectra, along with angular multipole range, bandpower window function-weighted multipole $\ell_\mathrm{eff}$, and associated uncertainty, $\sigma$.
The bandpowers and errors are quoted in units of $\mu$K$^2$.
The reported uncertainties are the square root of the diagonal elements of the covariance matrix and do not include beam or calibration uncertainties.
}
\label{tab:te_bandpowers_table}
\end{table*}

\clearpage
\bibliography{eete_2018}

\begin{thebibliography}{}
\expandafter\ifx\csname natexlab\endcsname\relax\def\natexlab#1{#1}\fi
\providecommand{\url}[1]{\href{#1}{#1}}
\providecommand{\dodoi}[1]{doi:~\href{http://doi.org/#1}{\nolinkurl{#1}}}
\providecommand{\doeprint}[1]{\href{http://ascl.net/#1}{\nolinkurl{http://ascl.net/#1}}}
\providecommand{\doarXiv}[1]{\href{https://arxiv.org/abs/#1}{\nolinkurl{https://arxiv.org/abs/#1}}}

\bibitem[{{Planck Collaboration} {et~al.}(2020{\natexlab{a}}){Planck
  Collaboration}, {Aghanim}, {Akrami}, {Ashdown}, {Aumont}, {Baccigalupi},
  {Ballardini}, {Banday}, {Barreiro}, {Bartolo}, {Basak}, {Benabed}, {Bernard},
  {Bersanelli}, {Bielewicz}, {Bock}, {Bond}, {Borrill}, {Bouchet}, {Boulanger},
  {Bucher}, {Burigana}, {Butler}, {Calabrese}, {Cardoso}, {Carron},
  {Casaponsa}, {Challinor}, {Chiang}, {Colombo}, {Combet}, {Crill}, {Cuttaia},
  {de Bernardis}, {de Rosa}, {de Zotti}, {Delabrouille}, {Delouis}, {Di
  Valentino}, {Diego}, {Dor{\'e}}, {Douspis}, {Ducout}, {Dupac}, {Dusini},
  {Efstathiou}, {Elsner}, {En{\ss}lin}, {Eriksen}, {Fantaye}, {Fernand
  ez-Cobos}, {Finelli}, {Frailis}, {Fraisse}, {Franceschi}, {Frolov},
  {Galeotta}, {Galli}, {Ganga}, {G{\'e}nova-Santos}, {Gerbino}, {Ghosh},
  {Giraud-H{\'e}raud}, {Gonz{\'a}lez-Nuevo}, {G{\'o}rski}, {Gratton},
  {Gruppuso}, {Gudmundsson}, {Hamann}, {Handley}, {Hansen}, {Herranz}, {Hivon},
  {Huang}, {Jaffe}, {Jones}, {Keih{\"a}nen}, {Keskitalo}, {Kiiveri}, {Kim},
  {Kisner}, {Krachmalnicoff}, {Kunz}, {Kurki-Suonio}, {Lagache}, {Lamarre},
  {Lasenby}, {Lattanzi}, {Lawrence}, {Le Jeune}, {Levrier}, {Lewis}, {Liguori},
  {Lilje}, {Lilley}, {Lindholm}, {L{\'o}pez-Caniego}, {Lubin}, {Ma},
  {Mac{\'\i}as-P{\'e}rez}, {Maggio}, {Maino}, {Mandolesi}, {Mangilli},
  {Marcos-Caballero}, {Maris}, {Martin}, {Mart{\'\i}nez-Gonz{\'a}lez},
  {Matarrese}, {Mauri}, {McEwen}, {Meinhold}, {Melchiorri}, {Mennella},
  {Migliaccio}, {Millea}, {Miville-Desch{\^e}nes}, {Molinari}, {Moneti},
  {Montier}, {Morgante}, {Moss}, {Natoli}, {N{\o}rgaard-Nielsen}, {Pagano},
  {Paoletti}, {Partridge}, {Patanchon}, {Peiris}, {Perrotta}, {Pettorino},
  {Piacentini}, {Polenta}, {Puget}, {Rachen}, {Reinecke}, {Remazeilles},
  {Renzi}, {Rocha}, {Rosset}, {Roudier}, {Rubi{\~n}o-Mart{\'\i}n},
  {Ruiz-Granados}, {Salvati}, {Sandri}, {Savelainen}, {Scott}, {Shellard},
  {Sirignano}, {Sirri}, {Spencer}, {Sunyaev}, {Suur-Uski}, {Tauber},
  {Tavagnacco}, {Tenti}, {Toffolatti}, {Tomasi}, {Trombetti}, {Valiviita}, {Van
  Tent}, {Vielva}, {Villa}, {Vittorio}, {Wandelt}, {Wehus}, {Zacchei}, \&
  {Zonca}}]{planck18-5}
{Planck Collaboration}, {Aghanim}, N., {Akrami}, Y., {et~al.}
\newblock {Planck 2018 results. V. CMB power spectra and likelihoods}.
  2020{\natexlab{a}}, \aap, 641, A5, \dodoi{10.1051/0004-6361/201936386}

\bibitem[{{Louis} {et~al.}(2017){Louis}, {Grace}, {Hasselfield}, {Lungu},
  {Maurin}, {Addison}, {Ade}, {Aiola}, {Allison}, {Amiri}, {Angile},
  {Battaglia}, {Beall}, {de Bernardis}, {Bond}, {Britton}, {Calabrese}, {Cho},
  {Choi}, {Coughlin}, {Crichton}, {Crowley}, {Datta}, {Devlin}, {Dicker},
  {Dunkley}, {D{\"u}nner}, {Ferraro}, {Fox}, {Gallardo}, {Gralla}, {Halpern},
  {Henderson}, {Hill}, {Hilton}, {Hilton}, {Hincks}, {Hlozek}, {Ho}, {Huang},
  {Hubmayr}, {Huffenberger}, {Hughes}, {Infante}, {Irwin}, {Muya Kasanda},
  {Klein}, {Koopman}, {Kosowsky}, {Li}, {Madhavacheril}, {Marriage}, {McMahon},
  {Menanteau}, {Moodley}, {Munson}, {Naess}, {Nati}, {Newburgh}, {Nibarger},
  {Niemack}, {Nolta}, {Nu{\~n}ez}, {Page}, {Pappas}, {Partridge}, {Rojas},
  {Schaan}, {Schmitt}, {Sehgal}, {Sherwin}, {Sievers}, {Simon}, {Spergel},
  {Staggs}, {Switzer}, {Thornton}, {Trac}, {Treu}, {Tucker}, {Van Engelen},
  {Ward}, \& {Wollack}}]{louis17}
{Louis}, T., {Grace}, E., {Hasselfield}, M., {et~al.}
\newblock {The Atacama Cosmology Telescope: two-season ACTPol spectra and
  parameters}. 2017, \jcap, 6, 031, \dodoi{10.1088/1475-7516/2017/06/031}

\bibitem[{{Reichardt} {et~al.}(2020){Reichardt}, {Patil}, {Ade}, {Anderson},
  {Austermann}, {Avva}, {Baxter}, {Beall}, {Bender}, {Benson}, {Bianchini},
  {Bleem}, {Carlstrom}, {Chang}, {Chaubal}, {Chiang}, {Chou}, {Citron},
  {Corbett Moran}, {Crawford}, {Crites}, {de Haan}, {Dobbs}, {Everett},
  {Gallicchio}, {George}, {Gilbert}, {Gupta}, {Halverson}, {Harrington},
  {Henning}, {Hilton}, {Holder}, {Holzapfel}, {Hrubes}, {Huang}, {Hubmayr},
  {Irwin}, {Knox}, {Lee}, {Li}, {Lowitz}, {Luong-Van}, {McMahon}, {Mehl},
  {Meyer}, {Millea}, {Mocanu}, {Mohr}, {Montgomery}, {Nadolski}, {Natoli},
  {Nibarger}, {Noble}, {Novosad}, {Omori}, {Padin}, {Pryke}, {Ruhl},
  {Saliwanchik}, {Sayre}, {Schaffer}, {Shirokoff}, {Sievers}, {Smecher},
  {Spieler}, {Staniszewski}, {Stark}, {Tucker}, {Vand erlinde}, {Veach},
  {Vieira}, {Wang}, {Whitehorn}, {Williamson}, {Wu}, \&
  {Yefremenko}}]{reichardt20}
{Reichardt}, C.~L., {Patil}, S., {Ade}, P.~A.~R., {et~al.}
\newblock {An Improved Measurement of the Secondary Cosmic Microwave Background
  Anisotropies from the SPT-SZ + SPTpol Surveys}. 2020, arXiv e-prints,
  arXiv:2002.06197.
\newblock \doarXiv{2002.06197}

\bibitem[{{Hu} \& {White}(1997)}]{hu97d}
{Hu}, W., \& {White}, M.
\newblock {A CMB polarization primer}. 1997, New Astronomy, 2, 323,
  \dodoi{10.1016/S1384-1076(97)00022-5}

\bibitem[{{Seljak} \& {Zaldarriaga}(1997)}]{seljak97}
{Seljak}, U., \& {Zaldarriaga}, M.
\newblock {Signature of Gravity Waves in the Polarization of the Microwave
  Background}. 1997, Physical Review Letters, 78, 2054,
  \dodoi{10.1103/PhysRevLett.78.2054}

\bibitem[{{Kamionkowski} {et~al.}(1997){Kamionkowski}, {Kosowsky}, \&
  {Stebbins}}]{kamionkowski97b}
{Kamionkowski}, M., {Kosowsky}, A., \& {Stebbins}, A.
\newblock {A Probe of Primordial Gravity Waves and Vorticity}. 1997, \prl, 78,
  2058, \dodoi{10.1103/PhysRevLett.78.2058}

\bibitem[{{Knox} \& {Song}(2002)}]{knox02}
{Knox}, L., \& {Song}, Y.
\newblock {Limit on the Detectability of the Energy Scale of Inflation}. 2002,
  Physical Review Letters, 89, 011303, \dodoi{10.1103/PhysRevLett.89.011303}

\bibitem[{{Galli} {et~al.}(2014){Galli}, {Benabed}, {Bouchet}, {Cardoso},
  {Elsner}, {Hivon}, {Mangilli}, {Prunet}, \& {Wandelt}}]{galli14}
{Galli}, S., {Benabed}, K., {Bouchet}, F., {Cardoso}, J.-F., {Elsner}, F.,
  {Hivon}, E., {Mangilli}, A., {Prunet}, S., \& {Wandelt}, B.
\newblock {CMB polarization can constrain cosmology better than CMB
  temperature}. 2014, \prd, 90, 063504, \dodoi{10.1103/PhysRevD.90.063504}

\bibitem[{{Gupta} {et~al.}(2019){Gupta}, {Reichardt}, {Ade}, {Anderson},
  {Archipley}, {Austermann}, {Avva}, {Beall}, {Bender}, {Benson}, {Bianchini},
  {Bleem}, {Carlstrom}, {Chang}, {Chiang}, {Citron}, {Moran}, {Crawford},
  {Crites}, {de Haan}, {Dobbs}, {Everett}, {Feng}, {Gallicchio}, {George},
  {Gilbert}, {Halverson}, {Harrington}, {Henning}, {Hilton}, {Holder},
  {Holzapfel}, {Hou}, {Hrubes}, {Huang}, {Hubmayr}, {Irwin}, {Knox}, {Lee},
  {Li}, {Lowitz}, {Luong-Van}, {Marrone}, {McMahon}, {Meyer}, {Mocanu}, {Mohr},
  {Montgomery}, {Nadolski}, {Natoli}, {Nibarger}, {Noble}, {Novosad}, {Padin},
  {Patil}, {Pryke}, {Ruhl}, {Saliwanchik}, {Sayre}, {Schaffer}, {Shirokoff},
  {Sievers}, {Smecher}, {Staniszewski}, {Stark}, {Story}, {Switzer}, {Tucker},
  {Vanderlinde}, {Veach}, {Vieira}, {Wang}, {Whitehorn}, {Williamson}, {Wu},
  {Yefremenko}, \& {Zhang}}]{gupta19}
{Gupta}, N., {Reichardt}, C.~L., {Ade}, P.~A.~R., {et~al.}
\newblock {Fractional polarization of extragalactic sources in the 500
  deg$^{2}$ SPTpol survey}. 2019, \mnras, 490, 5712,
  \dodoi{10.1093/mnras/stz2905}

\bibitem[{{Datta} {et~al.}(2019){Datta}, {Aiola}, {Choi}, {Devlin}, {Dunkley},
  {D{\"u}nner}, {Gallardo}, {Gralla}, {Halpern}, {Hasselfield}, {Hilton},
  {Hincks}, {Ho}, {Hubmayr}, {Huffenberger}, {Hughes}, {Kosowsky},
  {L{\'o}pez-Caraballo}, {Louis}, {Lungu}, {Marriage}, {Maurin}, {McMahon},
  {Moodley}, {Naess}, {Nati}, {Niemack}, {Page}, {Partridge}, {Prince},
  {Staggs}, {Switzer}, {Wollack}, \& {Farren}}]{datta19}
{Datta}, R., {Aiola}, S., {Choi}, S.~K., {et~al.}
\newblock {The Atacama Cosmology Telescope: two-season ACTPol extragalactic
  point sources and their polarization properties}. 2019, \mnras, 486, 5239,
  \dodoi{10.1093/mnras/sty2934}

\bibitem[{{Trombetti} {et~al.}(2018){Trombetti}, {Burigana}, {De Zotti},
  {Galluzzi}, \& {Massardi}}]{trombetti18}
{Trombetti}, T., {Burigana}, C., {De Zotti}, G., {Galluzzi}, V., \& {Massardi},
  M.
\newblock {Average fractional polarization of extragalactic sources at Planck
  frequencies}. 2018, \aap, 618, A29, \dodoi{10.1051/0004-6361/201732342}

\bibitem[{{Choi} {et~al.}(2020){Choi}, {Hasselfield}, {Ho}, {Koopman}, {Lungu},
  {Abitbol}, {Addison}, {Ade}, {Aiola}, {Alonso}, {Amiri}, {Amodeo}, {Angile},
  {Austermann}, {Baildon}, {Battaglia}, {Beall}, {Bean}, {Becker}, {Bond},
  {Bruno}, {Calabrese}, {Calafut}, {Campusano}, {Carrero}, {Chesmore}, {Cho.},
  {Clark}, {Cothard}, {Crichton}, {Crowley}, {Darwish}, {Datta}, {Denison},
  {Devlin}, {Duell}, {Duff}, {Duivenvoorden}, {Dunkley}, {D{\"u}nner},
  {Essinger-Hileman}, {Fankhanel}, {Ferraro}, {Fox}, {Fuzia}, {Gallardo},
  {Gluscevic}, {Golec}, {Grace}, {Gralla}, {Guan}, {Hall}, {Halpern}, {Han},
  {Hargrave}, {Henderson}, {Hensley}, {Hill}, {Hilton}, {Hilton}, {Hincks},
  {Hlo{\v{z}}ek}, {Hubmayr}, {Huffenberger}, {Hughes}, {Infante}, {Irwin},
  {Jackson}, {Klein}, {Knowles}, {Kosowsky}, {Lakey}, {Li}, {Li}, {Li},
  {Lokken}, {Louis}, {MacInnis}, {Madhavacheril}, {Maldonado}, {Mallaby-Kay},
  {Marsden}, {Maurin}, {McMahon}, {Menanteau}, {Moodley}, {Morton}, {Naess},
  {Namikawa}, {Nati}, {Newburgh}, {Nibarger}, {Nicola}, {Niemack}, {Nolta},
  {Orlowski-Sherer}, {Page}, {Pappas}, {Partridge}, {Phakathi}, {Prince},
  {Puddu}, {Qu}, {Rivera}, {Robertson}, {Rojas}, {Salatino}, {Schaan},
  {Schillaci}, {Schmitt}, {Sehgal}, {Sherwin}, {Sierra}, {Sievers}, {Sifon},
  {Sikhosana}, {Simon}, {Spergel}, {Staggs}, {Stevens}, {Storer}, {Sunder},
  {Switzer}, {Thorne}, {Thornton}, {Trac}, {Treu}, {Tucker}, {Vale}, {Van
  Engelen}, {Van Lanen}, {Vavagiakis}, {Wagoner}, {Wang}, {Ward}, {Wollack},
  {Xu}, {Zago}, \& {Zhu}}]{choi20}
{Choi}, S.~K., {Hasselfield}, M., {Ho}, S.-P.~P., {et~al.}
\newblock {The Atacama Cosmology Telescope: A Measurement of the Cosmic
  Microwave Background Power Spectra at 98 and 150 GHz}. 2020, arXiv e-prints,
  arXiv:2007.07289.
\newblock \doarXiv{2007.07289}

\bibitem[{{BICEP2 Collaboration} {et~al.}(2018){BICEP2 Collaboration}, {Keck
  Array Collaboration}, {Ade}, {Ahmed}, {Aikin}, {Alexand er}, {Barkats},
  {Benton}, {Bischoff}, {Bock}, {Bowens-Rubin}, {Brevik}, {Buder}, {Bullock},
  {Buza}, {Connors}, {Cornelison}, {Crill}, {Crumrine}, {Dierickx}, {Duband},
  {Dvorkin}, {Filippini}, {Fliescher}, {Grayson}, {Hall}, {Halpern},
  {Harrison}, {Hildebrand t}, {Hilton}, {Hui}, {Irwin}, {Kang}, {Karkare},
  {Karpel}, {Kaufman}, {Keating}, {Kefeli}, {Kernasovskiy}, {Kovac}, {Kuo},
  {Larsen}, {Lau}, {Leitch}, {Lueker}, {Megerian}, {Moncelsi}, {Namikawa},
  {Netterfield}, {Nguyen}, {O'Brient}, {Ogburn}, {Palladino}, {Pryke},
  {Racine}, {Richter}, {Schillaci}, {Schwarz}, {Sheehy}, {Soliman}, {St.
  Germaine}, {Staniszewski}, {Steinbach}, {Sudiwala}, {Teply}, {Thompson},
  {Tolan}, {Tucker}, {Turner}, {Umilt{\`a}}, {Vieregg}, {Wand ui}, {Weber},
  {Wiebe}, {Willmert}, {Wong}, {Wu}, {Yang}, {Yoon}, \& {Zhang}}]{bicep2keck18}
{BICEP2 Collaboration}, {Keck Array Collaboration}, {Ade}, P.~A.~R., {et~al.}
\newblock {Constraints on Primordial Gravitational Waves Using Planck, WMAP,
  and New BICEP2/Keck Observations through the 2015 Season}. 2018, \prl, 121,
  221301, \dodoi{10.1103/PhysRevLett.121.221301}

\bibitem[{{POLARBEAR Collaboration} {et~al.}(2019){POLARBEAR Collaboration},
  {Adachi}, {Aguilar Fa{\'u}ndez}, {Arnold}, {Baccigalupi}, {Barron}, {Beck},
  {Beckman}, {Bianchini}, {Boettger}, {Borrill}, {Carron}, {Chapman}, {Cheung},
  {Chinone}, {Crowley}, {Cukierman}, {Dobbs}, {El Bouhargani}, {Elleflot},
  {Errard}, {Fabbian}, {Feng}, {Fujino}, {Galitzki}, {Goeckner-Wald}, {Groh},
  {Hall}, {Halverson}, {Hamada}, {Hasegawa}, {Hazumi}, {Hill}, {Howe}, {Inoue},
  {Jaehnig}, {Jeong}, {Kaneko}, {Katayama}, {Keating}, {Keskitalo}, {Kikuchi},
  {Kisner}, {Krachmalnicoff}, {Kusaka}, {Lee}, {Leon}, {Linder}, {Lowry},
  {Mangu}, {Matsuda}, {Minami}, {Navaroli}, {Nishino}, {Pham}, {Poletti},
  {Puglisi}, {Reichardt}, {Segawa}, {Silva-Feaver}, {Siritanasak}, {Stebor},
  {Stompor}, {Suzuki}, {Tajima}, {Takakura}, {Takatori}, {Tanabe}, {Teply},
  {Tsai}, {Verges}, {Westbrook}, \& {Zhou}}]{polarbear19a}
{POLARBEAR Collaboration}, {Adachi}, S., {Aguilar Fa{\'u}ndez}, M.~A.~O.,
  {et~al.}
\newblock {A Measurement of the Degree Scale CMB B-mode Angular Power Spectrum
  with POLARBEAR}. 2019, arXiv e-prints, arXiv:1910.02608.
\newblock \doarXiv{1910.02608}

\bibitem[{{Adachi} {et~al.}(2020){Adachi}, {Aguilar Fa{\'u}ndez}, {Arnold},
  {Baccigalupi}, {Barron}, {Beck}, {Bianchini}, {Chapman}, {Cheung}, {Chinone},
  {Crowley}, {Dobbs}, {El Bouhargani}, {Elleflot}, {Errard}, {Fabbian}, {Feng},
  {Fujino}, {Galitzki}, {Goeckner-Wald}, {Groh}, {Hall}, {Hasegawa}, {Hazumi},
  {Hirose}, {Jaffe}, {Jeong}, {Kaneko}, {Katayama}, {Keating}, {Kikuchi},
  {Kisner}, {Kusaka}, {Lee}, {Leon}, {Linder}, {Lowry}, {Matsuda}, {Matsumura},
  {Minami}, {Navaroli}, {Nishino}, {Pham}, {Poletti}, {Reichardt}, {Segawa},
  {Siritanasak}, {Tajima}, {Takakura}, {Takatori}, {Tanabe}, {Teply}, {Tsai},
  {Verg{\`e}s}, {Westbrook}, \& {Zhou}}]{polarbear20}
{Adachi}, S., {Aguilar Fa{\'u}ndez}, M.~A.~O., {Arnold}, K., {et~al.}
\newblock {A measurement of the CMB E-mode angular power spectrum at subdegree
  scales from 670 square degrees of POLARBEAR data}. 2020, arXiv e-prints,
  arXiv:2005.06168.
\newblock \doarXiv{2005.06168}

\bibitem[{{Henning} {et~al.}(2018){Henning}, {Sayre}, {Reichardt}, {Ade},
  {Anderson}, {Austermann}, {Beall}, {Bender}, {Benson}, {Bleem}, {Carlstrom},
  {Chang}, {Chiang}, {Cho}, {Citron}, {Corbett Moran}, {Crawford}, {Crites},
  {de Haan}, {Dobbs}, {Everett}, {Gallicchio}, {George}, {Gilbert},
  {Halverson}, {Harrington}, {Hilton}, {Holder}, {Holzapfel}, {Hoover}, {Hou},
  {Hrubes}, {Huang}, {Hubmayr}, {Irwin}, {Keisler}, {Knox}, {Lee}, {Leitch},
  {Li}, {Lowitz}, {Manzotti}, {McMahon}, {Meyer}, {Mocanu}, {Montgomery},
  {Nadolski}, {Natoli}, {Nibarger}, {Novosad}, {Padin}, {Pryke}, {Ruhl},
  {Saliwanchik}, {Schaffer}, {Sievers}, {Smecher}, {Stark}, {Story}, {Tucker},
  {Vanderlinde}, {Veach}, {Vieira}, {Wang}, {Whitehorn}, {Wu}, \&
  {Yefremenko}}]{henning18}
{Henning}, J.~W., {Sayre}, J.~T., {Reichardt}, C.~L., {et~al.}
\newblock {Measurements of the Temperature and E-mode Polarization of the CMB
  from 500 Square Degrees of SPTpol Data}. 2018, \apj, 852, 97,
  \dodoi{10.3847/1538-4357/aa9ff4}

\bibitem[{{Sayre} {et~al.}(2020){Sayre}, {Reichardt}, {Henning}, {Ade},
  {Anderson}, {Austermann}, {Avva}, {Beall}, {Bender}, {Benson}, {Bianchini},
  {Bleem}, {Carlstrom}, {Chang}, {Chaubal}, {Chiang}, {Citron}, {Corbett
  Moran}, {Crawford}, {Crites}, {de Haan}, {Dobbs}, {Everett}, {Gallicchio},
  {George}, {Gilbert}, {Gupta}, {Halverson}, {Harrington}, {Hilton}, {Holder},
  {Holzapfel}, {Hrubes}, {Huang}, {Hubmayr}, {Irwin}, {Knox}, {Lee}, {Li},
  {Lowitz}, {McMahon}, {Meyer}, {Mocanu}, {Montgomery}, {Nadolski}, {Natoli},
  {Nibarger}, {Noble}, {Novosad}, {Padin}, {Patil}, {Pryke}, {Ruhl},
  {Saliwanchik}, {Schaffer}, {Sievers}, {Smecher}, {Stark}, {Tucker},
  {Vanderlinde}, {Veach}, {Vieira}, {Wang}, {Whitehorn}, {Wu}, {Yefremenko}, \&
  {SPTpol Collaboration}}]{sayre20}
{Sayre}, J.~T., {Reichardt}, C.~L., {Henning}, J.~W., {et~al.}
\newblock {Measurements of B-mode Polarization of the Cosmic Microwave
  Background from 500 Square Degrees of SPTpol Data}. 2020, \prd, 101, 122003,
  \dodoi{10.1103/PhysRevD.101.122003}

\bibitem[{{Henderson} {et~al.}(2016){Henderson}, {Allison}, {Austermann},
  {Baildon}, {Battaglia}, {Beall}, {Becker}, {De Bernardis}, {Bond},
  {Calabrese}, {Choi}, {Coughlin}, {Crowley}, {Datta}, {Devlin}, {Duff},
  {Dunkley}, {D{\"u}nner}, {van Engelen}, {Gallardo}, {Grace}, {Hasselfield},
  {Hills}, {Hilton}, {Hincks}, {Hlo{\^z}ek}, {Ho}, {Hubmayr}, {Huffenberger},
  {Hughes}, {Irwin}, {Koopman}, {Kosowsky}, {Li}, {McMahon}, {Munson}, {Nati},
  {Newburgh}, {Niemack}, {Niraula}, {Page}, {Pappas}, {Salatino}, {Schillaci},
  {Schmitt}, {Sehgal}, {Sherwin}, {Sievers}, {Simon}, {Spergel}, {Staggs},
  {Stevens}, {Thornton}, {Van Lanen}, {Vavagiakis}, {Ward}, \&
  {Wollack}}]{henderson16}
{Henderson}, S.~W., {Allison}, R., {Austermann}, J., {et~al.}
\newblock {Advanced ACTPol Cryogenic Detector Arrays and Readout}. 2016,
  Journal of Low Temperature Physics, 184, 772,
  \dodoi{10.1007/s10909-016-1575-z}

\bibitem[{{Ahmed} {et~al.}(2014){Ahmed}, {Amiri}, {Benton}, {Bock},
  {Bowens-Rubin}, {Buder}, {Bullock}, {Connors}, {Filippini}, {Grayson},
  {Halpern}, {Hilton}, {Hristov}, {Hui}, {Irwin}, {Kang}, {Karkare}, {Karpel},
  {Kovac}, {Kuo}, {Netterfield}, {Nguyen}, {O'Brient}, {Ogburn}, {Pryke},
  {Reintsema}, {Richter}, {Thompson}, {Turner}, {Vieregg}, {Wu}, \&
  {Yoon}}]{ahmed14}
{Ahmed}, Z., {Amiri}, M., {Benton}, S.~J., {et~al.}
\newblock {BICEP3: a 95GHz refracting telescope for degree-scale CMB
  polarization}. 2014, in \procspie, Vol. 9153, Society of Photo-Optical
  Instrumentation Engineers (SPIE) Conference Series, 1,
  \dodoi{10.1117/12.2057224}

\bibitem[{{Hui} {et~al.}(2018){Hui}, {Ade}, {Ahmed}, {Aikin}, {Alexander},
  {Barkats}, {Benton}, {Bischoff}, {Bock}, {Bowens-Rubin}, {Brevik}, {Buder},
  {Bullock}, {Buza}, {Connors}, {Cornelison}, {Crill}, {Crumrine}, {Dierickx},
  {Duband}, {Dvorkin}, {Filippini}, {Fliescher}, {Grayson}, {Hall}, {Halpern},
  {Harrison}, {Hildebrandt}, {Hilton}, {Irwin}, {Kang}, {Karkare}, {Karpel},
  {Kaufman}, {Keating}, {Kefeli}, {Kernasovskiy}, {Kovac}, {Kuo}, {Lau},
  {Larsen}, {Leitch}, {Lueker}, {Megerian}, {Moncelsi}, {Namikawa},
  {Netterfield}, {Nguyen}, {O'Brient}, {Ogburn}, {Palladino}, {Pryke},
  {Racine}, {Richter}, {Schwarz}, {Schillaci}, {Sheehy}, {Soliman},
  {St.~Germaine}, {Staniszewski}, {Steinbach}, {Sudiwala}, {Teply}, {Thompson},
  {Tolan}, {Tucker}, {Turner}, {Umilta}, {Vieregg}, {Wandui}, {Weber}, {Wiebe},
  {Willmert}, {Wong}, {Wu}, {Yang}, {Yoon}, \& {Zhang}}]{hui18}
{Hui}, H., {Ade}, P.~A.~R., {Ahmed}, Z., {et~al.}
\newblock {BICEP Array: a multi-frequency degree-scale CMB polarimeter}. 2018,
  in \procspie, Vol. 10708, Society of Photo-Optical Instrumentation Engineers
  (SPIE) Conference Series, 1070807, \dodoi{10.1117/12.2311725}

\bibitem[{{Suzuki} {et~al.}(2016){Suzuki}, {Ade}, {Akiba}, {Aleman}, {Arnold},
  {Baccigalupi}, {Barch}, {Barron}, {Bender}, {Boettger}, {Borrill}, {Chapman},
  {Chinone}, {Cukierman}, {Dobbs}, {Ducout}, {Dunner}, {Elleflot}, {Errard},
  {Fabbian}, {Feeney}, {Feng}, {Fujino}, {Fuller}, {Gilbert}, {Goeckner-Wald},
  {Groh}, {Haan}, {Hall}, {Halverson}, {Hamada}, {Hasegawa}, {Hattori},
  {Hazumi}, {Hill}, {Holzapfel}, {Hori}, {Howe}, {Inoue}, {Irie}, {Jaehnig},
  {Jaffe}, {Jeong}, {Katayama}, {Kaufman}, {Kazemzadeh}, {Keating}, {Kermish},
  {Keskitalo}, {Kisner}, {Kusaka}, {Jeune}, {Lee}, {Leon}, {Linder}, {Lowry},
  {Matsuda}, {Matsumura}, {Miller}, {Mizukami}, {Montgomery}, {Navaroli},
  {Nishino}, {Peloton}, {Poletti}, {Puglisi}, {Rebeiz}, {Raum}, {Reichardt},
  {Richards}, {Ross}, {Rotermund}, {Segawa}, {Sherwin}, {Shirley},
  {Siritanasak}, {Stebor}, {Stompor}, {Suzuki}, {Tajima}, {Takada}, {Takakura},
  {Takatori}, {Tikhomirov}, {Tomaru}, {Westbrook}, {Whitehorn}, {Yamashita},
  {Zahn}, \& {Zahn}}]{suzuki14}
{Suzuki}, A., {Ade}, P., {Akiba}, Y., {et~al.}
\newblock {The Polarbear-2 and the Simons Array Experiments}. 2016, Journal of
  Low Temperature Physics, 184, 805, \dodoi{10.1007/s10909-015-1425-4}

\bibitem[{{Simons Observatory Collaboration} {et~al.}(2019){Simons Observatory
  Collaboration}, {Ade}, {Aguirre}, {Ahmed}, {Aiola}, {Ali}, {Alonso},
  {Alvarez}, {Arnold}, {Ashton}, {Austermann}, {Awan}, {Baccigalupi},
  {Baildon}, {Barron}, {Battaglia}, {Battye}, {Baxter}, {Bazarko}, {Beall},
  {Bean}, {Beck}, {Beckman}, {Beringue}, {Bianchini}, {Boada}, {Boettger},
  {Bond}, {Borrill}, {Brown}, {Bruno}, {Bryan}, {Calabrese}, {Calafut},
  {Calisse}, {Carron}, {Challinor}, {Chesmore}, {Chinone}, {Chluba}, {Cho},
  {Choi}, {Coppi}, {Cothard}, {Coughlin}, {Crichton}, {Crowley}, {Crowley},
  {Cukierman}, {D'Ewart}, {D{\"u}nner}, {de Haan}, {Devlin}, {Dicker},
  {Didier}, {Dobbs}, {Dober}, {Duell}, {Duff}, {Duivenvoorden}, {Dunkley},
  {Dusatko}, {Errard}, {Fabbian}, {Feeney}, {Ferraro}, {Flux{\`a}}, {Freese},
  {Frisch}, {Frolov}, {Fuller}, {Fuzia}, {Galitzki}, {Gallardo}, {Tomas Galvez
  Ghersi}, {Gao}, {Gawiser}, {Gerbino}, {Gluscevic}, {Goeckner-Wald}, {Golec},
  {Gordon}, {Gralla}, {Green}, {Grigorian}, {Groh}, {Groppi}, {Guan},
  {Gudmundsson}, {Han}, {Hargrave}, {Hasegawa}, {Hasselfield}, {Hattori},
  {Haynes}, {Hazumi}, {He}, {Healy}, {Henderson}, {Hervias-Caimapo}, {Hill},
  {Hill}, {Hilton}, {Hilton}, {Hincks}, {Hinshaw}, {Hlo{\v{z}}ek}, {Ho}, {Ho},
  {Howe}, {Huang}, {Hubmayr}, {Huffenberger}, {Hughes}, {Ijjas}, {Ikape},
  {Irwin}, {Jaffe}, {Jain}, {Jeong}, {Kaneko}, {Karpel}, {Katayama}, {Keating},
  {Kernasovskiy}, {Keskitalo}, {Kisner}, {Kiuchi}, {Klein}, {Knowles},
  {Koopman}, {Kosowsky}, {Krachmalnicoff}, {Kuenstner}, {Kuo}, {Kusaka},
  {Lashner}, {Lee}, {Lee}, {Leon}, {Leung}, {Lewis}, {Li}, {Li}, {Limon},
  {Linder}, {Lopez-Caraballo}, {Louis}, {Lowry}, {Lungu}, {Madhavacheril},
  {Mak}, {Maldonado}, {Mani}, {Mates}, {Matsuda}, {Maurin}, {Mauskopf}, {May},
  {McCallum}, {McKenney}, {McMahon}, {Meerburg}, {Meyers}, {Miller},
  {Mirmelstein}, {Moodley}, {Munchmeyer}, {Munson}, {Naess}, {Nati},
  {Navaroli}, {Newburgh}, {Nguyen}, {Niemack}, {Nishino}, {Orlowski-Scherer},
  {Page}, {Partridge}, {Peloton}, {Perrotta}, {Piccirillo}, {Pisano},
  {Poletti}, {Puddu}, {Puglisi}, {Raum}, {Reichardt}, {Remazeilles},
  {Rephaeli}, {Riechers}, {Rojas}, {Roy}, {Sadeh}, {Sakurai}, {Salatino},
  {Sathyanarayana Rao}, {Schaan}, {Schmittfull}, {Sehgal}, {Seibert}, {Seljak},
  {Sherwin}, {Shimon}, {Sierra}, {Sievers}, {Sikhosana}, {Silva-Feaver},
  {Simon}, {Sinclair}, {Siritanasak}, {Smith}, {Smith}, {Spergel}, {Staggs},
  {Stein}, {Stevens}, {Stompor}, {Suzuki}, {Tajima}, {Takakura}, {Teply},
  {Thomas}, {Thorne}, {Thornton}, {Trac}, {Tsai}, {Tucker}, {Ullom},
  {Vagnozzi}, {van Engelen}, {Van Lanen}, {Van Winkle}, {Vavagiakis},
  {Verg{\`e}s}, {Vissers}, {Wagoner}, {Walker}, {Ward}, {Westbrook},
  {Whitehorn}, {Williams}, {Williams}, {Wollack}, {Xu}, {Yu}, {Yu}, {Zago},
  {Zhang}, {Zhu}, \& {The Simons Observatory collaboration}}]{ade19}
{Simons Observatory Collaboration}, {Ade}, P., {Aguirre}, J., {et~al.}
\newblock {The Simons Observatory: science goals and forecasts}. 2019, \jcap,
  2019, 056, \dodoi{10.1088/1475-7516/2019/02/056}

\bibitem[{{Benson} {et~al.}(2014){Benson}, {Ade}, {Ahmed}, {Allen}, {Arnold},
  {Austermann}, {Bender}, {Bleem}, {Carlstrom}, {Chang}, {Cho}, {Cliche},
  {Crawford}, {Cukierman}, {de Haan}, {Dobbs}, {Dutcher}, {Everett}, {Gilbert},
  {Halverson}, {Hanson}, {Harrington}, {Hattori}, {Henning}, {Hilton},
  {Holder}, {Holzapfel}, {Irwin}, {Keisler}, {Knox}, {Kubik}, {Kuo}, {Lee},
  {Leitch}, {Li}, {McDonald}, {Meyer}, {Montgomery}, {Myers}, {Natoli},
  {Nguyen}, {Novosad}, {Padin}, {Pan}, {Pearson}, {Reichardt}, {Ruhl},
  {Saliwanchik}, {Simard}, {Smecher}, {Sayre}, {Shirokoff}, {Stark}, {Story},
  {Suzuki}, {Thompson}, {Tucker}, {Vanderlinde}, {Vieira}, {Vikhlinin}, {Wang},
  {Yefremenko}, \& {Yoon}}]{benson14}
{Benson}, B.~A., {Ade}, P.~A.~R., {Ahmed}, Z., {et~al.}
\newblock {SPT-3G: A Next-Generation Cosmic Microwave Background Polarization
  Experiment on the South Pole Telescope}. 2014, in \procspie, Vol. 9153,
  Millimeter, Submillimeter, and Far-Infrared Detectors and Instrumentation for
  Astronomy VII, 91531P, \dodoi{10.1117/12.2057305}

\bibitem[{{Aylor} {et~al.}(2017){Aylor}, {Hou}, {Knox}, {Story}, {Benson},
  {Bleem}, {Carlstrom}, {Chang}, {Cho}, {Chown}, {Crawford}, {Crites}, {de
  Haan}, {Dobbs}, {Everett}, {George}, {Halverson}, {Harrington}, {Holder},
  {Holzapfel}, {Hrubes}, {Keisler}, {Lee}, {Leitch}, {Luong-Van}, {Marrone},
  {McMahon}, {Meyer}, {Millea}, {Mocanu}, {Mohr}, {Natoli}, {Omori}, {Padin},
  {Pryke}, {Reichardt}, {Ruhl}, {Sayre}, {Schaffer}, {Shirokoff},
  {Staniszewski}, {Stark}, {Vanderlinde}, {Vieira}, \& {Williamson}}]{aylor17}
{Aylor}, K., {Hou}, Z., {Knox}, L., {et~al.}
\newblock {A Comparison of Cosmological Parameters Determined from CMB
  Temperature Power Spectra from the South Pole Telescope and the Planck
  Satellite}. 2017, \apj, 850, 101, \dodoi{10.3847/1538-4357/aa947b}

\bibitem[{{Addison} {et~al.}(2016){Addison}, {Huang}, {Watts}, {Bennett},
  {Halpern}, {Hinshaw}, \& {Weiland}}]{addison16}
{Addison}, G.~E., {Huang}, Y., {Watts}, D.~J., {Bennett}, C.~L., {Halpern}, M.,
  {Hinshaw}, G., \& {Weiland}, J.~L.
\newblock {Quantifying Discordance in the 2015 Planck CMB Spectrum}. 2016,
  \apj, 818, 132, \dodoi{10.3847/0004-637X/818/2/132}

\bibitem[{{Planck Collaboration} {et~al.}(2020{\natexlab{b}}){Planck
  Collaboration}, {Aghanim}, {Akrami}, {Ashdown}, {Aumont}, {Baccigalupi},
  {Ballardini}, {Banday}, {Barreiro}, {Bartolo}, {Basak}, {Battye}, {Benabed},
  {Bernard}, {Bersanelli}, {Bielewicz}, {Bock}, {Bond}, {Borrill}, {Bouchet},
  {Boulanger}, {Bucher}, {Burigana}, {Butler}, {Calabrese}, {Cardoso},
  {Carron}, {Challinor}, {Chiang}, {Chluba}, {Colombo}, {Combet}, {Contreras},
  {Crill}, {Cuttaia}, {de Bernardis}, {de Zotti}, {Delabrouille}, {Delouis},
  {Di Valentino}, {Diego}, {Dor{\'e}}, {Douspis}, {Ducout}, {Dupac}, {Dusini},
  {Efstathiou}, {Elsner}, {En{\ss}lin}, {Eriksen}, {Fantaye}, {Farhang},
  {Fergusson}, {Fernandez-Cobos}, {Finelli}, {Forastieri}, {Frailis},
  {Fraisse}, {Franceschi}, {Frolov}, {Galeotta}, {Galli}, {Ganga},
  {G{\'e}nova-Santos}, {Gerbino}, {Ghosh}, {Gonz{\'a}lez-Nuevo}, {G{\'o}rski},
  {Gratton}, {Gruppuso}, {Gudmundsson}, {Hamann}, {Handley}, {Hansen},
  {Herranz}, {Hildebrandt}, {Hivon}, {Huang}, {Jaffe}, {Jones}, {Karakci},
  {Keih{\"a}nen}, {Keskitalo}, {Kiiveri}, {Kim}, {Kisner}, {Knox},
  {Krachmalnicoff}, {Kunz}, {Kurki-Suonio}, {Lagache}, {Lamarre}, {Lasenby},
  {Lattanzi}, {Lawrence}, {Le Jeune}, {Lemos}, {Lesgourgues}, {Levrier},
  {Lewis}, {Liguori}, {Lilje}, {Lilley}, {Lindholm}, {L{\'o}pez-Caniego},
  {Lubin}, {Ma}, {Mac{\'\i}as-P{\'e}rez}, {Maggio}, {Maino}, {Mandolesi},
  {Mangilli}, {Marcos-Caballero}, {Maris}, {Martin}, {Martinelli},
  {Mart{\'\i}nez-Gonz{\'a}lez}, {Matarrese}, {Mauri}, {McEwen}, {Meinhold},
  {Melchiorri}, {Mennella}, {Migliaccio}, {Millea}, {Mitra},
  {Miville-Desch{\^e}nes}, {Molinari}, {Montier}, {Morgante}, {Moss}, {Natoli},
  {N{\o}rgaard-Nielsen}, {Pagano}, {Paoletti}, {Partridge}, {Patanchon},
  {Peiris}, {Perrotta}, {Pettorino}, {Piacentini}, {Polastri}, {Polenta},
  {Puget}, {Rachen}, {Reinecke}, {Remazeilles}, {Renzi}, {Rocha}, {Rosset},
  {Roudier}, {Rubi{\~n}o-Mart{\'\i}n}, {Ruiz-Granados}, {Salvati}, {Sandri},
  {Savelainen}, {Scott}, {Shellard}, {Sirignano}, {Sirri}, {Spencer},
  {Sunyaev}, {Suur-Uski}, {Tauber}, {Tavagnacco}, {Tenti}, {Toffolatti},
  {Tomasi}, {Trombetti}, {Valenziano}, {Valiviita}, {Van Tent}, {Vibert},
  {Vielva}, {Villa}, {Vittorio}, {Wand elt}, {Wehus}, {White}, {White},
  {Zacchei}, \& {Zonca}}]{planck18-6}
{Planck Collaboration}, {Aghanim}, N., {Akrami}, Y., {et~al.}
\newblock {Planck 2018 results. VI. Cosmological parameters}.
  2020{\natexlab{b}}, \aap, 641, A6, \dodoi{10.1051/0004-6361/201833910}

\bibitem[{{Riess} {et~al.}(2019){Riess}, {Casertano}, {Yuan}, {Macri}, \&
  {Scolnic}}]{riess19}
{Riess}, A.~G., {Casertano}, S., {Yuan}, W., {Macri}, L.~M., \& {Scolnic}, D.
\newblock {Large Magellanic Cloud Cepheid Standards Provide a 1\% Foundation
  for the Determination of the Hubble Constant and Stronger Evidence for
  Physics beyond {\ensuremath{\Lambda}}CDM}. 2019, \apj, 876, 85,
  \dodoi{10.3847/1538-4357/ab1422}

\bibitem[{{Carlstrom} {et~al.}(2011){Carlstrom}, {Ade}, {Aird}, {Benson},
  {Bleem}, {Busetti}, {Chang}, {Chauvin}, {Cho}, {Crawford}, {Crites}, {Dobbs},
  {Halverson}, {Heimsath}, {Holzapfel}, {Hrubes}, {Joy}, {Keisler}, {Lanting},
  {Lee}, {Leitch}, {Leong}, {Lu}, {Lueker}, {Luong-van}, {McMahon}, {Mehl},
  {Meyer}, {Mohr}, {Montroy}, {Padin}, {Plagge}, {Pryke}, {Ruhl}, {Schaffer},
  {Schwan}, {Shirokoff}, {Spieler}, {Staniszewski}, {Stark}, {Tucker},
  {Vanderlinde}, {Vieira}, \& {Williamson}}]{carlstrom11}
{Carlstrom}, J.~E., {Ade}, P.~A.~R., {Aird}, K.~A., {et~al.}
\newblock {The 10 Meter South Pole Telescope}. 2011, \pasp, 123, 568,
  \dodoi{10.1086/659879}

\bibitem[{{Aiola} {et~al.}(2020){Aiola}, {Calabrese}, {Maurin}, {Naess},
  {Schmitt}, {Abitbol}, {Addison}, {Ade}, {Alonso}, {Amiri}, {Amodeo},
  {Angile}, {Austermann}, {Baildon}, {Battaglia}, {Beall}, {Bean}, {Becker},
  {Bond}, {Bruno}, {Calafut}, {Campusano}, {Carrero}, {Chesmore}, {Cho.},
  {Choi}, {Clark}, {Cothard}, {Crichton}, {Crowley}, {Darwish}, {Datta},
  {Denison}, {Devlin}, {Duell}, {Duff}, {Duivenvoorden}, {Dunkley},
  {D{\"u}nner}, {Essinger-Hileman}, {Fankhanel}, {Ferraro}, {Fox}, {Fuzia},
  {Gallardo}, {Gluscevic}, {Golec}, {Grace}, {Gralla}, {Guan}, {Hall},
  {Halpern}, {Han}, {Hargrave}, {Hasselfield}, {Helton}, {Henderson},
  {Hensley}, {Hill}, {Hilton}, {Hilton}, {Hincks}, {Hlo{\v{z}}ek}, {Ho},
  {Hubmayr}, {Huffenberger}, {Hughes}, {Infante}, {Irwin}, {Jackson}, {Klein},
  {Knowles}, {Koopman}, {Kosowsky}, {Lakey}, {Li}, {Li}, {Li}, {Lokken},
  {Louis}, {Lungu}, {MacInnis}, {Madhavacheril}, {Maldonado}, {Mallaby-Kay},
  {Marsden}, {McMahon}, {Menanteau}, {Moodley}, {Morton}, {Namikawa}, {Nati},
  {Newburgh}, {Nibarger}, {Nicola}, {Niemack}, {Nolta}, {Orlowski-Sherer},
  {Page}, {Pappas}, {Partridge}, {Phakathi}, {Prince}, {Puddu}, {Qu}, {Rivera},
  {Robertson}, {Rojas}, {Salatino}, {Schaan}, {Schillaci}, {Sehgal}, {Sherwin},
  {Sierra}, {Sievers}, {Sifon}, {Sikhosana}, {Simon}, {Spergel}, {Staggs},
  {Stevens}, {Storer}, {Sunder}, {Switzer}, {Thorne}, {Thornton}, {Trac},
  {Treu}, {Tucker}, {Vale}, {Van Engelen}, {Van Lanen}, {Vavagiakis},
  {Wagoner}, {Wang}, {Ward}, {Wollack}, {Xu}, {Zago}, \& {Zhu}}]{aiola20}
{Aiola}, S., {Calabrese}, E., {Maurin}, L., {et~al.}
\newblock {The Atacama Cosmology Telescope: DR4 Maps and Cosmological
  Parameters}. 2020, arXiv e-prints, arXiv:2007.07288.
\newblock \doarXiv{2007.07288}

\bibitem[{{Sobrin} {et~al.}(2018){Sobrin}, {Ade}, {Ahmed}, {Anderson}, {Avva},
  {Basu Thakur}, {Bender}, {Benson}, {Carlstrom}, {Carter}, {Cecil}, {Chang},
  {Cliche}, {Cukierman}, {de Haan}, {Ding}, {Dobbs}, {Dutcher}, {Everett},
  {Foster}, {Gallichio}, {Gilbert}, {Groh}, {Guns}, {Halverson},
  {Harke-Hosemann}, {Harrington}, {Henning}, {Holzapfel}, {Huang}, {Irwin},
  {Jeong}, {Jonas}, {Khaire}, {Kofman}, {Korman}, {Kubik}, {Kuhlmann}, {Kuo},
  {Lee}, {Lowitz}, {Meyer}, {Michalik}, {Montgomery}, {Nadolski}, {Natoli},
  {Nguyen}, {Noble}, {Novosad}, {Padin}, {Pan}, {Pearson}, {Posada}, {Quan},
  {Rahlin}, {Ruhl}, {Sayre}, {Shirokoff}, {Smecher}, {Stark}, {Story},
  {Suzuki}, {Thompson}, {Tucker}, {Vanderlinde}, {Vieira}, {Wang}, {Whitehorn},
  {Yefremenko}, {Yoon}, \& {Young}}]{sobrin18}
{Sobrin}, J.~A., {Ade}, P.~A.~R., {Ahmed}, Z., {et~al.}
\newblock {Design and characterization of the SPT-3G receiver}. 2018, in
  \procspie, Vol. 10708, \procspie, 107081H, \dodoi{10.1117/12.2314366}

\bibitem[{{Nadolski} {et~al.}(2020){Nadolski}, {Vieira}, {Sobrin}, {Kofman},
  {Ade}, {Ahmed}, {Anderson}, {Avva}, {Basu Thakur}, {Bender}, {Benson},
  {Bryant}, {Carlstrom}, {Carter}, {Cecil}, {Chang}, {Cheshire}, {Chesmore},
  {Cliche}, {Cukierman}, {de Haan}, {Dierickx}, {Ding}, {Dutcher}, {Everett},
  {Farwick}, {Ferguson}, {Florez}, {Foster}, {Fu}, {Gallicchio}, {Gambrel},
  {Gardner}, {Groh}, {Guns}, {Guyser}, {Halverson}, {Harke-Hosemann},
  {Harrington}, {Harris}, {Henning}, {Holzapfel}, {Howe}, {Huang}, {Irwin},
  {Jeong}, {Jonas}, {Jones}, {Korman}, {Kovac}, {Kubik}, {Kuhlmann}, {Kuo},
  {Lee}, {Lowitz}, {McMahon}, {Meier}, {Meyer}, {Michalik}, {Montgomery},
  {Natoli}, {Nguyen}, {Noble}, {Novosad}, {Padin}, {Pan}, {Paschos}, {Pearson},
  {Posada}, {Quan}, {Rahlin}, {Riebel}, {Ruhl}, {Sayre}, {Shirokoff},
  {Smecher}, {Stark}, {Stephen}, {Story}, {Suzuki}, {Tandoi}, {Thompson},
  {Tucker}, {Vanderlinde}, {Wang}, {Whitehorn}, {Yefremenko}, {Yoon}, \&
  {Young}}]{nadolski20}
{Nadolski}, A., {Vieira}, J.~D., {Sobrin}, J.~A., {et~al.}
\newblock {Broadband, millimeter-wave antireflection coatings for large-format,
  cryogenic aluminum oxide optics}. 2020, \ao, 59, 3285,
  \dodoi{10.1364/AO.383921}

\bibitem[{Suzuki {et~al.}(2012)Suzuki, Arnold, Edwards, Engargiola, Ghribi,
  Holzapfel, Lee, Meng, Myers, O'Brient, Quealy, Rebeiz, \&
  Richards}]{suzuki12}
Suzuki, A., Arnold, K., Edwards, J., Engargiola, G., Ghribi, A., Holzapfel, W.,
  Lee, A., Meng, X., Myers, M., O'Brient, R., Quealy, E., Rebeiz, G., \&
  Richards, P.
\newblock Multi-chroic Dual-Polarization Bolometric Focal Plane for Studies of
  the Cosmic Microwave Background. 2012, Journal of Low Temperature Physics,
  167, 852, \dodoi{10.1007/s10909-012-0602-y}

\bibitem[{{Suzuki} {et~al.}(2018){Suzuki}, {Ade}, {Akiba}, {Alonso}, {Arnold},
  {Aumont}, {Baccigalupi}, {Barron}, {Basak}, {Beckman}, {Borrill},
  {Boulanger}, {Bucher}, {Calabrese}, {Chinone}, {Cho}, {Crill}, {Cukierman},
  {Curtis}, {de Haan}, {Dobbs}, {Dominjon}, {Dotani}, {Duband}, {Ducout},
  {Dunkley}, {Duval}, {Elleflot}, {Eriksen}, {Errard}, {Fischer}, {Fujino},
  {Funaki}, {Fuskeland}, {Ganga}, {Goeckner-Wald}, {Grain}, {Halverson},
  {Hamada}, {Hasebe}, {Hasegawa}, {Hattori}, {Hattori}, {Hayes}, {Hazumi},
  {Hidehira}, {Hill}, {Hilton}, {Hubmayr}, {Ichiki}, {Iida}, {Imada}, {Inoue},
  {Inoue}, {Irwin}, {Ishino}, {Jeong}, {Kanai}, {Kaneko}, {Kashima},
  {Katayama}, {Kawasaki}, {Kernasovskiy}, {Keskitalo}, {Kibayashi}, {Kida},
  {Kimura}, {Kisner}, {Kohri}, {Komatsu}, {Komatsu}, {Kuo}, {Kurinsky},
  {Kusaka}, {Lazarian}, {Lee}, {Li}, {Linder}, {Maffei}, {Mangilli}, {Maki},
  {Matsumura}, {Matsuura}, {Meilhan}, {Mima}, {Minami}, {Mitsuda}, {Montier},
  {Nagai}, {Nagasaki}, {Nagata}, {Nakajima}, {Nakamura}, {Namikawa}, {Naruse},
  {Nishino}, {Nitta}, {Noguchi}, {Ogawa}, {Oguri}, {Okada}, {Okamoto},
  {Okamura}, {Otani}, {Patanchon}, {Pisano}, {Rebeiz}, {Remazeilles},
  {Richards}, {Sakai}, {Sakurai}, {Sato}, {Sato}, {Sawada}, {Segawa},
  {Sekimoto}, {Seljak}, {Sherwin}, {Shimizu}, {Shinozaki}, {Stompor}, {Sugai},
  {Sugita}, {Suzuki}, {Tajima}, {Takada}, {Takaku}, {Takakura}, {Takatori},
  {Tanabe}, {Taylor}, {Thompson}, {Thorne}, {Tomaru}, {Tomida}, {Tomita},
  {Tristram}, {Tucker}, {Turin}, {Tsujimoto}, {Uozumi}, {Utsunomiya}, {Uzawa},
  {Vansyngel}, {Wehus}, {Westbrook}, {Willer}, {Whitehorn}, {Yamada},
  {Yamamoto}, {Yamasaki}, {Yamashita}, \& {Yoshida}}]{suzuki18}
{Suzuki}, A., {Ade}, P.~A.~R., {Akiba}, Y., {et~al.}
\newblock {The LiteBIRD Satellite Mission: Sub-Kelvin Instrument}. 2018,
  Journal of Low Temperature Physics, 193, 1048,
  \dodoi{10.1007/s10909-018-1947-7}

\bibitem[{{Galitzki} {et~al.}(2018){Galitzki}, {Ali}, {Arnold}, {Ashton},
  {Austermann}, {Baccigalupi}, {Baildon}, {Barron}, {Beall}, {Beckman},
  {Bruno}, {Bryan}, {Calisse}, {Chesmore}, {Chinone}, {Choi}, {Coppi},
  {Crowley}, {Crowley}, {Cukierman}, {Devlin}, {Dicker}, {Dober}, {Duff},
  {Dunkley}, {Fabbian}, {Gallardo}, {Gerbino}, {Goeckner-Wald}, {Golec},
  {Gudmundsson}, {Healy}, {Henderson}, {Hill}, {Hilton}, {Ho}, {Howe},
  {Hubmayr}, {Jeong}, {Keating}, {Koopman}, {Kiuchi}, {Kusaka}, {Lashner},
  {Lee}, {Li}, {Limon}, {Lungu}, {Matsuda}, {Mauskopf}, {May}, {McCallum},
  {McMahon}, {Nati}, {Niemack}, {Orlowski-Scherer}, {Parshley}, {Piccirillo},
  {Sathyanarayana Rao}, {Raum}, {Salatino}, {Seibert}, {Sierra},
  {Silva-Feaver}, {Simon}, {Staggs}, {Stevens}, {Suzuki}, {Teply}, {Thornton},
  {Tsai}, {Ullom}, {Vavagiakis}, {Vissers}, {Westbrook}, {Wollack}, {Xu}, \&
  {Zhu}}]{galitzki18}
{Galitzki}, N., {Ali}, A., {Arnold}, K.~S., {et~al.}
\newblock {The Simons Observatory: instrument overview}. 2018, in \procspie,
  Vol. 10708, \procspie, 1070804, \dodoi{10.1117/12.2312985}

\bibitem[{{Posada} {et~al.}(2015){Posada}, {Ade}, {Ahmed}, {Arnold},
  {Austermann}, {Bender}, {Bleem}, {Benson}, {Byrum}, {Carlstrom}, {Chang},
  {Cho}, {Ciocys}, {Cliche}, {Crawford}, {Cukierman}, {Czaplewski}, {Ding},
  {Divan}, {de Haan}, {Dobbs}, {Dutcher}, {Everett}, {Gilbert}, {Halverson},
  {Harrington}, {Hattori}, {Henning}, {Hilton}, {Holzapfel}, {Hubmayr},
  {Irwin}, {Jeong}, {Keisler}, {Kubik}, {Kuo}, {Lee}, {Leitch}, {Lendinez},
  {Meyer}, {Miller}, {Montgomery}, {Myers}, {Nadolski}, {Natoli}, {Nguyen},
  {Novosad}, {Padin}, {Pan}, {Pearson}, {Ruhl}, {Saliwanchik}, {Smecher},
  {Sayre}, {Shirokoff}, {Stan}, {Stark}, {Sobrin}, {Story}, {Suzuki},
  {Thompson}, {Tucker}, {Vanderlinde}, {Vieira}, {Wang}, {Whitehorn},
  {Yefremenko}, {Yoon}, \& {Ziegler}}]{posada15}
{Posada}, C.~M., {Ade}, P.~A.~R., {Ahmed}, Z., {et~al.}
\newblock {Fabrication of large dual-polarized multichroic TES bolometer arrays
  for CMB measurements with the SPT-3G camera}. 2015, Superconductor Science
  Technology, 28, 094002, \dodoi{10.1088/0953-2048/28/9/094002}

\bibitem[{{Posada} {et~al.}(2018){Posada}, {Ade}, {Ahmed}, {Anderson},
  {Austermann}, {Avva}, {Thakur}, {Bender}, {Benson}, {Carlstrom}, {Carter},
  {Cecil}, {Chang}, {Cliche}, {Cukierman}, {Denison}, {de Haan}, {Ding},
  {Divan}, {Dobbs}, {Dutcher}, {Everett}, {Foster}, {Gannon}, {Gilbert},
  {Groh}, {Halverson}, {Harke-Hosemann}, {Harrington}, {Henning}, {Hilton},
  {Holzapfel}, {Huang}, {Irwin}, {Jeong}, {Jonas}, {Khaire}, {Kofman},
  {Korman}, {Kubik}, {Kuhlmann}, {Kuo}, {Lee}, {Lowitz}, {Meyer}, {Michalik},
  {Miller}, {Montgomery}, {Nadolski}, {Natoli}, {Nguyen}, {Noble}, {Novosad},
  {Padin}, {Pan}, {Pearson}, {Rahlin}, {Ruhl}, {Saunders}, {Sayre}, {Shirley},
  {Shirokoff}, {Smecher}, {Sobrin}, {Stan}, {Stark}, {Story}, {Suzuki}, {Tang},
  {Thompson}, {Tucker}, {Vale}, {Vanderlinde}, {Vieira}, {Wang}, {Whitehorn},
  {Yefremenko}, {Yoon}, \& {Young}}]{posada18}
---.
\newblock {Fabrication of Detector Arrays for the SPT-3G Receiver}. 2018,
  Journal of Low Temperature Physics, \dodoi{10.1007/s10909-018-1924-1}

\bibitem[{{Dutcher} {et~al.}(2018){Dutcher}, {Ade}, {Ahmed}, {Anderson},
  {Avva}, {Thakur}, {Bender}, {Benson}, {Carlstrom}, {Carter}, {Cecil},
  {Chang}, {Cliche}, {Cukierman}, {de Haan}, {Ding}, {Dobbs}, {Everett},
  {Foster}, {Gallicchio}, {Gilbert}, {Groh}, {Harke-Hosemann}, {Guns},
  {Halverson}, {Harrington}, {Henning}, {Holzapfel}, {Huang}, {Irwin}, {Jeong},
  {Jonas}, {Khaire}, {Kofman}, {Korman}, {Kubik}, {Kuhlmann}, {Kuo}, {Lowitz},
  {Lee}, {Meyer}, {Michalik}, {Montgomery}, {Nadolski}, {Natoli}, {Nguyen},
  {Noble}, {Novosad}, {Padin}, {Pan}, {Pearson}, {Posada}, {Quan}, {Rahlin},
  {Ruhl}, {Sayre}, {Shirokoff}, {Smecher}, {Sobrin}, {Stark}, {Story},
  {Suzuki}, {Thompson}, {Tucker}, {Vanderlinde}, {Vieira}, {Wang}, {Whitehorn},
  {Yefremenko}, {Yoon}, \& {Young}}]{dutcher18}
{Dutcher}, D., {Ade}, P.~A.~R., {Ahmed}, Z., {et~al.}
\newblock {Characterization and performance of the second-year SPT-3G focal
  plane}. 2018, in \procspie, Vol. 10708, \procspie, 107081Z,
  \dodoi{10.1117/12.2312451}

\bibitem[{{Bender} {et~al.}(2014){Bender}, {Cliche}, {de Haan}, {Dobbs},
  {Gilbert}, {Montgomery}, {Rowlands}, {Smecher}, {Smith}, \&
  {Wilson}}]{bender14}
{Bender}, A.~N., {Cliche}, J.-F., {de Haan}, T., {Dobbs}, M.~A., {Gilbert},
  A.~J., {Montgomery}, J., {Rowlands}, N., {Smecher}, G.~M., {Smith}, K., \&
  {Wilson}, A.
\newblock {Digital frequency domain multiplexing readout electronics for the
  next generation of millimeter telescopes}. 2014, in \procspie, Vol. 9153,
  Millimeter, Submillimeter, and Far-Infrared Detectors and Instrumentation for
  Astronomy VII, 91531A, \dodoi{10.1117/12.2054949}

\bibitem[{{Bender} {et~al.}(2016){Bender}, {Ade}, {Anderson}, {Avva}, {Ahmed},
  {Arnold}, {Austermann}, {Basu Thakur}, {Benson}, {Bleem}, {Byrum},
  {Carlstrom}, {Carter}, {Chang}, {Cho}, {Cliche}, {Crawford}, {Cukierman},
  {Czaplewski}, {Ding}, {Divan}, {de Haan}, {Dobbs}, {Dutcher}, {Everett},
  {Gilbert}, {Groh}, {Guyser}, {Halverson}, {Harke-Hosemann}, {Harrington},
  {Hattori}, {Henning}, {Hilton}, {Holzapfel}, {Huang}, {Irwin}, {Jeong},
  {Khaire}, {Korman}, {Kubik}, {Kuo}, {Lee}, {Leitch}, {Lendinez}, {Meyer},
  {Miller}, {Montgomery}, {Nadolski}, {Natoli}, {Nguyen}, {Novosad}, {Padin},
  {Pan}, {Pearson}, {Posada}, {Rahlin}, {Reichardt}, {Ruhl}, {Saliwanchik},
  {Sayre}, {Shariff}, {Shirley}, {Shirokoff}, {Smecher}, {Sobrin}, {Stan},
  {Stark}, {Story}, {Suzuki}, {Tang}, {Thompson}, {Tucker}, {Vanderlinde},
  {Vieira}, {Wang}, {Whitehorn}, {Yefremenko}, \& {Yoon}}]{bender16}
{Bender}, A.~N., {Ade}, P.~A.~R., {Anderson}, A.~J., {et~al.}
\newblock {Integrated performance of a frequency domain multiplexing readout in
  the SPT-3G receiver}. 2016, in \procspie, Vol. 9914, Millimeter,
  Submillimeter, and Far-Infrared Detectors and Instrumentation for Astronomy
  VIII, 99141D, \dodoi{10.1117/12.2232146}

\bibitem[{{Planck Collaboration} {et~al.}(2016{\natexlab{a}}){Planck
  Collaboration}, {Adam}, {Ade}, {Aghanim}, {Alves}, {Arnaud}, {Ashdown},
  {Aumont}, {Baccigalupi}, {Banday}, {Barreiro}, {Bartlett}, {Bartolo},
  {Battaner}, {Benabed}, {Beno{\^\i}t}, {Benoit-L{\'e}vy}, {Bernard},
  {Bersanelli}, {Bielewicz}, {Bock}, {Bonaldi}, {Bonavera}, {Bond}, {Borrill},
  {Bouchet}, {Boulanger}, {Bucher}, {Burigana}, {Butler}, {Calabrese},
  {Cardoso}, {Catalano}, {Challinor}, {Chamballu}, {Chary}, {Chiang},
  {Christensen}, {Clements}, {Colombi}, {Colombo}, {Combet}, {Couchot},
  {Coulais}, {Crill}, {Curto}, {Cuttaia}, {Danese}, {Davies}, {Davis}, {de
  Bernardis}, {de Rosa}, {de Zotti}, {Delabrouille}, {D{\'e}sert}, {Dickinson},
  {Diego}, {Dole}, {Donzelli}, {Dor{\'e}}, {Douspis}, {Ducout}, {Dupac},
  {Efstathiou}, {Elsner}, {En{\ss}lin}, {Eriksen}, {Falgarone}, {Fergusson},
  {Finelli}, {Forni}, {Frailis}, {Fraisse}, {Franceschi}, {Frejsel},
  {Galeotta}, {Galli}, {Ganga}, {Ghosh}, {Giard}, {Giraud-H{\'e}raud},
  {Gjerl{\o}w}, {Gonz{\'a}lez-Nuevo}, {G{\'o}rski}, {Gratton}, {Gregorio},
  {Gruppuso}, {Gudmundsson}, {Hansen}, {Hanson}, {Harrison}, {Helou},
  {Henrot-Versill{\'e}}, {Hern{\'a}ndez-Monteagudo}, {Herranz}, {Hildebrandt},
  {Hivon}, {Hobson}, {Holmes}, {Hornstrup}, {Hovest}, {Huffenberger}, {Hurier},
  {Jaffe}, {Jaffe}, {Jones}, {Juvela}, {Keih{\"a}nen}, {Keskitalo}, {Kisner},
  {Kneissl}, {Knoche}, {Kunz}, {Kurki-Suonio}, {Lagache},
  {L{\"a}hteenm{\"a}ki}, {Lamarre}, {Lasenby}, {Lattanzi}, {Lawrence}, {Le
  Jeune}, {Leahy}, {Leonardi}, {Lesgourgues}, {Levrier}, {Liguori}, {Lilje},
  {Linden-V{\o}rnle}, {L{\'o}pez-Caniego}, {Lubin}, {Mac{\'\i}as-P{\'e}rez},
  {Maggio}, {Maino}, {Mandolesi}, {Mangilli}, {Maris}, {Marshall}, {Martin},
  {Mart{\'\i}nez-Gonz{\'a}lez}, {Masi}, {Matarrese}, {McGehee}, {Meinhold},
  {Melchiorri}, {Mendes}, {Mennella}, {Migliaccio}, {Mitra},
  {Miville-Desch{\^e}nes}, {Moneti}, {Montier}, {Morgante}, {Mortlock}, {Moss},
  {Munshi}, {Murphy}, {Naselsky}, {Nati}, {Natoli}, {Netterfield},
  {N{\o}rgaard-Nielsen}, {Noviello}, {Novikov}, {Novikov}, {Orlando},
  {Oxborrow}, {Paci}, {Pagano}, {Pajot}, {Paladini}, {Paoletti}, {Partridge},
  {Pasian}, {Patanchon}, {Pearson}, {Perdereau}, {Perotto}, {Perrotta},
  {Pettorino}, {Piacentini}, {Piat}, {Pierpaoli}, {Pietrobon}, {Plaszczynski},
  {Pointecouteau}, {Polenta}, {Pratt}, {Pr{\'e}zeau}, {Prunet}, {Puget},
  {Rachen}, {Reach}, {Rebolo}, {Reinecke}, {Remazeilles}, {Renault}, {Renzi},
  {Ristorcelli}, {Rocha}, {Rosset}, {Rossetti}, {Roudier},
  {Rubi{\~n}o-Mart{\'\i}n}, {Rusholme}, {Sandri}, {Santos}, {Savelainen},
  {Savini}, {Scott}, {Seiffert}, {Shellard}, {Spencer}, {Stolyarov}, {Stompor},
  {Strong}, {Sudiwala}, {Sunyaev}, {Sutton}, {Suur-Uski}, {Sygnet}, {Tauber},
  {Terenzi}, {Toffolatti}, {Tomasi}, {Tristram}, {Tucci}, {Tuovinen}, {Umana},
  {Valenziano}, {Valiviita}, {Van Tent}, {Vielva}, {Villa}, {Wade}, {Wandelt},
  {Wehus}, {Wilkinson}, {Yvon}, {Zacchei}, \& {Zonca}}]{planck15-10}
{Planck Collaboration}, {Adam}, R., {Ade}, P.~A.~R., {et~al.}
\newblock {Planck 2015 results. X. Diffuse component separation: Foreground
  maps}. 2016{\natexlab{a}}, \aap, 594, A10,
  \dodoi{10.1051/0004-6361/201525967}

\bibitem[{{Story} {et~al.}(2013){Story}, {Reichardt}, {Hou}, {Keisler}, {Aird},
  {Benson}, {Bleem}, {Carlstrom}, {Chang}, {Cho}, {Crawford}, {Crites}, {de
  Haan}, {Dobbs}, {Dudley}, {Follin}, {George}, {Halverson}, {Holder},
  {Holzapfel}, {Hoover}, {Hrubes}, {Joy}, {Knox}, {Lee}, {Leitch}, {Lueker},
  {Luong-Van}, {McMahon}, {Mehl}, {Meyer}, {Millea}, {Mohr}, {Montroy},
  {Padin}, {Plagge}, {Pryke}, {Ruhl}, {Sayre}, {Schaffer}, {Shaw}, {Shirokoff},
  {Spieler}, {Staniszewski}, {Stark}, {van Engelen}, {Vanderlinde}, {Vieira},
  {Williamson}, \& {Zahn}}]{story13}
{Story}, K.~T., {Reichardt}, C.~L., {Hou}, Z., {et~al.}
\newblock {A Measurement of the Cosmic Microwave Background Damping Tail from
  the 2500-Square-Degree SPT-SZ Survey}. 2013, \apj, 779, 86,
  \dodoi{10.1088/0004-637X/779/1/86}

\bibitem[{{Crites} {et~al.}(2015){Crites}, {Henning}, {Ade}, {Aird},
  {Austermann}, {Beall}, {Bender}, {Benson}, {Bleem}, {Carlstrom}, {Chang},
  {Chiang}, {Cho}, {Citron}, {Crawford}, {de Haan}, {Dobbs}, {Everett},
  {Gallicchio}, {Gao}, {George}, {Gilbert}, {Halverson}, {Hanson},
  {Harrington}, {Hilton}, {Holder}, {Holzapfel}, {Hoover}, {Hou}, {Hrubes},
  {Huang}, {Hubmayr}, {Irwin}, {Keisler}, {Knox}, {Lee}, {Leitch}, {Li},
  {Liang}, {Luong-Van}, {McMahon}, {Mehl}, {Meyer}, {Mocanu}, {Montroy},
  {Natoli}, {Nibarger}, {Novosad}, {Padin}, {Pryke}, {Reichardt}, {Ruhl},
  {Saliwanchik}, {Sayre}, {Schaffer}, {Smecher}, {Stark}, {Story}, {Tucker},
  {Vanderlinde}, {Vieira}, {Wang}, {Whitehorn}, {Yefremenko}, \&
  {Zahn}}]{crites15}
{Crites}, A.~T., {Henning}, J.~W., {Ade}, P.~A.~R., {et~al.}
\newblock {Measurements of E-Mode Polarization and Temperature-E-Mode
  Correlation in the Cosmic Microwave Background from 100 Square Degrees of
  SPTpol Data}. 2015, \apj, 805, 36, \dodoi{10.1088/0004-637X/805/1/36}

\bibitem[{{Keisler} {et~al.}(2015){Keisler}, {Hoover}, {Harrington}, {Henning},
  {Ade}, {Aird}, {Austermann}, {Beall}, {Bender}, {Benson}, {Bleem},
  {Carlstrom}, {Chang}, {Chiang}, {Cho}, {Citron}, {Crawford}, {Crites}, {de
  Haan}, {Dobbs}, {Everett}, {Gallicchio}, {Gao}, {George}, {Gilbert},
  {Halverson}, {Hanson}, {Hilton}, {Holder}, {Holzapfel}, {Hou}, {Hrubes},
  {Huang}, {Hubmayr}, {Irwin}, {Knox}, {Lee}, {Leitch}, {Li}, {Luong-Van},
  {Marrone}, {McMahon}, {Mehl}, {Meyer}, {Mocanu}, {Natoli}, {Nibarger},
  {Novosad}, {Padin}, {Pryke}, {Reichardt}, {Ruhl}, {Saliwanchik}, {Sayre},
  {Schaffer}, {Shirokoff}, {Smecher}, {Stark}, {Story}, {Tucker},
  {Vanderlinde}, {Vieira}, {Wang}, {Whitehorn}, {Yefremenko}, \&
  {Zahn}}]{keisler15}
{Keisler}, R., {Hoover}, S., {Harrington}, N., {et~al.}
\newblock {Measurements of Sub-degree B-mode Polarization in the Cosmic
  Microwave Background from 100 Square Degrees of SPTpol Data}. 2015, \apj,
  807, 151, \dodoi{10.1088/0004-637X/807/2/151}

\bibitem[{{Jones} {et~al.}(2007){Jones}, {Montroy}, {Crill}, {Contaldi},
  {Kisner}, {Lange}, {MacTavish}, {Netterfield}, \& {Ruhl}}]{jones07}
{Jones}, W.~C., {Montroy}, T.~E., {Crill}, B.~P., {Contaldi}, C.~R., {Kisner},
  T.~S., {Lange}, A.~E., {MacTavish}, C.~J., {Netterfield}, C.~B., \& {Ruhl},
  J.~E.
\newblock {Instrumental and analytic methods for bolometric polarimetry}. 2007,
  \aap, 470, 771, \dodoi{10.1051/0004-6361:20065911}

\bibitem[{{Zaldarriaga}(2001)}]{zaldarriaga01}
{Zaldarriaga}, M.
\newblock {Nature of the E-B decomposition of CMB polarization}. 2001, \prd,
  64, 103001, \dodoi{10.1103/PhysRevD.64.103001}

\bibitem[{{Tristram} {et~al.}(2005){Tristram}, {Mac{\'{\i}}as-P{\'e}rez},
  {Renault}, \& {Santos}}]{tristram05}
{Tristram}, M., {Mac{\'{\i}}as-P{\'e}rez}, J.~F., {Renault}, C., \& {Santos},
  D.
\newblock {XSPECT, estimation of the angular power spectrum by computing
  cross-power spectra with analytical error bars}. 2005, \mnras, 358, 833,
  \dodoi{10.1111/j.1365-2966.2005.08760.x}

\bibitem[{{Polenta} {et~al.}(2005){Polenta}, {Marinucci}, {Balbi}, {de
  Bernardis}, {Hivon}, {Masi}, {Natoli}, \& {Vittorio}}]{polenta05}
{Polenta}, G., {Marinucci}, D., {Balbi}, A., {de Bernardis}, P., {Hivon}, E.,
  {Masi}, S., {Natoli}, P., \& {Vittorio}, N.
\newblock {Unbiased estimation of an angular power spectrum}. 2005, \jcap, 11,
  1, \dodoi{10.1088/1475-7516/2005/11/001}

\bibitem[{{Hivon} {et~al.}(2002){Hivon}, {G{\'o}rski}, {Netterfield}, {Crill},
  {Prunet}, \& {Hansen}}]{hivon02}
{Hivon}, E., {G{\'o}rski}, K.~M., {Netterfield}, C.~B., {Crill}, B.~P.,
  {Prunet}, S., \& {Hansen}, F.
\newblock {MASTER of the Cosmic Microwave Background Anisotropy Power Spectrum:
  A Fast Method for Statistical Analysis of Large and Complex Cosmic Microwave
  Background Data Sets}. 2002, \apj, 567, 2, \dodoi{10.1086/338126}

\bibitem[{{G{\'o}rski} {et~al.}(2005){G{\'o}rski}, {Hivon}, {Banday},
  {Wandelt}, {Hansen}, {Reinecke}, \& {Bartelmann}}]{gorski05}
{G{\'o}rski}, K.~M., {Hivon}, E., {Banday}, A.~J., {Wandelt}, B.~D., {Hansen},
  F.~K., {Reinecke}, M., \& {Bartelmann}, M.
\newblock {HEALPix: A Framework for High-Resolution Discretization and Fast
  Analysis of Data Distributed on the Sphere}. 2005, \apj, 622, 759,
  \dodoi{10.1086/427976}

\bibitem[{Zonca {et~al.}(2019)Zonca, Singer, Lenz, Reinecke, Rosset, Hivon, \&
  Gorski}]{zonca19}
Zonca, A., Singer, L., Lenz, D., Reinecke, M., Rosset, C., Hivon, E., \&
  Gorski, K.
\newblock healpy: equal area pixelization and spherical harmonics transforms
  for data on the sphere in Python. 2019, Journal of Open Source Software, 4,
  1298, \dodoi{10.21105/joss.01298}

\bibitem[{{George} {et~al.}(2015){George}, {Reichardt}, {Aird}, {Benson},
  {Bleem}, {Carlstrom}, {Chang}, {Cho}, {Crawford}, {Crites}, {de Haan},
  {Dobbs}, {Dudley}, {Halverson}, {Harrington}, {Holder}, {Holzapfel}, {Hou},
  {Hrubes}, {Keisler}, {Knox}, {Lee}, {Leitch}, {Lueker}, {Luong-Van},
  {McMahon}, {Mehl}, {Meyer}, {Millea}, {Mocanu}, {Mohr}, {Montroy}, {Padin},
  {Plagge}, {Pryke}, {Ruhl}, {Schaffer}, {Shaw}, {Shirokoff}, {Spieler},
  {Staniszewski}, {Stark}, {Story}, {van Engelen}, {Vanderlinde}, {Vieira},
  {Williamson}, \& {Zahn}}]{george15}
{George}, E.~M., {Reichardt}, C.~L., {Aird}, K.~A., {et~al.}
\newblock {A Measurement of Secondary Cosmic Microwave Background Anisotropies
  from the 2500-Square-degree SPT-SZ Survey}. 2015, \apj, 799, 177,
  \dodoi{10.1088/0004-637X/799/2/177}

\bibitem[{{De Zotti} {et~al.}(2005){De Zotti}, {Ricci}, {Mesa}, {Silva},
  {Mazzotta}, {Toffolatti}, \& {Gonz{\'a}lez-Nuevo}}]{dezotti05}
{De Zotti}, G., {Ricci}, R., {Mesa}, D., {Silva}, L., {Mazzotta}, P.,
  {Toffolatti}, L., \& {Gonz{\'a}lez-Nuevo}, J.
\newblock {Predictions for high-frequency radio surveys of extragalactic
  sources}. 2005, \aap, 431, 893, \dodoi{10.1051/0004-6361:20042108}

\bibitem[{{B{\'e}thermin} {et~al.}(2012){B{\'e}thermin}, {Daddi}, {Magdis},
  {Sargent}, {Hezaveh}, {Elbaz}, {Le Borgne}, {Mullaney}, {Pannella}, {Buat},
  {Charmandaris}, {Lagache}, \& {Scott}}]{bethermin12}
{B{\'e}thermin}, M., {Daddi}, E., {Magdis}, G., {Sargent}, M.~T., {Hezaveh},
  Y., {Elbaz}, D., {Le Borgne}, D., {Mullaney}, J., {Pannella}, M., {Buat}, V.,
  {Charmandaris}, V., {Lagache}, G., \& {Scott}, D.
\newblock {A Unified Empirical Model for Infrared Galaxy Counts Based on the
  Observed Physical Evolution of Distant Galaxies}. 2012, \apjl, 757, L23,
  \dodoi{10.1088/2041-8205/757/2/L23}

\bibitem[{{Everett} {et~al.}(2020){Everett}, {Zhang}, {Crawford}, {Vieira},
  {Aravena}, {Archipley}, {Austermann}, {Benson}, {Bleem}, {Carlstrom},
  {Chang}, {Chapman}, {Crites}, {de Haan}, {Dobbs}, {George}, {Halverson},
  {Harrington}, {Holder}, {Holzapfel}, {Hrubes}, {Knox}, {Lee}, {Luong-Van},
  {Mangian}, {Marrone}, {McMahon}, {Meyer}, {Mocanu}, {Mohr}, {Natoli},
  {Padin}, {Pryke}, {Reichardt}, {Reuter}, {Ruhl}, {Sayre}, {Schaffer},
  {Shirokoff}, {Spilker}, {Stalder}, {Staniszewski}, {Stark}, {Story},
  {Switzer}, {Vanderlinde}, {Wei{\ss}}, \& {Williamson}}]{everett20}
{Everett}, W.~B., {Zhang}, L., {Crawford}, T.~M., {et~al.}
\newblock {Millimeter-wave Point Sources from the 2500 Square Degree SPT-SZ
  Survey: Catalog and Population Statistics}. 2020, \apj, 900, 55,
  \dodoi{10.3847/1538-4357/ab9df7}

\bibitem[{{Keisler} {et~al.}(2011){Keisler}, {Reichardt}, {Aird}, {Benson},
  {Bleem}, {Carlstrom}, {Chang}, {Cho}, {Crawford}, {Crites}, {de Haan},
  {Dobbs}, {Dudley}, {George}, {Halverson}, {Holder}, {Holzapfel}, {Hoover},
  {Hou}, {Hrubes}, {Joy}, {Knox}, {Lee}, {Leitch}, {Lueker}, {Luong-Van},
  {McMahon}, {Mehl}, {Meyer}, {Millea}, {Mohr}, {Montroy}, {Natoli}, {Padin},
  {Plagge}, {Pryke}, {Ruhl}, {Schaffer}, {Shaw}, {Shirokoff}, {Spieler},
  {Staniszewski}, {Stark}, {Story}, {van Engelen}, {Vanderlinde}, {Vieira},
  {Williamson}, \& {Zahn}}]{keisler11}
{Keisler}, R., {Reichardt}, C.~L., {Aird}, K.~A., {et~al.}
\newblock {A Measurement of the Damping Tail of the Cosmic Microwave Background
  Power Spectrum with the South Pole Telescope}. 2011, \apj, 743, 28,
  \dodoi{10.1088/0004-637X/743/1/28}

\bibitem[{{Lueker} {et~al.}(2010){Lueker}, {Reichardt}, {Schaffer}, {Zahn},
  {Ade}, {Aird}, {Benson}, {Bleem}, {Carlstrom}, {Chang}, {Cho}, {Crawford},
  {Crites}, {de Haan}, {Dobbs}, {George}, {Hall}, {Halverson}, {Holder},
  {Holzapfel}, {Hrubes}, {Joy}, {Keisler}, {Knox}, {Lee}, {Leitch}, {McMahon},
  {Mehl}, {Meyer}, {Mohr}, {Montroy}, {Padin}, {Plagge}, {Pryke}, {Ruhl},
  {Shaw}, {Shirokoff}, {Spieler}, {Stalder}, {Staniszewski}, {Stark},
  {Vanderlinde}, {Vieira}, \& {Williamson}}]{lueker10}
{Lueker}, M., {Reichardt}, C.~L., {Schaffer}, K.~K., {et~al.}
\newblock {Measurements of Secondary Cosmic Microwave Background Anisotropies
  with the South Pole Telescope}. 2010, \apj, 719, 1045,
  \dodoi{10.1088/0004-637X/719/2/1045}

\bibitem[{{Lewis} \& {Bridle}(2002)}]{lewis02b}
{Lewis}, A., \& {Bridle}, S.
\newblock {Cosmological parameters from CMB and other data: A Monte Carlo
  approach}. 2002, \prd, 66, 103511

\bibitem[{{Lewis} {et~al.}(2000){Lewis}, {Challinor}, \& {Lasenby}}]{lewis00}
{Lewis}, A., {Challinor}, A., \& {Lasenby}, A.
\newblock {Efficient Computation of Cosmic Microwave Background Anisotropies in
  Closed Friedmann-Robertson-Walker Models}. 2000, \apj, 538, 473,
  \dodoi{10.1086/309179}

\bibitem[{{Jeong} {et~al.}(2014){Jeong}, {Chluba}, {Dai}, {Kamionkowski}, \&
  {Wang}}]{jeong14}
{Jeong}, D., {Chluba}, J., {Dai}, L., {Kamionkowski}, M., \& {Wang}, X.
\newblock {Effect of aberration on partial-sky measurements of the cosmic
  microwave background temperature power spectrum}. 2014, \prd, 89, 023003,
  \dodoi{10.1103/PhysRevD.89.023003}

\bibitem[{{Manzotti} {et~al.}(2014){Manzotti}, {Hu}, \&
  {Benoit-L{\'e}vy}}]{manzotti14}
{Manzotti}, A., {Hu}, W., \& {Benoit-L{\'e}vy}, A.
\newblock {Super-sample CMB lensing}. 2014, \prd, 90, 023003,
  \dodoi{10.1103/PhysRevD.90.023003}

\bibitem[{{Natale} {et~al.}(2020){Natale}, {Pagano}, {Lattanzi}, {Migliaccio},
  {Colombo}, {Gruppuso}, {Natoli}, \& {Polenta}}]{natale20}
{Natale}, U., {Pagano}, L., {Lattanzi}, M., {Migliaccio}, M., {Colombo}, L.~P.,
  {Gruppuso}, A., {Natoli}, P., \& {Polenta}, G.
\newblock {A novel CMB polarization likelihood package for large angular scales
  built from combined WMAP and Planck LFI legacy maps}. 2020, arXiv e-prints,
  arXiv:2005.05600.
\newblock \doarXiv{2005.05600}

\bibitem[{{Planck Collaboration} {et~al.}(2016{\natexlab{b}}){Planck
  Collaboration}, {Adam}, {Ade}, {Aghanim}, {Arnaud}, {Aumont}, {Baccigalupi},
  {Banday}, {Barreiro}, {Bartlett}, {Bartolo}, {Battaner}, {Benabed},
  {Benoit-L{\'e}vy}, {Bernard}, {Bersanelli}, {Bielewicz}, {Bonaldi},
  {Bonavera}, {Bond}, {Borrill}, {Bouchet}, {Boulanger}, {Bracco}, {Bucher},
  {Burigana}, {Butler}, {Calabrese}, {Cardoso}, {Catalano}, {Challinor},
  {Chamballu}, {Chary}, {Chiang}, {Christensen}, {Clements}, {Colombi},
  {Colombo}, {Combet}, {Couchot}, {Coulais}, {Crill}, {Curto}, {Cuttaia},
  {Danese}, {Davies}, {Davis}, {de Bernardis}, {de Zotti}, {Delabrouille},
  {Delouis}, {D{\'e}sert}, {Dickinson}, {Diego}, {Dolag}, {Dole}, {Donzelli},
  {Dor{\'e}}, {Douspis}, {Ducout}, {Dunkley}, {Dupac}, {Efstathiou}, {Elsner},
  {En{\ss}lin}, {Eriksen}, {Falgarone}, {Finelli}, {Forni}, {Frailis},
  {Fraisse}, {Franceschi}, {Frejsel}, {Galeotta}, {Galli}, {Ganga}, {Ghosh},
  {Giard}, {Giraud-H{\'e}raud}, {Gjerl{\o}w}, {Gonz{\'a}lez-Nuevo},
  {G{\'o}rski}, {Gratton}, {Gregorio}, {Gruppuso}, {Guillet}, {Hansen},
  {Hanson}, {Harrison}, {Helou}, {Henrot-Versill{\'e}},
  {Hern{\'a}ndez-Monteagudo}, {Herranz}, {Hivon}, {Hobson}, {Holmes},
  {Huffenberger}, {Hurier}, {Jaffe}, {Jaffe}, {Jewell}, {Jones}, {Juvela},
  {Keih{\"a}nen}, {Keskitalo}, {Kisner}, {Kneissl}, {Knoche}, {Knox},
  {Krachmalnicoff}, {Kunz}, {Kurki-Suonio}, {Lagache}, {Lamarre}, {Lasenby},
  {Lattanzi}, {Lawrence}, {Leahy}, {Leonardi}, {Lesgourgues}, {Levrier},
  {Liguori}, {Lilje}, {Linden-V{\o}rnle}, {L{\'o}pez-Caniego}, {Lubin},
  {Mac{\'\i}as-P{\'e}rez}, {Maffei}, {Maino}, {Mandolesi}, {Mangilli}, {Maris},
  {Martin}, {Mart{\'\i}nez-Gonz{\'a}lez}, {Masi}, {Matarrese}, {Mazzotta},
  {Meinhold}, {Melchiorri}, {Mendes}, {Mennella}, {Migliaccio}, {Mitra},
  {Miville-Desch{\^e}nes}, {Moneti}, {Montier}, {Morgante}, {Mortlock}, {Moss},
  {Munshi}, {Murphy}, {Naselsky}, {Nati}, {Natoli}, {Netterfield},
  {N{\o}rgaard-Nielsen}, {Noviello}, {Novikov}, {Novikov}, {Pagano}, {Pajot},
  {Paladini}, {Paoletti}, {Partridge}, {Pasian}, {Patanchon}, {Pearson},
  {Perdereau}, {Perotto}, {Perrotta}, {Pettorino}, {Piacentini}, {Piat},
  {Pierpaoli}, {Pietrobon}, {Plaszczynski}, {Pointecouteau}, {Polenta},
  {Ponthieu}, {Popa}, {Pratt}, {Prunet}, {Puget}, {Rachen}, {Reach}, {Rebolo},
  {Remazeilles}, {Renault}, {Renzi}, {Ricciardi}, {Ristorcelli}, {Rocha},
  {Rosset}, {Rossetti}, {Roudier}, {Rouill{\'e} d'Orfeuil},
  {Rubi{\~n}o-Mart{\'\i}n}, {Rusholme}, {Sandri}, {Santos}, {Savelainen},
  {Savini}, {Scott}, {Soler}, {Spencer}, {Stolyarov}, {Stompor}, {Sudiwala},
  {Sunyaev}, {Sutton}, {Suur-Uski}, {Sygnet}, {Tauber}, {Terenzi},
  {Toffolatti}, {Tomasi}, {Tristram}, {Tucci}, {Tuovinen}, {Valenziano},
  {Valiviita}, {Van Tent}, {Vibert}, {Vielva}, {Villa}, {Wade}, {Wandelt},
  {Watson}, {Wehus}, {White}, {White}, {Yvon}, {Zacchei}, \&
  {Zonca}}]{planck12-30}
{Planck Collaboration}, {Adam}, R., {Ade}, P.~A.~R., {et~al.}
\newblock {Planck intermediate results. XXX. The angular power spectrum of
  polarized dust emission at intermediate and high Galactic latitudes}.
  2016{\natexlab{b}}, \aap, 586, A133, \dodoi{10.1051/0004-6361/201425034}

\bibitem[{{Planck Collaboration} {et~al.}(2020{\natexlab{c}}){Planck
  Collaboration}, {Akrami}, {Ashdown}, {Aumont}, {Baccigalupi}, {Ballardini},
  {Band ay}, {Barreiro}, {Bartolo}, {Basak}, {Benabed}, {Bernard},
  {Bersanelli}, {Bielewicz}, {Bond}, {Borrill}, {Bouchet}, {Boulanger},
  {Bracco}, {Bucher}, {Burigana}, {Calabrese}, {Cardoso}, {Carron}, {Chiang},
  {Combet}, {Crill}, {de Bernardis}, {de Zotti}, {Delabrouille}, {Delouis}, {Di
  Valentino}, {Dickinson}, {Diego}, {Ducout}, {Dupac}, {Efstathiou}, {Elsner},
  {En{\ss}lin}, {Falgarone}, {Fantaye}, {Ferri{\`e}re}, {Finelli},
  {Forastieri}, {Frailis}, {Fraisse}, {Franceschi}, {Frolov}, {Galeotta},
  {Galli}, {Ganga}, {G{\'e}nova-Santos}, {Ghosh}, {Gonz{\'a}lez-Nuevo},
  {G{\'o}rski}, {Gruppuso}, {Gudmundsson}, {Guillet}, {Handley}, {Hansen},
  {Herranz}, {Huang}, {Jaffe}, {Jones}, {Keih{\"a}nen}, {Keskitalo}, {Kiiveri},
  {Kim}, {Krachmalnicoff}, {Kunz}, {Kurki-Suonio}, {Lamarre}, {Lasenby}, {Le
  Jeune}, {Levrier}, {Liguori}, {Lilje}, {Lindholm}, {L{\'o}pez-Caniego},
  {Lubin}, {Ma}, {Mac{\'\i}as-P{\'e}rez}, {Maggio}, {Maino}, {Mandolesi},
  {Mangilli}, {Martin}, {Mart{\'\i}nez-Gonz{\'a}lez}, {Matarrese}, {McEwen},
  {Meinhold}, {Melchiorri}, {Migliaccio}, {Miville-Desch{\^e}nes}, {Molinari},
  {Moneti}, {Montier}, {Morgante}, {Natoli}, {Pagano}, {Paoletti}, {Pettorino},
  {Piacentini}, {Polenta}, {Puget}, {Rachen}, {Reinecke}, {Remazeilles},
  {Renzi}, {Rocha}, {Rosset}, {Roudier}, {Rubi{\~n}o-Mart{\'\i}n},
  {Ruiz-Granados}, {Salvati}, {Sandri}, {Savelainen}, {Scott}, {Soler},
  {Spencer}, {Tauber}, {Tavagnacco}, {Toffolatti}, {Tomasi}, {Trombetti},
  {Valiviita}, {Vansyngel}, {Van Tent}, {Vielva}, {Villa}, {Vittorio}, {Wehus},
  {Zacchei}, \& {Zonca}}]{planck18-11}
{Planck Collaboration}, {Akrami}, Y., {Ashdown}, M., {et~al.}
\newblock {Planck 2018 results. XI. Polarized dust foregrounds}.
  2020{\natexlab{c}}, \aap, 641, A11, \dodoi{10.1051/0004-6361/201832618}

\bibitem[{{Mocanu} {et~al.}(2019){Mocanu}, {Crawford}, {Aylor}, {Benson},
  {Bleem}, {Carlstrom}, {Chang}, {Cho}, {Chown}, {Crites}, {de Haan}, {Dobbs},
  {Everett}, {George}, {Halverson}, {Harrington}, {Henning}, {Holder},
  {Holzapfel}, {Hou}, {Hrubes}, {Knox}, {Lee}, {Luong-Van}, {Marrone},
  {McMahon}, {Meyer}, {Millea}, {Mohr}, {Natoli}, {Omori}, {Padin}, {Pryke},
  {Reichardt}, {Ruhl}, {Sayre}, {Schaffer}, {Shirokoff}, {Staniszewski},
  {Stark}, {Story}, {Vanderlinde}, {Vieira}, {Williamson}, \& {Wu}}]{mocanu19}
{Mocanu}, L.~M., {Crawford}, T.~M., {Aylor}, K., {et~al.}
\newblock {Consistency of cosmic microwave background temperature measurements
  in three frequency bands in the 2500-square-degree SPT-SZ survey}. 2019,
  \jcap, 2019, 038, \dodoi{10.1088/1475-7516/2019/07/038}

\bibitem[{{Gratton} \& {Challinor}(2019)}]{gratton19}
{Gratton}, S., \& {Challinor}, A.
\newblock {Understanding parameter differences between analyses employing
  nested data subsets}. 2019, arXiv e-prints, arXiv:1911.07754.
\newblock \doarXiv{1911.07754}

\bibitem[{{Freedman} {et~al.}(2019){Freedman}, {Madore}, {Hatt}, {Hoyt},
  {Jang}, {Beaton}, {Burns}, {Lee}, {Monson}, {Neeley}, {Phillips}, {Rich}, \&
  {Seibert}}]{freedman19}
{Freedman}, W.~L., {Madore}, B.~F., {Hatt}, D., {Hoyt}, T.~J., {Jang}, I.-S.,
  {Beaton}, R.~L., {Burns}, C.~R., {Lee}, M.~G., {Monson}, A.~J., {Neeley},
  J.~R., {Phillips}, M.~M., {Rich}, J.~A., \& {Seibert}, M.
\newblock {The Carnegie-Chicago Hubble Program. VIII. An Independent
  Determination of the Hubble Constant Based on the Tip of the Red Giant
  Branch}. 2019, arXiv e-prints, arXiv:1907.05922.
\newblock \doarXiv{1907.05922}

\bibitem[{{Wong} {et~al.}(2020){Wong}, {Suyu}, {Chen}, {Rusu}, {Millon},
  {Sluse}, {Bonvin}, {Fassnacht}, {Taubenberger}, {Auger}, {Birrer}, {Chan},
  {Courbin}, {Hilbert}, {Tihhonova}, {Treu}, {Agnello}, {Ding}, {Jee},
  {Komatsu}, {Shajib}, {Sonnenfeld}, {Bland ford}, {Koopmans}, {Marshall}, \&
  {Meylan}}]{wong19}
{Wong}, K.~C., {Suyu}, S.~H., {Chen}, G. C.~F., {et~al.}
\newblock {H0LiCOW XIII. A 2.4\% measurement of H$_{0}$ from lensed quasars:
  5.3{\ensuremath{\sigma}} tension between early and late-Universe probes}.
  2020, \mnras, \dodoi{10.1093/mnras/stz3094}

\bibitem[{{Birrer} {et~al.}(2020){Birrer}, {Shajib}, {Galan}, {Millon}, {Treu},
  {Agnello}, {Auger}, {Chen}, {Christensen}, {Collett}, {Courbin}, {Fassnacht},
  {Koopmans}, {Marshall}, {Park}, {Rusu}, {Sluse}, {Spiniello}, {Suyu},
  {Wagner-Carena}, {Wong}, {Barnab{\`e}}, {Bolton}, {Czoske}, {Ding},
  {Frieman}, \& {Van de Vyvere}}]{birrer20}
{Birrer}, S., {Shajib}, A.~J., {Galan}, A., {et~al.}
\newblock {TDCOSMO IV: Hierarchical time-delay cosmography -- joint inference
  of the Hubble constant and galaxy density profiles}. 2020, arXiv e-prints,
  arXiv:2007.02941.
\newblock \doarXiv{2007.02941}

\bibitem[{{Bianchini} {et~al.}(2020){Bianchini}, {Wu}, {Ade}, {Anderson},
  {Austermann}, {Avva}, {Beall}, {Bender}, {Benson}, {Bleem}, {Carlstrom},
  {Chang}, {Chaubal}, {Chiang}, {Citron}, {Moran}, {Crawford}, {Crites}, {de
  Haan}, {Dobbs}, {Everett}, {Gallicchio}, {George}, {Gilbert}, {Gupta},
  {Halverson}, {Harrington}, {Henning}, {Hilton}, {Holder}, {Holzapfel},
  {Hrubes}, {Huang}, {Hubmayr}, {Irwin}, {Knox}, {Lee}, {Li}, {Lowitz},
  {Manzotti}, {McMahon}, {Meyer}, {Millea}, {Mocanu}, {Montgomery}, {Nadolski},
  {Natoli}, {Nibarger}, {Noble}, {Novosad}, {Omori}, {Padin}, {Patil}, {Pryke},
  {Reichardt}, {Ruhl}, {Saliwanchik}, {Sayre}, {Schaffer}, {Sievers}, {Simard},
  {Smecher}, {Stark}, {Story}, {Tucker}, {Vanderlinde}, {Veach}, {Vieira},
  {Wang}, {Whitehorn}, \& {Yefremenko}}]{bianchini20}
{Bianchini}, F., {Wu}, W.~L.~K., {Ade}, P.~A.~R., {et~al.}
\newblock {Constraints on Cosmological Parameters from the 500 deg$^{2}$ SPTPOL
  Lensing Power Spectrum}. 2020, \apj, 888, 119,
  \dodoi{10.3847/1538-4357/ab6082}

\bibitem[{{Heymans} {et~al.}(2020){Heymans}, {Tr{\"o}ster}, {Asgari}, {Blake},
  {Hildebrandt}, {Joachimi}, {Kuijken}, {Lin}, {S{\'a}nchez}, {van den Busch},
  {Wright}, {Amon}, {Bilicki}, {de Jong}, {Crocce}, {Dvornik}, {Erben},
  {Getman}, {Giblin}, {Glazebrook}, {Hoekstra}, {Joudaki}, {Kannawadi},
  {Lidman}, {K{\"o}hlinger}, {Miller}, {Napolitano}, {Parkinson}, {Schneider},
  {Shan}, \& {Wolf}}]{heymans20}
{Heymans}, C., {Tr{\"o}ster}, T., {Asgari}, M., {et~al.}
\newblock {KiDS-1000 Cosmology: Multi-probe weak gravitational lensing and
  spectroscopic galaxy clustering constraints}. 2020, arXiv e-prints,
  arXiv:2007.15632.
\newblock \doarXiv{2007.15632}

\bibitem[{{DES Collaboration} {et~al.}(2017){DES Collaboration}, {Abbott},
  {Abdalla}, {Alarcon}, {Aleksi{\'c}}, {Allam}, {Allen}, {Amara}, {Annis},
  {Asorey}, {Avila}, {Bacon}, {Balbinot}, {Banerji}, {Banik}, {Barkhouse},
  {Baumer}, {Baxter}, {Bechtol}, {Becker}, {Benoit-L{\'e}vy}, {Benson},
  {Bernstein}, {Bertin}, {Blazek}, {Bridle}, {Brooks}, {Brout}, {Buckley-Geer},
  {Burke}, {Busha}, {Capozzi}, {Carnero Rosell}, {Carrasco Kind}, {Carretero},
  {Castander}, {Cawthon}, {Chang}, {Chen}, {Childress}, {Choi}, {Conselice},
  {Crittenden}, {Crocce}, {Cunha}, {D'Andrea}, {da Costa}, {Das}, {Davis},
  {Davis}, {De Vicente}, {DePoy}, {DeRose}, {Desai}, {Diehl}, {Dietrich},
  {Dodelson}, {Doel}, {Drlica-Wagner}, {Eifler}, {Elliott}, {Elsner},
  {Elvin-Poole}, {Estrada}, {Evrard}, {Fang}, {Fernandez}, {Fert{\'e}},
  {Finley}, {Flaugher}, {Fosalba}, {Friedrich}, {Frieman},
  {Garc{\'{\i}}a-Bellido}, {Garcia-Fernandez}, {Gatti}, {Gaztanaga}, {Gerdes},
  {Giannantonio}, {Gill}, {Glazebrook}, {Goldstein}, {Gruen}, {Gruendl},
  {Gschwend}, {Gutierrez}, {Hamilton}, {Hartley}, {Hinton}, {Honscheid},
  {Hoyle}, {Huterer}, {Jain}, {James}, {Jarvis}, {Jeltema}, {Johnson},
  {Johnson}, {Kacprzak}, {Kent}, {Kim}, {King}, {Kirk}, {Kokron}, {Kovacs},
  {Krause}, {Krawiec}, {Kremin}, {Kuehn}, {Kuhlmann}, {Kuropatkin}, {Lacasa},
  {Lahav}, {Li}, {Liddle}, {Lidman}, {Lima}, {Lin}, {MacCrann}, {Maia},
  {Makler}, {Manera}, {March}, {Marshall}, {Martini}, {McMahon}, {Melchior},
  {Menanteau}, {Miquel}, {Miranda}, {Mudd}, {Muir}, {M{\"o}ller}, {Neilsen},
  {Nichol}, {Nord}, {Nugent}, {Ogando}, {Palmese}, {Peacock}, {Peiris},
  {Peoples}, {Percival}, {Petravick}, {Plazas}, {Porredon}, {Prat}, {Pujol},
  {Rau}, {Refregier}, {Ricker}, {Roe}, {Rollins}, {Romer}, {Roodman},
  {Rosenfeld}, {Ross}, {Rozo}, {Rykoff}, {Sako}, {Salvador}, {Samuroff},
  {S{\'a}nchez}, {Sanchez}, {Santiago}, {Scarpine}, {Schindler}, {Scolnic},
  {Secco}, {Serrano}, {Sevilla-Noarbe}, {Sheldon}, {Smith}, {Smith}, {Smith},
  {Soares-Santos}, {Sobreira}, {Suchyta}, {Tarle}, {Thomas}, {Troxel},
  {Tucker}, {Tucker}, {Uddin}, {Varga}, {Vielzeuf}, {Vikram}, {Vivas},
  {Walker}, {Wang}, {Wechsler}, {Weller}, {Wester}, {Wolf}, {Yanny}, {Yuan},
  {Zenteno}, {Zhang}, {Zhang}, \& {Zuntz}}]{des17}
{DES Collaboration}, {Abbott}, T.~M.~C., {Abdalla}, F.~B., {et~al.}
\newblock {Dark Energy Survey Year 1 Results: Cosmological Constraints from
  Galaxy Clustering and Weak Lensing}. 2017, ArXiv e-prints.
\newblock \doarXiv{1708.01530}

\bibitem[{{Bocquet} {et~al.}(2019){Bocquet}, {Dietrich}, {Schrabback}, {Bleem},
  {Klein}, {Allen}, {Applegate}, {Ashby}, {Bautz}, {Bayliss}, {Benson},
  {Brodwin}, {Bulbul}, {Canning}, {Capasso}, {Carlstrom}, {Chang}, {Chiu},
  {Cho}, {Clocchiatti}, {Crawford}, {Crites}, {de Haan}, {Desai}, {Dobbs},
  {Foley}, {Forman}, {Garmire}, {George}, {Gladders}, {Gonzalez}, {Grandis},
  {Gupta}, {Halverson}, {Hlavacek-Larrondo}, {Hoekstra}, {Holder}, {Holzapfel},
  {Hou}, {Hrubes}, {Huang}, {Jones}, {Khullar}, {Knox}, {Kraft}, {Lee}, {von
  der Linden}, {Luong-Van}, {Mantz}, {Marrone}, {McDonald}, {McMahon}, {Meyer},
  {Mocanu}, {Mohr}, {Morris}, {Padin}, {Patil}, {Pryke}, {Rapetti},
  {Reichardt}, {Rest}, {Ruhl}, {Saliwanchik}, {Saro}, {Sayre}, {Schaffer},
  {Shirokoff}, {Stalder}, {Stanford}, {Staniszewski}, {Stark}, {Story},
  {Strazzullo}, {Stubbs}, {Vanderlinde}, {Vieira}, {Vikhlinin}, {Williamson},
  \& {Zenteno}}]{bocquet19}
{Bocquet}, S., {Dietrich}, J.~P., {Schrabback}, T., {et~al.}
\newblock {Cluster Cosmology Constraints from the 2500 deg$^{2}$ SPT-SZ Survey:
  Inclusion of Weak Gravitational Lensing Data from Magellan and the Hubble
  Space Telescope}. 2019, \apj, 878, 55, \dodoi{10.3847/1538-4357/ab1f10}

\bibitem[{{Alam} {et~al.}(2017){Alam}, {Ata}, {Bailey}, {Beutler}, {Bizyaev},
  {Blazek}, {Bolton}, {Brownstein}, {Burden}, {Chuang}, {Comparat}, {Cuesta},
  {Dawson}, {Eisenstein}, {Escoffier}, {Gil-Mar{\'{\i}}n}, {Grieb}, {Hand},
  {Ho}, {Kinemuchi}, {Kirkby}, {Kitaura}, {Malanushenko}, {Malanushenko},
  {Maraston}, {McBride}, {Nichol}, {Olmstead}, {Oravetz}, {Padmanabhan},
  {Palanque-Delabrouille}, {Pan}, {Pellejero-Ibanez}, {Percival}, {Petitjean},
  {Prada}, {Price-Whelan}, {Reid}, {Rodr{\'{\i}}guez-Torres}, {Roe}, {Ross},
  {Ross}, {Rossi}, {Rubi{\~n}o-Mart{\'{\i}}n}, {Saito}, {Salazar-Albornoz},
  {Samushia}, {S{\'a}nchez}, {Satpathy}, {Schlegel}, {Schneider},
  {Sc{\'o}ccola}, {Seo}, {Sheldon}, {Simmons}, {Slosar}, {Strauss}, {Swanson},
  {Thomas}, {Tinker}, {Tojeiro}, {Maga{\~n}a}, {Vazquez}, {Verde}, {Wake},
  {Wang}, {Weinberg}, {White}, {Wood-Vasey}, {Y{\`e}che}, {Zehavi}, {Zhai}, \&
  {Zhao}}]{alam17}
{Alam}, S., {Ata}, M., {Bailey}, S., {et~al.}
\newblock {The clustering of galaxies in the completed SDSS-III Baryon
  Oscillation Spectroscopic Survey: cosmological analysis of the DR12 galaxy
  sample}. 2017, \mnras, 470, 2617, \dodoi{10.1093/mnras/stx721}

\bibitem[{{Blomqvist} {et~al.}(2019){Blomqvist}, {du Mas des Bourboux},
  {Busca}, {de Sainte Agathe}, {Rich}, {Balland}, {Bautista}, {Dawson},
  {Font-Ribera}, {Guy}, {Le Goff}, {Palanque-Delabrouille}, {Percival},
  {P{\'e}rez-R{\`a}fols}, {Pieri}, {Schneider}, {Slosar}, \&
  {Y{\`e}che}}]{blomqvist19}
{Blomqvist}, M., {du Mas des Bourboux}, H., {Busca}, N.~G., {et~al.}
\newblock {Baryon acoustic oscillations from the cross-correlation of
  Ly{\ensuremath{\alpha}} absorption and quasars in eBOSS DR14}. 2019, \aap,
  629, A86, \dodoi{10.1051/0004-6361/201935641}

\bibitem[{{Calabrese} {et~al.}(2008){Calabrese}, {Slosar}, {Melchiorri},
  {Smoot}, \& {Zahn}}]{calabrese08}
{Calabrese}, E., {Slosar}, A., {Melchiorri}, A., {Smoot}, G.~F., \& {Zahn}, O.
\newblock {Cosmic microwave weak lensing data as a test for the dark universe}.
  2008, \prd, 77, 123531, \dodoi{10.1103/PhysRevD.77.123531}

\bibitem[{{Knox} \& {Millea}(2020)}]{knox19}
{Knox}, L., \& {Millea}, M.
\newblock {Hubble constant hunter's guide}. 2020, \prd, 101, 043533,
  \dodoi{10.1103/PhysRevD.101.043533}

\bibitem[{Pordes {et~al.}(2007)}]{pordes07}
Pordes, R., {et~al.}
\newblock {The Open Science Grid}. 2007, J. Phys. Conf. Ser., 78, 012057,
  \dodoi{10.1088/1742-6596/78/1/012057}

\bibitem[{{Sfiligoi} {et~al.}(2009){Sfiligoi}, {Bradley}, {Holzman},
  {Mhashilkar}, {Padhi}, \& {Wurthwein}}]{sfiligoi09}
{Sfiligoi}, I., {Bradley}, D.~C., {Holzman}, B., {Mhashilkar}, P., {Padhi}, S.,
  \& {Wurthwein}, F.
\newblock The Pilot Way to Grid Resources Using glideinWMS. 2009, in 2, Vol.~2,
  2009 WRI World Congress on Computer Science and Information Engineering,
  428--432, \dodoi{10.1109/CSIE.2009.950}

\bibitem[{Hunter(2007)}]{hunter07}
Hunter, J.~D.
\newblock Matplotlib: A 2D graphics environment. 2007, Computing In Science \&
  Engineering, 9, 90, \dodoi{10.1109/MCSE.2007.55}

\bibitem[{Jones {et~al.}(2001)Jones, Oliphant, Peterson, {et~al.}}]{jones01}
Jones, E., Oliphant, T., Peterson, P., {et~al.} 2001, {SciPy}: Open source
  scientific tools for {Python}.
\newblock \url{http://www.scipy.org/}

\bibitem[{van~der Walt {et~al.}(2011)van~der Walt, Colbert, \&
  Varoquaux}]{vanDerWalt11}
van~der Walt, S., Colbert, S., \& Varoquaux, G.
\newblock The NumPy Array: A Structure for Efficient Numerical Computation.
  2011, Computing in Science Engineering, 13, 22, \dodoi{10.1109/MCSE.2011.37}

\end{thebibliography}

\end{document}